%% file: regularize_SCM.tex
\documentclass[graybox]{svmult}
\usepackage{amsmath, amssymb, amsfonts}
\usepackage[colorlinks=true, pdfstartview=FitV, linkcolor=black, citecolor=blue, breaklinks=true, urlcolor=red]{hyperref}
\usepackage{tikz}
\usepackage{pgfplots}
\usepackage{subfig}
\pgfplotsset{compat=1.13}
\usepackage{mathtools}
\DeclarePairedDelimiterX{\ip}[2]{\langle}{\rangle_\Fr}{#1, #2}

\usepackage{type1cm}        
\usepackage{makeidx}         
\usepackage{graphicx}        
\usepackage{multicol}        
\usepackage[bottom]{footmisc}

\usepackage{newtxtext}       %
\usepackage[varvw]{newtxmath}       

\makeindex             

\pgfplotsset{mystyleportfolio/.style={%
    width=5.8cm, 
    height=5.2cm,
    xlabel near ticks,
    ylabel near ticks,
    tick label style={font=\footnotesize},
    label style={font=\footnotesize},
    xmajorgrids,
ymajorgrids,
	yticklabel style={%
		/pgf/number format/precision=3,
		/pgf/number format/fixed},
	every axis y label/.style={%
		at={(rel axis cs:-0.27,0.5)},			
		rotate=90,								
		font=\footnotesize},					
    legend style={font=\footnotesize,
                  draw=none,					
                  overlay,at={(-0.3567,-.5)},	
                  anchor=north west},			
    legend columns=-1,							
}}
\pgfplotsset{mystyle/.style={%
     width=5.8cm, 
    height=5.2cm,
    xlabel near ticks,
    ylabel near ticks,
    tick label style={font=\footnotesize},
    label style={font=\footnotesize},
    xmajorgrids,
ymajorgrids,
	yticklabel style={%
		/pgf/number format/precision=3,
		/pgf/number format/fixed},
	every axis y label/.style={%
		at={(rel axis cs:-0.25,0.5)},			
		rotate=90,								
		font=\footnotesize},					
    legend style={font=\footnotesize,
                  draw=none,					
                  overlay,at={(-0.3567,-.45)},	
                  anchor=north west},			
    legend columns=-1,							
}}

\newcommand{\var}{\mathsf{var}}
\newcommand{\kurt}{\mathsf{kurt}}
\newcommand{\bias}{\mathsf{bias}}
\newcommand\cv{{\mathbf c}}

\newcommand{\expec}{\mathsf{E}}

\newcommand{\pdim}{p}
\newcommand{\ndim}{n}
\newcommand{\M}{\boldsymbol{\Sigma}} 
\newcommand{\bo}[1]{\mathbf{#1}}              
\newcommand{\bom}[1]{\boldsymbol{#1}}    

\newcommand{\be}{\beta}
\newcommand{\al}{\alpha}  
\newcommand{\ka}{\kappa}

\newcommand{\iidsim}{\overset{iid}{\sim}}

\newcommand{\beq}{\begin{equation}}
\newcommand{\eeq}{\end{equation}}
\newcommand{\bmat}{\begin{pmatrix}}
\newcommand{\emat}{\end{pmatrix}}

\newcommand{\V}{\mathbf{V}}
\renewcommand{\S}{\mathbf{S}}
\newcommand{\R}{\mathbb{R}}

\newcommand{\A}{\mathbf{A}}
\newcommand{\W}{\mathbf{W}}
\newcommand{\C}{\bo C}
\newcommand{\T}{\mathbf{T}}

\newcommand{\fR}{\mathbb{R}}

\newcommand{\x}{\bo x}

\renewcommand{\a}{\bo a}
\newcommand{\w}{\mathbf{w}}

\newcommand{\Fr}{\mathrm{F}}
\newcommand{\MSE}{\mathsf{MSE}}
\newcommand{\NMSE}{\mathsf{NMSE}}

\DeclareMathOperator{\tr}{tr}
\DeclareMathOperator{\Tr}{tr}
\DeclareMathOperator{\diag}{diag}
\newcommand{\cov}{\mathsf{cov}} 
\newcommand{\Sym}[1]{\mathbb{R}_{\mathrm{Sym}}^{#1 \times #1}}
\renewcommand{\time}{t}
\newcommand{\tabasco}{\textsc{Tabasco}{}}
\newlength\fwidth

\begin{document}

\title*{Linear shrinkage of sample covariance matrix or matrices under elliptical distributions: a review}
\author{Esa Ollila}
\institute{Esa Ollila \at Aalto University, Department of Information and Communications Engineering, Finland \email{esa.ollila@aalto.fi}}
%
%
\maketitle

\abstract*{This chapter reviews methods for linear shrinkage of the sample covariance matrix (SCM)  and matrices (SCM-s) under elliptical distributions in single and multiple populations settings, respectively. In the single sample setting  a popular  linear shrinkage estimator is defined as a linear combination of the sample covariance matrix (SCM) with a scaled identity matrix. The optimal shrinkage coefficients minimizing the mean squared error (MSE) under elliptical sampling are shown to be functions of few key parameters only, such as elliptical kurtosis and sphericity parameter.  Similar results and estimators are derived for multiple population setting and applications of the studied shrinkage estimators are illustrated  in portfolio optimization.}

\abstract{This chapter reviews methods for linear shrinkage of the sample covariance matrix (SCM)  and matrices (SCM-s) under elliptical distributions in single and multiple populations settings, respectively. In the single sample setting  a popular  linear shrinkage estimator is defined as a linear combination of the sample covariance matrix (SCM) with a scaled identity matrix. The optimal shrinkage coefficients minimizing the mean squared error (MSE) under elliptical sampling are shown to be functions of few key parameters only, such as elliptical kurtosis and sphericity parameter.  Similar results and estimators are derived for multiple population setting and applications of the studied shrinkage estimators are illustrated  in portfolio optimization.}

\section{Introduction} 

Consider a set of $p$-dimensional (real-valued) vectors $\mathcal X=\{ \x_i \}_{i=1}^n$ sampled from a distribution of a random vector $\x$ with unknown mean vector $\bom \mu=\expec[\x]$ and unknown positive definite symmetric (PDS) $p \times p$ covariance matrix $\M \equiv \mbox{cov}(\x)=\expec[(\x-\bom \mu)(\x - \bom \mu)^\top]$. 
A popular estimate of $\M$ is the \emph{sample covariance matrix (SCM)}, defined by  
\beq \label{eq:SCM}
\bo S = \frac{1}{n-1} \sum_{i=1}^n ( \x_i - \bar \x)(\x_i - \bar \x)^\top . 
\eeq 
where $\bar \x= \frac 1 n \sum_{i=1}^n \x_i$ denotes the sample mean vector.   It has some favourable properties such as being unbiased.  i.e., $\expec[\S] = \M$, and its scaled version 
$\S_{\textup{ML}} = [(n-1)/n] \cdot \S$ is the maximum likelihood estimator of the covariance matrix when the samples are  independent and identically distributed (i.i.d.) from a multivariate normal (MVN) distribution $\mathcal N_p(\boldsymbol{\mu},\M)$. 

In many applications, the estimation accuracy (or another performance criterion) can alternatively be improved by using a so-called \textit{tapered} SCM.
Such estimate is defined as $ \bo W \circ \bo{S}$, where $\circ$ denotes the Hadamard (or Schur) element-wise product, and where $\bo W$ is a \textit{tapering} matrix (also referred to as covariance matrix taper), i.e., a template that imposes some additional structure to the SCM.  Note that above $(\bo{W} \circ \bo{S})_{ij} = w_{ij} s_{ij}$ for $(\bo{W})_{ij} = w_{ij} $ and $ (\bo{S})_{ij} = s_{ij} $. 
Covariance matrix tapers have found applications in diverse fields. 
For example, the true covariance matrix may be known to have a diagonally dominant structure (e.g., in autoregressive models). 
This means that the variables have a natural order in the sense that $|i -j |$ large implies that the correlation between the $i$th and the $j$th variables is close to zero. 
In this settings, popular estimation approaches are to use a banding-type tapering matrices such as thresholding \cite{bickel2008regularized,bickel2008covariance}: 
\beq \label{eq:Wband}
(\bo{W})_{ij} = \begin{cases} 
 1, & | i - j | < k \\ 
0, & | i - j | \geq k
\end{cases} 
\eeq 
for some integer $k \in [\![1,p]\!] = \{1,\ldots,p\}$ called the \emph{bandwidth} parameter.  Other types of template matrices are also possible, see \cite{ollila2022regularized}. 

Let $\hat \M$ denote an estimator of  $\M$  based on a 
sample $\mathcal X$.  It is now well-known that an estimator that performs better than $\hat \M$ can be easily constructed  using the concept called regularization or shrinkage  which leverages on the concept called \emph{bias-variance tradeoff}.  The key idea  in shrinkage/regularization is to shift (or shrink) the estimator towards a predetermined target or model. The principle is  to decrease the variance of the estimator while introducing some bias,  and thus improving the overall performance of the estimation by reducing its \emph{mean squared error (MSE)},  defined as 
\beq \label{eq:MSE} 
\MSE(\hat\M)   = \expec\left[ \| \hat \M - \M \|^2_{\Fr} \right]   ,
\eeq 
where $\| \cdot \|_{\Fr}$ denotes the Frobenius matrix norm,  $\| \A \|_{\Fr} = \sqrt{\tr(\A^\top \A)}$  for any matrix $\A$ and  $\tr(\cdot)$ denotes the matrix trace, $\tr(\A)=\sum_{i=1}^p a_{ii}$, for any square matrix $\A$. 
Recall that bias of $\hat \M$ is defined as 
\[
\mathsf{bias}(\hat \M)= \M - \expec[\hat \M]
\]
and an estimator is called unbiased iff $\bias(\hat \M)=\bo 0$. This reduction in MSE can be understood via the bias-variance decomposition of 
the MSE:
\begin{align} \label{eq:bias_variance}
\MSE(\hat \M) = \expec\left[\|\hat \M - \expec[\hat\M] \|_{\Fr}^2\right] + \| \mathsf{bias}(\hat \M) \| ^2_{\Fr},
\end{align}
where the first term on the right-hand side is the \emph{total variance} and the second
term is the squared \emph{total bias} of the estimator.  If the estimator $\hat \M$ is unbiased, then its MSE
is equal to its total variance. By using a shrinkage estimator, say $\hat \M({\be})$, where $\be>0$ is some tuning parameter that introduces some bias to the estimator $\hat \M$, it is
possible to reduce its MSE significantly given that the total variance is reduced in larger extent. This will be illustrated in detail in Section~\ref{sec:bias_var_tradeoff}. 

In order to be able to derive MSE-optimal shrinkage parameters and their estimates under the assumption that data $\mathcal X$ is generated from an elliptically symmetric (ES) distribution, one needs to derive the moments of the SCM or tapered SCM, such as its normalized MSE (NMSE). These results as well as some key parameters, the elliptical kurtosis and a measure of sphericity, are defined and elaborated in Section~\ref{sec:NMSE_SCM}. 

Shrinkage estimation was introduced by Stein in the context of improved estimation of the mean in  his  seminal
works \cite{stein1981estimation,stein1956some}. These ideas were further studied in \cite{james1961estimation,efron1973stein}. 
This chapter reviews linear shrinkage estimators of SCM(-s) in single and multiple covariance matrices estimation problems. One of the earliest reference studying a linear shrinkage estimator  is \cite{haff1980empirical}.  
 A linear shrinkage estimator can often  be represented in the form  
 \beq \label{eq:lin_shrink}
 \hat \M(\beta,\alpha)= \beta \mathbf{S} +  \alpha  \hat \eta \mathbf{T}
 \eeq 
  where $\mathbf{T}$ is positive definite symmetric target matrix, $\alpha$ and $\beta$ are tuning parameters, while $\hat \eta$ is a \emph{scale statistics}\footnote{Formally, $\eta \equiv \eta(\M)$ is  a \emph{scale} parameter if it verifies $\eta(\mathbf{I}) = 1$ and  $\eta(a\boldsymbol{\Sigma}) = a \eta(\boldsymbol{\Sigma})$ for all $a>0$ \cite{paindaveine2008canonical}. Then $\hat \eta$ is statistic that estimates this parameter based on data $\mathcal X$.}
  such as $\hat \eta = \tr(\S)/p$ or $\hat \eta = p/\tr(\S^{-1})$. In \eqref{eq:lin_shrink} the SCM is pulled or shrunk toward a predetermined or estimated target structure $\mathbf{T}$, which may be 
chosen based on prior assumptions about the data at hand.  Choosing $\mathbf{T}$ as the dentity matrix ($\mathbf{T} = \mathbf{I}$)  implies having no a priori knowledge of the shape of the data cloud. One such  estimator, defined as 
 $\hat \M(\beta,\alpha)= \beta \mathbf{S} +  \alpha  \hat \eta \mathbf{I}$ with $\hat \eta = \tr(\S)/p$ was proposed in \cite{ledoit2004well}.   This estimator will be described in more detail in Section~\ref{sec:RegSCM}, where the  MSE optimal estimator is considered when $\mathcal X$  follows an unspecified ES distribution. 
     Shrinkage estimation of the form  $ \mathbf{S} +  \alpha \mathbf{I}$ (so $\beta=\hat \eta=1$, $\mathbf{T} = \mathbf{I}$)   is often referred to as "diagonal loading"  in signal processing literature \cite{carlson1988covariance,li2003robust,du2010fully}. 
  
   Different target matrices $\T$ have been considered in the literature. For example, \cite{ledoit2003improved} 
used  a  target matrix following a single-index market factor  model whose motivation stems from portfolio optimization and capital asset pricing model (CAPM),  
while a constant correlation model was adopted as the target  matrix in \cite{ledoit2004honey}.   
It is also possible to shrink toward multiple target matrices simultaneously as proposed in \cite{lancewicki2014multi,bartz2014multi,tong2018linear,raninen2021linear}.  
Such multi-target shrinkage
covariance matrix estimators are defined by
\begin{equation}\label{eq:MTS}
    \hat \M(\a) = a_0 \S + \sum_{k=1}^K a_k \T_k,
\end{equation}
where $\T_k$, $k=1,\ldots,K$, are {linearly independent} target PDS matrices
and $a_j$, $j=0,\ldots,K$, are the regularization coefficients. It is also common to impose some restrictions on the parameters 
such as non-negativity $a_k \geq 0$,  and scale constraints, such as $\sum_{k=1}^K a_k \leq
1$ for $k=1,\ldots,K$ and $a_0 = 1-\sum_{k=1}^K a_k$, as in \cite{lancewicki2014multi,bartz2014multi}.

In the multiple population setting, regularization via pooling the information
in the different class samples is also possible.
 For example,~\cite{besson2020maximum} considered covariance matrix estimation from two independent data sets, whose covariance matrices are
different but close to each other.  In discriminant analysis classification,
the pooled SCM, $\S_{\text{pool}} = \frac{1}{n}\sum_{k=1}^K n_k \S_k$, $n =
\sum_{k=1}^K n_k$, is often used as a shrinkage target and the class covariance
matrices are estimated via a convex combination $\hat \M_k = a \S_k + (1-a)
\S_{\text{pool}}$, where $a \in [0,1]$. This was studied in a Bayesian framework
in~\cite{greene1989partially} and~\cite{rayens1991covariance}, and in the
Regularized Discriminant Analysis (RDA) framework
in~\cite{friedman1989regularized}. 
In this chapter, we consider more general multiple population linear shrinkage settings. First we consider the coupled linear shrinkage approach  \cite{raninen2021coupled}, where the SCM  of $k$th sample is first linearly shrinked with pooled SCM  $\S_{\text{pool}}$, and this estimator is then shrinked towards scaled identity matrix to guarantee positive-definiteness.  The optimal coefficients are estimated that minimize the MSE under the assumption that data are sampled from unknown (unspecified) elliptical distributions. Then we consider more general approach, where  the covariance matrix estimator of the $k$th class is formed as linear combination of all class SCM-s where coefficients that minimize the MSE are estimated  similarly under the elliptical distribution assumption. These developments are discussed in Section~\ref{sec:multiple_SCM}.   
Application to portfolio selection in finance is provided in \autoref{sec:portfolio}. 
Finally, Section~\ref{sec:concl} concludes. 

\section{Bias-variance tradeoff and shrinkage} \label{sec:bias_var_tradeoff}

To illustrate the idea of shrinkage estimators of covariance matrix, consider the simplest possible shrinkage estimator 
\[
\hat \M(\be) = \be \hat \M, 
\]
where $\beta>0$ is a shrinkage parameter that can be optimally tuned and  
$\hat \M$ is some unbiased estimator of  $\M$ such as the SCM, so verifying $\expec[\hat \M] = \M$.  
First note that $\hat \M({\be})$ is obviously biased for any $\be \neq 1$, the bias being
\beq \label{eq:bias_of_Mbe} 
\mathsf{bias}[\hat \M({\be})]=   \M - \expec[ \be\hat \M] = (1-\be) \M.
\eeq 
It is yet possible to improve on the MSE by seeking  an optimal constant $\be_{\textup{o}}$ such that $\hat \M_{\textup{o}} = \beta_o \hat \M$ attains a smaller MSE than $\hat \M$, i.e., 
\beq \label{eq:smaller_MSE} 
\MSE(\hat \M_{\textup{o}} ) < \MSE(\hat \M) \quad \mbox{for any $\M \succ 0 $ } .
 \eeq 
 This is equivalent to saying that $\hat \M_{\textup{o}}$ is more efficient estimator than $\hat \M$ (regardless of the structure of the true underlying covariance matrix $\M$). 
Now consider finding the optimal scaling term as
\[
\be_{\textup{o}} = \arg \min_{\beta > 0} \expec \big[ \| \be \hat \M  - \M \|^2_{\Fr} \big].
\] 
Due to \eqref{eq:bias_variance} and \eqref{eq:bias_of_Mbe}, we have that  
 \beq \label{eq:optimal_beta_apu}
 \MSE(\hat \M (\be))= \expec \big[  \| \be \hat \M  - \M \|^2_{\Fr}    \big] =  \be^2 \MSE( \hat \M) + (1-\be)^2 \| \M \|_{\Fr}^2. 
 \eeq 
Since $f(\be)=\MSE(\hat \M_\be)$ is a  strictly convex quadratic function, we can easily find the minimum $\beta_{\textup{o}}$ of $f(\be)$ as solution of $f'(\be)=0$, which gives
 \begin{align}
 \be_{\textup{o}} &= \frac{ \| \M \|^2_{\Fr}}{ \MSE(\hat \M) + \| \M \|^2_{\Fr}} 
 = \frac{ 1}{1+ \NMSE(\hat \M)} , \label{eq:optimal_beta_eq2}
 \end{align}
 where 
\beq \label{eq:NMSE} 
  \mathrm{NMSE}(\hat \M) = \frac{\expec \big[ \| \hat \M - \M \|_\Fr^2 \big]}{\| \M \|^2_\Fr} 
\eeq 
is  the \emph{normalized MSE (NMSE)} of $\hat \M$. \index{normalized MSE} Equation \eqref{eq:optimal_beta_eq2}  shows that  $\beta_{\textup{o}}<1$ since 
$\NMSE(\hat \M)>0$. It not yet clear, however, if \eqref{eq:smaller_MSE} holds. We prove this next. 

First, note from \eqref{eq:optimal_beta_apu} that 
\beq \label{eq:optimal_MSE_eq1} 
\MSE(\hat \M_{\textup{o}}) = \be_{\textup{o}}^2 \MSE(\hat \M) + (1- \be_{\textup{o}})^2 \|\M \|_{\Fr}^2.
\eeq 
Then subsituting  
\[
1- \be_{\textup{o}} = 1-  \frac{ 1}{1+ \NMSE(\hat \M)} = \frac{ \NMSE(\hat \M)}{1+ \NMSE(\hat \M)} 
= \be_{\textup{o}} \NMSE(\hat \M) 
\]
into \eqref{eq:optimal_MSE_eq1} yields 
\begin{align} 
\MSE(\hat \M_{\textup{o}})  
&=  \be_{\textup{o}}^2 \MSE(\hat \M) +  \be_{\textup{o}}^2 \{\NMSE(\hat \M)\}^2  \cdot \| \M \|_{\Fr}^{2} \notag \\ 
&= \be_{\textup{o}}^2 \MSE(\hat \M) + \be_{\textup{o}}^2 \NMSE(\hat \M) \cdot \MSE(\hat \M) \notag \\ 
&= \be_{\textup{o}}^2 \MSE(\hat \M) \big(1+ \NMSE(\hat \M) \big) \notag \\ 
&= \be_{\textup{o}} \MSE(\hat \M) \label{eq:MSE_Mo}
\end{align} 
where the last identity follows from  $1/\be_{\textup{o}} = 1+ \NMSE(\hat \M) $ due to \eqref{eq:optimal_beta_eq2}. 
Since $\be_{\textup{o}} <1$ for all $\M \succ 0$, it thus follows that \eqref{eq:smaller_MSE}  holds, and thus $ \be_{\textup{o}} \hat \M$ is more efficient estimator than 
$\hat \M$. It is important to observe that this does not hold just for SCM, but for any unbiased estimator $\hat \M$ of $\M$.

We now illustrate this fundamental result in the  $1$-dimensional case ($p=1$). In this case the covariance matrix $\M$ is equal to variance $\sigma^2 = \var(x)$ of a random variable $x \in \mathbb{R}$. 
Suppose we have a random sample $x_1,\ldots,x_n$ distributed as $x$. The \emph{sample variance} is defined as
\beq \label{eq:Ssq}
s^2 = \frac{1}{\ndim-1} \sum_{i=1}^\ndim (x_i - \bar x)^2 
\eeq 
where $\bar x = \frac{1}{\ndim} \sum_{i=1}^{\ndim} x_i$ denotes the sample mean.  Since $s^2$ is an unbiased estimator of $\sigma^2$, we have that 
\beq \label{eq:var_s2}
\MSE(s^2)=\var(s^2)  =  \sigma^4 \Big( \frac{\kurt(x)}{\ndim} + \frac{2}{\ndim-1} \Big) ,
\eeq 
where $\kurt(x)$ denotes the (excess) \emph{kurtosis}  of a random variable $x$, defined as 
\beq \label{eq:kurtosis} 
\kurt(x) = \frac{\expec[ (x-\mu)^4]}{\sigma^4} - 3 .
\eeq 
Let us now consider the shrinkage estimator $\hat \sigma^2(\be)= \be s^2$. Due to \eqref{eq:MSE_Mo} we know that 
$\hat \sigma^2_{\textup{o}} = \beta_{\textup{o}} s^2$ where $\beta_{\textup{o}} < 1$, is always more efficient estimator than the sample variance since 
\[
\MSE(\hat \sigma^2_{\textup{o}} ) = \be_{\textup{o}} \MSE(s^2) < \MSE(s^2)  \quad \mbox{for any $\sigma^2>0$.} 
\]
Using \eqref{eq:optimal_beta_eq2} and \eqref{eq:var_s2}, 
the optimal scaling constant $\be_{\textup{o}}$ that minimizes  $\expec[ ( \be s^2 - \sigma^2)^2]$ can be expressed compactly as 
\begin{align*}
\be_{\textup{o}}
&= \frac{\sigma^4}{ \var(s^2) + \sigma^4} =  \frac{n(n-1)}{ \kurt(x) (n-1) +n(n+1)}.
\end{align*}
 For example, if the data is from a Gaussian distribution ($x \sim \mathcal N(\mu,\sigma^2$)), then $\kurt(x)=0$, and $\be_{\textup{o}} = (n-1)/(n+1)$, and hence
\[
\hat \sigma^2_{\textup{o}} =  \frac{1}{\ndim+1} \sum_{i=1}^\ndim (x_i - \bar x)^2 
\]
is always more efficient estimator than the sample variance $s^2$ for Gaussian samples. 

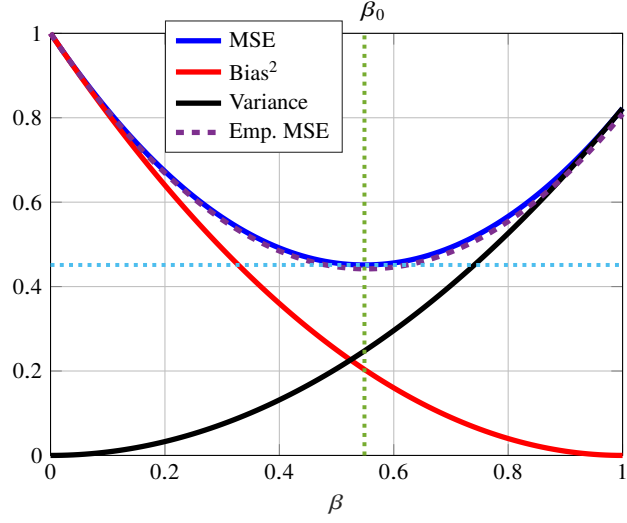
\begin{figure}[!t]
\setlength\fwidth{\textwidth}
\centering
\input{tikz/bias_var.tex}
\caption{The graphs of MSE,  squared bias and the variance of  a shrinkage estimator $\hat \sigma^2(\be)= \be s^2$ when sampling  from a $t$-distribution of unit variance and d.o.f. $\nu=5$. Sample size is $n=10$. The minimum  $\MSE(\hat \sigma^2(\beta_{\textup{o}}))$  is indicated via dotted horizontal line and the value of the optimum $\be_{\textup{o}}$ via dotted vertical line.} \label{fig:bias_var}
\end{figure}

For Gaussian data, $\beta_{\textup{o}} \approx 1$, but if the kurtosis is large and positive and $n$ is small,  the optimal shrinkage factor $\be_{\textup{o}}$  can be significantly smaller than 1.   For example, consider the case that  data is from a  standard ($\mu=0,\sigma=1$)  $t$-distribution with $\nu>4$ degrees of freedom (d.o.f.) and unit variance.  In this case the kurtosis is  $\kurt(x) = 6/(\nu-4)$.  
\autoref{fig:bias_var} depicts the graphs of MSE, squared bias and variance of $\hat \sigma^2(\be)$ 
as a function of $\beta\in [0,1]$ when  $\ndim=10$ and $\nu=5$. 
Recall the connections between these quantities through the bias-variance decomposition, 
\[
\MSE(\hat \sigma^2(\be))= \var( \hat \sigma^2(\be))+ \bias( \hat \sigma^2(\be))^2. 
\] 
The minimum MSE of $\hat \sigma_{\textup{o}}^2$ 
is identified as dotted horizontal line and the optimum $\be_{\textup{o}}$ as a dotted vertical line in the plot.   
We also computed the empirical MSE averaged over 20000 MC trials. The following conclusions  can be drawn. The sample variance $s^2$ needs to be shrunked nearly by a factor $\be_{\textup{o}} \approx 1/2$ which is substantial scaling. For $\be$ close to 1, the bias goes to zero (as expected) while the bias increases when $\be$ descends towards 0. The opposite effect is seen in the variance. 
Optimal tradeoff is obtained by using $\hat \sigma^2_{\textup{o}}=\be_{\textup{o}} s^2$. 
Morever, one notices that a significant improvement  in MSE can be attained by using the MSE-optimal scaled estimator $\be_{\textup{o}} s^2$. One can also notice that the empirical MSE curve has a good match  with the theoretical MSE curve.

\section{NMSE of SCM under elliptical distributions}  \label{sec:NMSE_SCM}

We remind the reader that  a 
random vector is said to have an \emph{elliptically symmetric (ES) distribution} if and only if admits \emph{stochastic representation} ~\cite{fang1990symmetric},
\begin{equation}\label{eq:stochasticrepresentation}
    \x = \boldsymbol{\mu} + r \M^{1/2} \mathbf{u},
\end{equation}
with $\mathbf{u}$ having  a uniform distribution on the
unit sphere  $S^{p-1}=\mathbf{u} \in \{\mathbf{z} \in \fR^p :
\|\mathbf{z}\|=1\}$  and  $r\geq 0$ being  a random variable independent of $\mathbf{u}$.  The variable $r$  is  called the \emph{modular variate} and 
due \eqref{eq:stochasticrepresentation} it verifies 
\beq \label{eq:r2} 
r^2 = (\x - \boldsymbol{\mu})^\top
\M^{-1} (\x - \boldsymbol{\mu}). 
\eeq 
 The parameter  $\boldsymbol{\mu} \in \mathbb{R}^p$ is the \emph{symmetry center} and $\M$ is a PDS $p \times p$ matrix  parameter, called the \emph{scatter matrix}.  We  assume $\x$ is an absolutely
continuous random vector $\x \in \fR^p$ and has finite 4th order moments. Thus it has a probability density function  (p.d.f.) up to a constant of the form
\begin{equation*}
    |\M|^{-1/2} g( (\x - \boldsymbol{\mu})^\top \M^{-1} (\x - \boldsymbol{\mu}) ),
\end{equation*}
where $g:\fR_{\geq 0} \to \fR_{> 0}$ is called the \emph{density
generator}   which we without any loss of generality assume to verify  $C^{-1} \int_0^\infty t^{p/2} g(t) \mathrm{d} t = p$, where 
$C = \int_0^\infty t^{p/2-1} g(t) \mathrm{d} t$ which is equivalent\footnote{This can be done due to scaling ambiguity of  \eqref{eq:stochasticrepresentation}:  the  scale of $r$ can absorbed in $\M$, and thus a scale constraint on $r$ (or $\M$) should  be imposed for uniquely parametrizing the elliptical distribution when $g$ is not specified.} to assuming that $\expec[r^2]=p$. 
We write $\x \sim \mathcal E_p(\boldsymbol{\mu},\M,g)$ to denote this case.  

The symmetry center $\boldsymbol{\mu}$ is equal to the mean vector $ \boldsymbol{\mu} = \expec[\x]$ and  $\M$ represents the covariance matrix  $\M = \cov(\x)$. 
 For example, the MVN distribution $\mathcal N_p(\boldsymbol{\mu},\M)$  is a particular instance of
the elliptical distribution with $g(t) = \exp(-t/2)$.  Sometimes we are only interested in the covariance matrix up to a scaling
constant. Hence, we define the \emph{shape matrix} as 
\begin{equation*}
    \boldsymbol{\Lambda} = p \frac{\M}{\tr(\M)}, 
\end{equation*}
which verifies $\tr(\boldsymbol{\Lambda} )=p$.   

Two key scalar population parameters in this chapter regarding $\M$ are
the \emph{scale} and the \emph{sphericity}. The scale 
\beq \label{eq:scale} 
\eta= \frac{\tr(\M)}{p}  =   \frac 1 p \sum_{i=1}^p \lambda_i  
\eeq 
 is the mean of the eigenvalues $\lambda_1, \ldots, \lambda_p$ of $\M$. 
The sphericity  is  defined as 
\begin{equation}\label{eq:gamma}
    \gamma = \frac{p \tr(\boldsymbol{\Sigma}^2)}{\tr(\boldsymbol{\Sigma})^2} 
    =  \frac{\| \boldsymbol{\Lambda} \|^2_{\Fr}}{p} =    \frac{ \frac 1 p \sum_{i=1}^p \lambda_i^2 }{ \left( \frac  1 p \sum_{i=1}^p \lambda_i \right)^2} .
\end{equation}
 Thus the sphericity measure \eqref{eq:gamma} is  the ratio of the mean of the squared eigenvalues of $\boldsymbol{\Sigma}$ relative to the
mean of its eigenvalues squared.  Letting  $s^2_{\lambda} = \frac 1 p \sum_{i=1}^p (\lambda_i - \eta)^2$  denote the sample variance of the eigenvalues, we may express $\gamma$ as  
\begin{align*} 
\gamma &= 1+  \frac{s^2_{\lambda}}{\eta^2}  = 1+  \frac{1}{p} \| \boldsymbol{\Lambda} - \mathbf{I} \|_{\Fr}^2 .
\end{align*} 
Thus the sphericity measures how close $\M$ is to a scaled identity matrix  
or how concentrated the eigenvalues are around their mean value $\eta$.  In fact, $\gamma \in [1,p]$,  where $\gamma=1$ if and only if $\M \propto \mathbf{I}$ and $\gamma = p$ if and only if $\M$ has its rank equal to 1.  The fact that  $\gamma$  is lower bounded by $\gamma \leq p$ is easiest seen by recalling the submultiplicativity of the matrix trace; namely, for any positive semidefinite matrices $\mathbf{A}$  and $\mathbf{B}$, it holds that $\tr(\mathbf{A}\mathbf{B}) \leq \tr(\mathbf{A}) \tr(\mathbf{B})$. Thus $\tr(\boldsymbol{\Sigma}^2) \leq \tr(\boldsymbol{\Sigma})^2$ and consequently $\gamma = p \tr(\M^2)/\tr(\M)^2 \leq p$. 

A statistical variable describing the heavy-tailedness of the elliptical distribution is  \emph{elliptical kurtosis}~\cite{muirhead1982aspects}  which is defined as 
\beq \label{eq:kappa} 
\kappa =  \frac{ \expec[ r^4]}{  \big(\expec[ r^2]\big)^2}  \frac{p}{p+2} - 1  = 
 \frac{ \expec[ r^4]}{ p(p+2)} - 1 
\eeq 
where $r^2$ is the 2nd-order modular variate  defined in \eqref{eq:r2}. 
The latter identity in \eqref{eq:kappa}  follows due to assumption $\expec[r^2]=p$. 
For kurtosis to exists, we need to assume that the elliptical distribution has finite fourth order moments. 
The elliptical kurtosis shares properties similar to the kurtosis of a real
random variable. Especially, if $\x \sim \mathcal N_\pdim(\boldsymbol{\mu},\M)$, then
$\ka=0$. This follows by noticing that  the quadratic form $r^2$ has a chi-squared distribution
with $\pdim$ degrees of freedom ($r^2 \sim \chi^2_\pdim$) and hence
$\expec[r^4] = \pdim(\pdim+2)$. 
This result becomes more obvious when one notices the following relationship of $\ka$ with the marginal (excess) kurtosis, 
$\kurt(x_i)$,  of any component of $x_i$ of $\x \sim \mathcal E_p(\boldsymbol{\mu},\M,g)$ \cite{ollila2019optimal}, \cite[Lemma 3]{ollila2021shrinking}:
\beq \label{eq:kappa_kurt}
\kappa = \frac 1 3 \cdot \kurt(x_i) , \mbox{ for any $i \in \{1, \ldots,p\}$. }
\eeq

\subsection{NMSE of SCM} 

We are now ready to derive important results moments of SCM under the elliptical distribution. 
Before stating the NMSE we recall the following result. 

\begin{lemma} \cite[Lemma~2]{ollila2019optimal}   \label{lem:EtrS2}  Let $\x_1,\ldots, \x_n \iidsim \mathcal E_p(\boldsymbol{\mu},\M,g)$ with $\M=\cov(\x)$ and assume
that finite fourth-order moments exist. Then 
\begin{align}
\expec \left[ \left \| \S \right\|_{\Fr}^2 \right]  &= \left( 1 + \tau_1 + \tau_2 \right) \| \M \|_{\Fr}^2 +  \tau_1 \tr(\M)^2   ,  \label{eq:trS2} \\ 
\expec \left [\tr(\bo S)^2 \right] 	&= 2 \tau_1  \| \M \|_{\Fr}^2 +   \big(1+ \tau_2 \big) \tr(\M)^2 \label{eq:trS_2} , 
\end{align}
where the scalars are defined by
\beq \label{eq:tau_1and2}
\tau_1 = \frac{1}{n-1} + \frac{\kappa}{n} \qquad  \mbox{and} \quad  \tau_2 =  \frac{\kappa}{n}
\eeq 
\end{lemma} 

It is important to notice that these expectations depend on the underlying ES distribution (and hence on the density generator $g$) only via its kurtosis parameter $\kappa$.  The NMSE of SCM  is given next. 

\begin{lemma} \cite[Lemma~1]{ollila2019optimal}   \label{lem:NMSE_of_SCM} Let $\x_1,\ldots, \x_n \iidsim \mathcal E_p(\boldsymbol{\mu},\M,g)$ with $\M=\cov(\x)$ and assume
that finite fourth-order moments exist. Then 
\beq \label{eq:NMSE_of_SCM} 
\NMSE(\S) =  \Big( 1 + \frac{p}{\gamma} \Big) \Big( \frac{1}{n-1} + \frac{\kappa}{n} \Big) +  \frac{\kappa}{n} 
\eeq 
where $\gamma$ denotes the sphericity parameter. 
\end{lemma}

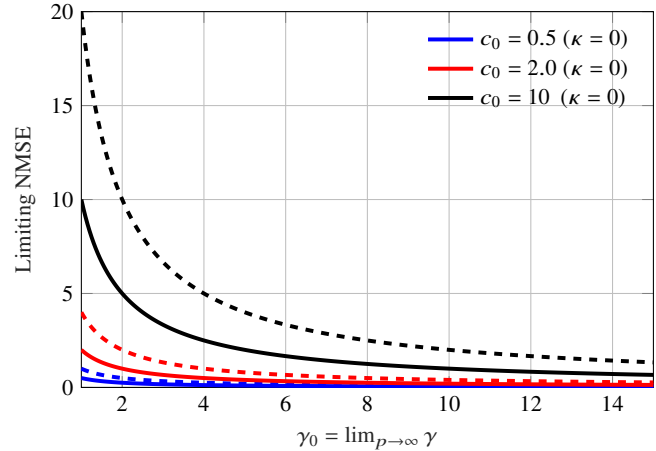
\begin{figure}[!t]
\centerline{\input{tikz/AR1_NMSE_large_p.tex}}
\caption{Limiting NMSE in \eqref{eq:NMSE_as} as a function of limiting sphericity $\gamma_0$  when $p/n \to c_0$ as $p,n \to \infty$. The solid line corresponds to the case $\kappa=0$ and dotted line $\kappa=1$. } \label{fig:NMSE_AR1_large_p}
\end{figure}

Sphericity parameter plays crucial role in determining the accuracy of the SCM. 
Consider the doubly asymptotic regime, 
\beq \label{eq:RMT} 
 c= \frac{p}{n}  \to c_{0}, \quad  0 < c_0  < \infty, \quad \mbox{ as  $p, n \to \infty$}.
 \eeq 
Assume that $\gamma$ remains bounded,   $ \gamma  \to  \gamma_0 $ as $p \to \infty$.  
Then using \eqref{eq:NMSE_of_SCM}, it immediately follows that  the limiting NMSE under the doubly asymptotic regime \eqref{eq:RMT} is
\beq \label{eq:NMSE_as}
\mathrm{NMSE}(\S)  \to  \frac{ 1+ \kappa }{\gamma_0}  c_0
\eeq 
which shows that $\S$ {\it is not a consistent estimator} of $\boldsymbol{\Sigma}$  unless $c=p/n \to 0$. This is illustrated in \autoref{fig:NMSE_AR1_large_p} which displays the limiting NMSE as a function of $\gamma_0$ for different cases of $c_0$. Again the  limiting NMSE is largest when $\boldsymbol{\Sigma}$ is close to being spherical ($\gamma \approx 1$). Moreover, if $c_0 >1$ (undersampled case), the limiting NMSE can be very large. The solid lines are for case $\kappa = 0$ (which holds for MVN distribution) and the dotted lines for the case $\kappa =1$. For example, a multivariate $t$-distribution (MVT) with d.o.f. $\nu=6$  has $\kappa=1$. Figure also illustrates that when the distribution is heavy-tailed ($\kappa=1$) and close to spherical, then the limiting NMSE of SCM can be very large. Finally,  we point out that the effect of sphericity in finite sample case  is  illustrated later  in \autoref{fig:NMSE_AR1gau}a. 
Since sphericity plays a crucial role in determining the accuracy of the SCM, it is of interest to find an accurate estimator of sphericity. This is the topic of subsection~\ref{subsec:sphericity}. 
 
  \subsection{NMSE of tapered SCM} 
  
Let us now derive the MSE of the tapered SCM. For this purpose, assume that template matrix $\W  \in  \mathcal W^+$, where 
\beq
\label{eq:Set_W}
 \mathcal W^+ = \{ \W \in \Sym{p} : w_{ii}=1,  w_{ij} \geq 0  \,  \forall i, j  \in [\![1,p]\!]\}
\eeq
and with $ \Sym{p}$ denoting the set of all symmetric $p \times p$ matrices. 
Write $\diag(\A)  \equiv \diag(a_{11}, \ldots, a_{pp})$ for any matrix $\A = (a_{ij})_{p\times p}$,
Then we have the following result. 

  \begin{lemma} \cite[Lemma~1]{ollila2022regularized}   \label{lem:EtrTaper} Let $\{\x_i \}_{i=1}^n$ be an i.i.d. random sample from  $\mathcal E_p(\boldsymbol{\mu}, \M,g)$ with finite 4th order moments. Then for any $\W \in \mathcal W^+$, it holds that 
\[
\expec \left[  \left\| \W \circ \S \right \|^2_{\Fr} \right]  =  (1+ \tau_1 + \tau_2) \| \W \circ \M \|^2_{\Fr}  +  \tau_1 \tr( (\mathbf{D}_{\M} \W)^2). 
\]
where $\mathbf{D}_{\M} = \diag(\M)$  and $\tau_1, \tau_2$ are defined in \eqref{eq:tau_1and2}. 
\end{lemma}
  
  Notice that the  MSE of the tapered SCM is 
\begin{align}
\mathrm{MSE}(\W \circ \S) &= \expec \big[ \| \W \circ \S - \M \|_{\Fr}^2  \big]  \notag   \\ 
&= \expec \left[  \right\| \W \circ \S \left \|^2_{\Fr} \right] + \| \M \|^2_{\Fr} - 2 \| \mathbf{V} \circ \M \|^2_{\Fr},  \label{eq:MSEtaper} 
\end{align}
where 
\beq \label{eq:VsqrtW}
\mathbf{V} = (v_{ij})_{p \times p}  \text{~with~} v_{ij} = \sqrt{w_{ij}} \text{~for~} \W \in \mathcal W^+.
\eeq 
Thus plugging in the expression from \autoref{lem:EtrTaper} into \eqref{eq:MSEtaper} provides us the MSE of the tapered SCM $\W \circ \S$ when sampling from an ES distribution.  The NMSE is then obtained from this formula via \eqref{eq:NMSE}. 
  
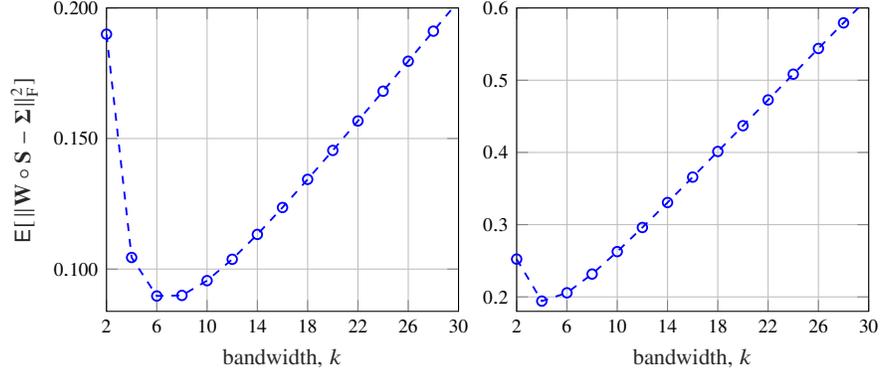
\begin{figure}[!t]
\centering
\setlength\fwidth{1.0\textwidth}
\subfloat{\input{tikz/taper_NMSE_versus_k_nuinf_alpha0dot1_p250_n100_ell1_uusi.tex}}
\subfloat{\input{tikz/taper_NMSE_versus_k_nu5_alpha0dot1_p250_n100_ell1.tex}}
\vspace{-0.25cm}
\caption{NMSE curve of tapered SCM $\W \circ \S$ as a function of used bandwidth $k$  of $\W$   
 when sampling from a MVN distribution (left panel) and MVT distribution (right panel) with d.o.f. $\nu=5$, 
 $\M$  has structure \eqref{eq:Cai_model} with $\alpha = 0.1$, $n=100$ and $p=250$.} \label{fig:tapNMSE_vs_k}
\end{figure}

  \autoref{fig:tapNMSE_vs_k} displays the NMSE curve of tapered SCM $\W \circ \S$ when $\W$ is of the form \eqref{eq:Wband}  and the bandwidth parameter $k$ of $\W$  varies. The data is sampled from a MVN distribution (left panel) and MVT distribution (right panel) with $\nu=5$ d.o.f., sample size is $n=100$ and the dimension is $p=250$.   In this example, the true covariance matrix $\M$ has a following structure 
  \beq\label{eq:Cai_model}
(\M)_{ij} =  
\begin{cases} 
1 &,  i=j  \\ 
\rho | i - j |^{-(\alpha +1 )} &, i \neq j, 
\end{cases}  
\eeq
where $\alpha$ is a decay parameter and $\rho$ is a correlation parameter which are set to 
 $\alpha=0.1$ and $\rho = 0.6$, respectively.     
 \autoref{fig:tapNMSE_vs_k} shows the important point. Since the banding template  in \eqref{eq:Wband} with suitable chosen bandwidth parameter $k$ is well adapted to the true model of $\M$ in \eqref{eq:Cai_model}, the NMSE can be significantly reduced with tapered SCM. For MVN data,  the best bandwidth  $k=6$ yields the NMSE of  
$ 0.089$. Note that bandwidth $k=p$ implies $\W = \mathbf{1} \mathbf{1}^\top$ and the tapered SCM reduces to SCM (i.e., $\W \circ \S = \S$). 
This worst case bandwidth  $k=p$ gives  1.082 as the NMSE.  Thus tje tapered SCM improves the MSE of SCM significantly (more than a factor of ten). 
Performance improvement is even more significant when the data is from a heavy-tailed ES distribution as is illustrated from the more steeply increasing NMSE curve on the right hand side panel of \autoref{fig:tapNMSE_vs_k}.

\subsection{Estimator of sphericity}  \label{subsec:sphericity}

The spatial sign covariance matrix (SSCM) \cite{visuri2000sign} is an estimate of the shape matrix $\boldsymbol{\Lambda}$.  The scaled\footnote{The common definition is without the multiplier $p$} SSCM is defined as 
\beq \label{eq:SSGN} 
\hat{\boldsymbol{\Lambda}} = \frac{p}{n} \sum_{i=1}^{n}  \frac{ (\x_{i} - \hat{\bom \mu})(\x_{i} - \hat{\bom \mu})^\top}{\|\x_{i} - \hat{\bom \mu} \|^{2}},  
\eeq 
where $\hat{\bom{\mu}} = \arg \min_{\bom \mu} \sum_{i=1}^{n} \| \x_{i} - \bom \mu \|$ is
the sample spatial median~\cite{brown1983statistical}.
When $\boldsymbol{\mu}$ is known ($\boldsymbol{\mu} = \bo 0$), the SSCM is defined as 
$$
\hat{\boldsymbol{\Lambda}} = \frac{p}{n} \sum_{i=1}^{n} \frac{ \x_{i}\x_{i}^\top}{\|\x_{i} \|^{2}}.  
$$
One of the major selling points of SSCM are its impeccable robustness properties: 
it possesses the highest possible breakdown point of 1 with fixed location~\cite{magyar2014asymptotic} and breakdown point of 1/2 when using the spatial
median to estimate the location~\cite{croux2010k}. This can be contrasted to M-estimators of scatter  for which the best possible breakdown point
is $1/p$ and obtained by Tyler's M-estimator \cite{dumbgen2005breakdown}.

An estimate of  sphericity based on the SSCM, defined by  
\begin{equation}\label{eq:gammahat}
    \hat \gamma =  \frac{n}{n-1}\left(  \frac{\| \hat{\boldsymbol{\Lambda}} \|_{\Fr}^2}{p}  -    \frac{p}{n}\right), 
\end{equation}
has been studied in many papers (e.g., \cite{zou2014multivariate, zhang2016automatic,raninen2021linear}). 
In \cite{raninen2021linear} it was shown that \eqref{eq:gammahat}  is asymptotically (as $p \to \infty$) unbiased when sampling from ES distribution under the  assumption   $\gamma/p \to 0$ as $p\to \infty$. This assumption is sufficiently general and holds for many scatter matrix models  \cite[Prop.~3]{raninen2021linear}.  
For example, if $\M$ has an autoregressive model (\textbf{AR(1)}) structure,
\beq \label{eq:AR1cov}
(\boldsymbol{\Sigma})_{ij} = \eta \varrho^{|i-j|},
\eeq 
where $\eta$ is the scale \eqref{eq:scale} and $\varrho$ is the correlation parameter, $\varrho \in (-1,1)$, then   
\begin{align} \label{eq:gammaAR1}
 \gamma        &= \frac{p - p \varrho^4 - 2 \varrho^2 + 2 (\varrho^2)^{p+1}}{p (\varrho^2-1)^2}. 
\end{align}
Note that $\gamma = O(1) = o(p)$. 

Another estimator proposed in \cite[Sect. IV-B]{ollila2019optimal} is defined by 
 \begin{equation}\label{eq:hatgamma_ell2}
	\hat \gamma
	= \hat b_n \left( \frac{p\tr(\S^2)}{\tr(\S)^2} - \hat a_n \frac{p}{n}  \right), 
\end{equation}
where 
\begin{align*}
	\hat a_n &=  \left(\frac{n}{n+ \hat \kappa} \right) \left(  \frac{n}{n-1} +  \hat \kappa \right)  \quad \mbox{and} \quad 
\hat 	b_n &= \frac{  \hat (\ka  + n)(n-1)^2}{ (n-2)(3 \kappa (n-1) + n(n+1))}. 
\end{align*}
and $\hat \kappa$ is an estimate of the elliptical kurtosis. In \cite{ollila2021shrinking} estimators of $\gamma$ based on robust M-estimators of scatter were constructed under the assumption that $n>p$ (oversampled case). A comparative study of different estimators of sphericity were recently conducted in \cite{
ollila2022robust}. 

Slightly modified Ell1 or Ell2-estimators of the sphericity parameter of tapered covariance matrix,  
\beq\label{eq:gamma_tap} 
\gamma_{\W} \equiv \gamma(\W \circ \M) =   \frac{p \tr\left((\W \circ \M)^2\right)}{\tr(\M)^2},  \ \W \in \mathcal W^+
\eeq 
can be constructed as shown in \cite[Section IV]{ollila2022regularized}.

\section{Linear shrinkage of  SCM}  \label{sec:RegSCM}

In this section we consider the single sample setting and linear shrinkage estimators of the SCM $\S$ or the tapered SCM $\W \circ \S$ in Subsection~\ref{subsec:RegSCM} and~\ref{subsec:regtapered}, respectively. 

\subsection{Regularized SCM (RSCM)}   \label{subsec:RegSCM}

The regularized SCM (RSCM) considered in \cite{ollila2019optimal} is defined as 
\begin{equation} \label{eq:regSCM}
	\hat \M(\al, \be) = \be \S + \al \mathbf{I},
\end{equation}
where $\S$ is the unbiased SCM defined in \eqref{eq:SCM},  and  $\alpha,\beta \geq 0$ are are tuning or regularization parameters.  The MSE of RSCM can be written as  
\cite[Appendix~A]{ollila2019optimal} 
\beq \label{eq:MSEapu}
\MSE(\hat \M(\al, \be) ) = \beta^2 \MSE(\S) + \| \be \S + \al \mathbf{I} - \M \|_{\Fr}^2. 
\eeq 
Then assuming a sample $\x_1,\ldots, \x_n$ from an arbitrary distribution with finite 4th-order moments, the optimal tuning parameters that minimize the MSE are 
 \cite[Theorem~1]{ollila2019optimal} 
\begin{align}
	\al_{\textup{o}} = (1 - \be_{\textup{o}}) \eta \quad \mbox{and} \qquad \be_{\textup{o}}
	= \dfrac{(\gamma-1)}{(\gamma-1) + \gamma \cdot \mathrm{NMSE}(\S)} \label{eq:beta0ver2b}
\end{align}
where the scale $\eta$ and sphericity $\gamma$ are defined in \eqref{eq:scale} and \eqref{eq:gamma}, respectively.   
Note that the NMSE$(\S)$ for elliptical data is given in  \autoref{lem:NMSE_of_SCM}.

Let $\hat \M_{\textup{o}} = \hat \M(\al_{\textup{o}},\be_{\textup{o}})$ denote the optimal or oracle RSCM that has the knowledge of these optimal parameters. Then 
\begin{align*}
\NMSE(\hat \M_{\textup{o}}) &=   (1- \beta_{\textup{o}}) \frac{ \| \M  - \eta \mathbf{I} \|_{\Fr}^2}{\| \M \|_{\Fr}^2}  = (1- \beta_{\textup{o}})   \frac{\gamma-1}{\gamma} . 
\end{align*} 
Next we give an  instructive example illustrating the power of regularization. 

\begin{figure}[!t] 
\centerline{\input{tikz/AR1_NMSE_SCM} \input{tikz/AR1_NMSE_LW} }
\caption{The effect of sphericity  $\gamma$ on the NMSE of  SCM $\S$ (left panel)  and optimal oracle RSCM $\hat \M_{\textup{o}}$ (right panel). 
Samples are from  MVN distribution with $\boldsymbol{\Sigma}$ having an AR$(1)$ structure; $p=50$. } \label{fig:NMSE_AR1gau}
\end{figure}
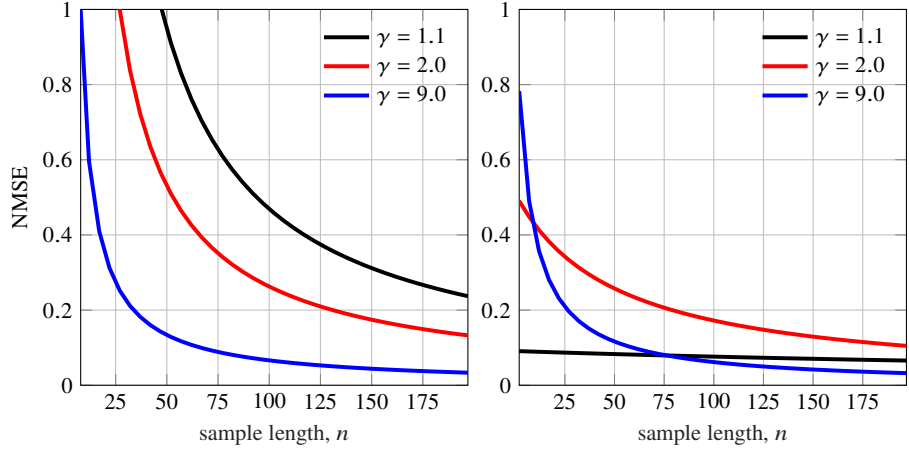

\begin{example}{Comparing the NMSE of SCM and RSCM}The samples are generated from a  $p=50$ dimensional MVN distribution,  $\mathcal N_p(\boldsymbol{\mu}, \boldsymbol{\Sigma})$,  
with AR(1) covariance structure in \eqref{eq:AR1cov}. 
The left panel of \autoref{fig:NMSE_AR1gau} displays the NMSE  of SCM for varying sample lengths $n$.  As can be noted, the accuracy of SCM $\S$ depends heavily on the value of $\gamma$. When $\gamma \approx 1$ (i.e., the distribution  is close to being spherical, so $\varrho \approx 0$), the NMSE is largest, and rises steeply when $n < p$.   The right panel of \autoref{fig:NMSE_AR1gau} displays the NMSE of the optimal RSCM $\hat \M_{\textup{o}}$.  The performance improvement is drastic in the cases when the covariance matrix is close to being spherical (black and red lines) and/or when $n \leq p$.  
\end{example}

In practise one does not have access to the true $\al_{\textup{o}}$ or $\be_{\textup{o}}$ and thus the oracle RSCM is not computable. However, as can be inferred from \eqref{eq:beta0ver2b} and the NMSE expression in \autoref{lem:NMSE_of_SCM}, the optimal parameter  $\beta_{\textup{o}}$  
depends on the  sphericity $\gamma$ and the elliptical kurtosis parameter $\kappa$, i.e., $\beta_{\textup{o}} \equiv \beta_{\textup{o}}(\kappa,\gamma)$.  One may compute an estimate 
$\hat \kappa$  using the empirical average of  the kurtosis parameters (scaled by $1/3$) due to \eqref{eq:kappa_kurt} as detailed in \cite[Sect.~IV]{ollila2019optimal} while for an estimate of sphericity one may use the estimator defined in  \eqref{eq:gammahat}. 
This gives $\hat \beta_{\textup{o}}  = \beta_{\textup{o}} (\hat \kappa, \hat \gamma)$ 
as the estimate of $\beta_{\textup{o}}$.   To estimate $\eta$ 
one uses $\hat{\eta} = \mathrm{tr}(\S)/p$, and then sets $\hat \alpha_{\textup{o}} =  (1- \hat \beta_{\textup{o}}) \hat \eta$ (recall \eqref{eq:beta0ver2b}).    After estimating these parameters, we can compute the regularized SCM as 
\beq \label{eq:RSCM_final}
\hat \M_{\textup{RSCM}}= \hat  \be_{\textup{o}} \S + (1- \hat \beta_{\textup{o}}) \hat \eta \mathbf{I},
\eeq 
This estimator was referred to as {\bf RSCM-Ell1}. The estimator using \eqref{eq:hatgamma_ell2} as the estimate of sphericity was referred to as {\bf RSCM-Ell2}. MATLAB package is available at \url{http://users.spa.aalto.fi/esollila/regscm/} to compute these estimators. 

\subsection{Regularized tapered SCM}  \label{subsec:regtapered}

Let $\mathbb{W} = \{ \bo W (k) \}_{k=1}^K$ be a finite set of possible template matrices, i.e., matrices satisfying $\W (k) \in  \mathcal W^+ ~ \forall k\in [\![1,K ]\!]$, where $k$ is  an index that identifies the matrix $\W$ in the set $\mathbb{W}$.   For example, the set $\mathbb{W}$ can be the set of all banding matrices $\W(k)$, $k=1,\ldots, p$ as defined in  \eqref{eq:Wband} or a union of different type of template matrices. 
Then, \cite{ollila2022regularized} proposed an estimator, referred to as \tabasco{} (TApered or BAnded Shrinkage COvariance matrix), defined as
\beq
\hat \M(\be,k)  =  \be (\bo W (k) \circ \bo{S} )  + (1-\be) \frac{\Tr(\S)}{p} \mathbf{I} , \label{eq:TAPSHRINK}
\eeq
which benefits both from shrinkage and exploitation of structure via tapering templates $\W \in \mathbb{W}$.  Above $\be \in [0,1]$ is the shrinkage parameter and $k \in \{1,\ldots,K\}$ is the index that identifies the tapering matrix in the set $\mathbb{W}$.  Note that $\hat \M(\be,k) $  preserves the original scale of the SCM since $\Tr( \W \circ \S)= \Tr(\S)$ $\forall \W \in \mathcal W^+$. 
Obviously, the success of banding and/or tapering depends on one's ability to choose the parameters $\beta$ and $k$ correctly.  Since both the RSCM in \eqref{eq:RSCM_final} (if $\W= \bo 1 \bo 1^\top \in \mathbb{W}$ where $\bo 1$ denotes a $\pdim$-vector of ones) and the tapered SCM (if $\beta=1$) appear as special cases of \eqref{eq:TAPSHRINK}, \tabasco{} performs never worse than these two estimators in terms of MSE independent of the underlying structure of the true covariance matrix $\M$.  Indeed in the simulation study reported  in \cite{ollila2022regularized} \tabasco{}  outperformed  these estimators as well as many commonly used shrinkage or banding/tapering estimators.

For a given fixed index $k$, let $\W\equiv \W(k)$ denote the associated template matrix and    $\hat \M(\be) \equiv \hat \M(\be,k)$ the associated \tabasco{} estimator. 
Then it was shown that 
\begin{align}
\be_{\textup{o}}  &=\underset{\be\in [0,1]}{\arg \min}   \ \expec \Big[ \big\| \hat{\M}(\be) - \M \|^2_{\mathrm{F}} \Big]  \\ 
	    &= \frac{p (\gamma_{\V}-1) \eta^2}{\expec \left[  \| \W \circ \S  \|^2_{\Fr} \right] - p^{-1} \expec[ \tr(\bo S)^2]} 
	    \label{eq:beta0id1}  
\end{align}
where $\mathbf{V} = (v_{ij})$ with $v_{ij} = \sqrt{w_{ij}}$ (as in \eqref{eq:VsqrtW}),  $\gamma_{\V}$ is the sphericity parameter of  $\V \circ \M$, 
defined via \eqref{eq:gamma_tap}, and $\eta = \tr(\M)/p$ is the scale of $\M$. 
Under the assumption that data is from an ES distribution, one can derive an  explicit analytical expression for $\be_{\textup{o}}$ using 
expressions for $\expec \left[  \| \W \circ \S  \|^2_{\Fr} \right] $ and $\expec[ \tr(\bo S)^2]$ given in  \autoref{lem:EtrTaper}  and \autoref{lem:EtrS2}, respectively; see \cite[Theorem~ 2]{ollila2022regularized} in particularly. 

When $k$ is not fixed, then  $\beta_{\textup{o}}=\beta_{\textup{o}}(k)$ depends  on  $k$ via $\W=\W(k)$ and $\V= \V(k)$. 
 Then,  as shown in \cite{ollila2022regularized}, the MSE optimal index $k$ can be chosen as 
\beq \label{eq:optim_bandwidth}
k_{\textup{o}} = \arg \underset{k}{\min}  \, \beta_0(k) (1-\gamma_{\V}(k)), 
\eeq 
where $\gamma_{\V}(k)$ is the sphericity parameter in \eqref{eq:gamma_tap} for $\V=\V(k)$. 

Naturally, in practise we need to replace the oracle $\beta_{\textup{o}}(k)$ by its estimate $\hat \beta_{\textup{o}}(k)$. 
Finally, given  $\hat \beta_{\textup{o}}(k)$ and an estimate of  sphericity $\hat \gamma_{\V}(k)$, one can choose the best index $k$  (and the associated template $\W= \W(k)$) as 
$\hat k_{\textup{o}} = \arg \min_ k \,  \hat \beta_{\textup{o}}(k) (1- \hat \gamma_{\V}(k))$ as in  \eqref{eq:optim_bandwidth}. These values are then used to obtain the final optimal \tabasco{} estimator  $\hat \M_{\tabasco{}}= \hat \M(\hat \beta_{\textup{o}},\hat k_{\textup{o}})$ via equation \eqref{eq:TAPSHRINK}, where $\hat \beta_{\textup{o}} = \hat \beta_{\textup{o}}(\hat k_{\textup{o}})$.  We refer to \cite{ollila2022regularized} for more details of the calculations.
Efficient MATLAB toolbox for computing the \tabasco{} estimator is available at \url{https://github.com/esollila/Tabasco}. 

\section{Multiple class estimation problem} \label{sec:multiple_SCM}

In this section, we consider the case where we have $K$
different classes or populations, and we have observed $n_k$, $k=1,\ldots,K$,
i.i.d. $p$-dimensional samples from these populations. 
The covariance matrix of class $k \in \{1,\ldots,K\}$ is defined as
\begin{equation*}
	\M_k 
	= 
	\mathsf{E}[(\x_{ik}-\boldsymbol{\mu}_k)( \x_{ik}-\boldsymbol{\mu}_k)^\top], 
\end{equation*}
where $\x_{ik}$ denotes the $i$th sample from class $k$ and $\boldsymbol{\mu}_k =
\mathsf{E}[\x_{ik}]$ is the mean of class $k$. The conventional estimate for
the covariance matrix is the unbiased SCM defined
for class $k$ by
\begin{equation*}
	\S_k =
	\frac{1}{n_k-1}\sum_{i=1}^{n_k}{ (\x_{ik}-\overline\x_k)
	(\x_{ik}-\overline\x_k)^\top}, 
\end{equation*}
where $\overline \x_k = (1/n_k)\sum_{i=1}^{n_k} \x_{ik}$ is the sample mean of class
$k$.  

The estimators that are considered in this section combine or pool the information from
the other classes in order to reduce the MSE of the estimator of a given class. The
underlying rationale for pooling comes from the often plausible assumption 
that the class populations share a somewhat similar structure.  
This is because the same variables that are measured under slightly different population conditions are often
positively correlated, and thus, share a similar correlation/covariance structure. Thus the information available in another class should be used for improving the estimation in the target class. 

Since the classes can be assumed to have a similar
covariance structure, it is beneficial to shrink the individual class
covariance matrix estimates toward the pooled (average) SCM of the classes,
using the pooled SCM defined by
\begin{equation} \label{eq:pooledS}
    \S_{\textup{pool}} = \sum_{k=1}^K \pi_k \S_k, \qquad
    \pi_k = \frac{n_k}{ \sum_{j=1}^K n_j }.
\end{equation}
Often better choise is to use a convex combination of the SCM and the pooled SCM; 
For example,~\cite{friedman1989regularized} proposed to use the convex combination 
\begin{equation}
    \hat \M_k(\beta)
    = \beta \S_k + (1-\beta)\S_{\textup{pool}},
    \label{eq:partiallypooledSCM}
\end{equation} as an estimate for the class covariance matrix, where $\beta \in [0,1]$ is the tuning parameter. 
This partially pooled estimator is then further regularized toward a scaled identity matrix in order
to stabilize its eigenvalues  and guarantee positive definiteness of the estimator in  low sample size settings ($p_k > n_k$ for some $k$): 
\begin{align}
    \hat\M_k(\alpha,\beta) = 
    \alpha \hat\M_k(\beta)
    +
    (1-\alpha)
    \mathbf{I}_{\hat \M_k(\beta)},
    \label{eq:RSCMRDA}
\end{align}
where $\hat \M_k(\beta)$ is given in~\eqref{eq:partiallypooledSCM}, 
$\mathbf{I}_{\mathbf{A}} = (\tr(\mathbf{A})/p)\mathbf{I}$ and $\al,\be  \in [0,1]$ are tuning parameters.
The author of~\cite{friedman1989regularized} then proposed RDA framework based on this estimator.  
Similar ideas but from Bayesian perspectives were developed in \cite{greene1989partially,rayens1991covariance}.

\subsection{Coupled RSCM} 

We call  the estimator  in \eqref{eq:RSCMRDA}  as the \emph{coupled RSCM} estimator as it couples two different types
of regularization. The task that remains is to determine the optimal tuning parameters $(\alpha_k, \beta_k ) \in [0,1] \times [0,1]$, for $k=1,\ldots,K$.  
In RDA \cite{friedman1989regularized}, one uses $\beta  \equiv \beta_k$ and $\alpha \equiv \alpha_k$, i.e., same parameter pair  is used {\it for all} classes $k=1,\ldots,K$, and then one picks up the best pair $(\alpha,\beta)$ from a grid of values using cross-validation.  It is easy to criticise that such an approach is suboptimal but also computer intensive.  As a remedy \cite{raninen2021coupled} proposed a data-adaptive approach for 
 choosing  class-specific choices $(\alpha_k, \beta_k )$ that  minimize the $\MSE(\hat \M_k(\alpha,\beta))$ for each $k=1, \ldots,K.$  This method is described in this section in more detail. 

Before proceeding, it is worthwhile to point out 4 special instances of the estimator~\eqref{eq:RSCMRDA}:
\begin{description} 
    \item[(C1)] \emph{The unpooled regularized SCM estimator} omits the
	pooled SCM and only shrinks toward the scaled identity matrix: 
        \[
	    \hat \M_k(\alpha_k,\beta_k=1) 
	    = \alpha_k \S_k + (1-\alpha_k) \mathbf{I}_{\S_k}.
        \]
	This type of shrinkage is typically considered in single class
	covariance matrix estimation (see
	e.g.,~\cite{ledoit2004well} and~\cite{ollila2019optimal}).

    \item[(C2)] \emph{The partially pooled estimator} omits regularization
	toward the scaled identity and only shrinks toward the pooled SCM:
        \[
            \hat \M_k(\alpha_k=1,\beta_k) 
            = \hat \M_k(\beta_k) = \beta_k \S_k + (1-\beta_k) \S.
        \]

    \item[(C3)] \emph{The fully pooled estimator} uses the pooled SCM for
	every class $k$ and shrinks it toward the scaled identity matrix:
        \[
            \hat \M_k(\alpha_k,\beta_k=0) 
			= \alpha_k \S_{\textup{pool}} + (1-\alpha_k) \mathbf{I}_{\S_{\textup{pool}}}.
        \]
        Such shrinkage can be considered if all classes have an identical
        distribution.

    \item[(C4)] \emph{The scaled identity estimator} uses the partially pooled
	estimator to scale the identity matrix:
        \[
            \hat \M_k(\alpha_k=0,\beta_k) 
			= \mathbf{I}_{(\beta_k \S_k + (1-\beta_k) \S_{\textup{pool}})}.
        \]
\end{description} 
Since it is clear that the tuning parameters are class-specific, we drop the
subscripts from $\alpha_k$ and $\beta_k$ and denote them from now on simply by
$\alpha$ and $\beta$.

\begin{example}{The NMSE of coupled RSCM and estimates of tuning parameters}We adopt the Setup~A from \cite{raninen2021coupled} consisting of $K=4$ classes, which all follow an AR$(1)$ covariance model 
in \eqref{eq:AR1cov} with correlations  $\varrho_k = (0.2,0.3,0.4,0.5)$, sample sizes $n_k = (25,50,75,100)$, and scales $\eta_k \equiv 1$ $\forall k$. 
The data are generated from MVT distribution with d.o.f.   $\nu=8$. The dimension is $p=200$. 
Figure~\ref{fig:abplane} displays the NMSE of the $4$th class  $\hat \M_4(\al,\be)$  in  \eqref{eq:RSCMRDA}. The gray dots depict the estimated tuning parameters (showing 400 realizations of 4000 Monte Carlo trials) using the estimation method proposed in \cite{raninen2021coupled}. The
black triangle ($\blacktriangle$) identifies the optimal tuning parameter pair, and the
blue square  ({\color{blue}$\blacksquare$}) depicts the mean of the estimated tuning parameters. One
can notice that using the estimator (C3) would be beneficial in this case and using the estimated tuning parameters one obtains an estimator with MSE that is very close to the best possible oracle estimator. 
\end{example}

An alternative, streamlined estimator to~\eqref{eq:RSCMRDA} was further proposed in \cite{raninen2021coupled} by changing 
the $\alpha$-regularization target, and by defining the estimator as 
\begin{equation}
    \tilde \M_k(\alpha,\beta) = \alpha \hat \M_k(\beta) + (1-\alpha) \mathbf{I}_{\T},
    \label{eq:RSCMstreamlined}
\end{equation}
where $\T \in \{\S_k, \S\}$ and $\hat \M_k(\beta)$ is defined
in~\eqref{eq:partiallypooledSCM}. This simplifies the expression for the MSE
and allows for an analytical solution for the tuning parameters as given below. 

\begin{theorem} \cite[Theorem~ 3]{raninen2021coupled}  \label{thm:polysimple}
    The theoretical MSE of   estimator~\eqref{eq:RSCMstreamlined} is a bivariate polynomial of the form
    \begin{align*}
	\MSE(\tilde \M_k(\alpha,\beta))
	&= 
	\alpha^2 \beta^2 B_{22}
	+ \alpha^2 \beta B_{21}
	+ \alpha^2 B_{20}
	+ \alpha \beta B_{11}
	+ \alpha B_{10}
	+ B_{00},
    \end{align*}
    where the coefficients $B_{ij}$ depend on the scalars $ \eta_j = \tr(\M_j)/p$,
    $\expec[\|\S_j\|_\Fr^2]$, $\expec[\|\mathbf{I}_{\S_j}\|_\Fr^2]$, and $\ip{\M_i}{\M_j} = \tr(\M_i \M_j)$. If
    $(\alpha^\star,\beta^\star) \in (0,1) \times (0,1)$, the optimal tuning
    parameters $(\alpha^\star,\beta^\star)$ minimizing the MSE are
    \begin{align*}
	\alpha^\star
	= \frac{2 B_{10} B_{22} - B_{11}B_{21}}
	{B_{21}^2 - 4B_{20} B_{22}}
	~\text{and}~
	\beta^\star
	= \frac{2 B_{11} B_{20} - B_{10} B_{21}}
	{2B_{10} B_{22} - B_{11} B_{21}}. 
    \end{align*}
    Otherwise, the optimal parameters are on the boundary of the feasible set
    $[0,1] \times [0,1]$, and are given by one of the following options
    \begin{enumerate}
	\item[i)]
	    $\alpha^\star = \bigg[-\dfrac{1}{2}\dfrac{B_{10}}{B_{20}}
	    \bigg]_0^1$ and $\beta^\star = 0$,
	\item[ii)]
	    $\alpha^\star = \bigg[-\dfrac{1}{2}\dfrac{B_{10} + B_{11}}{B_{22}
	    + B_{21} + B_{20}} \bigg]_0^1$ and $\beta^\star = 1$,
	\item[iii)]
	    $\alpha^\star = 1$ and $\beta^\star =
	    \left[-\dfrac{1}{2}\dfrac{B_{21} + B_{11}}{B_{22}}\right]_0^1$,
	\item[iv)] 
	    $\alpha^\star = 0$, which implies $\tilde \M = \mathbf{I}_\T$ and that the
	    MSE does not depend on $\beta$.
    \end{enumerate} 
Above    the \emph{clip} function $[c]_a^b = \max\{a,\min\{b,c\}\}$  projects $c$
on to the interval $[a,b]$. 
\end{theorem}

The unknown constants $B_{ij}$ are replaced by their estimated values when constructing the streamlined estimator, again assuming that data are generated from unspecified elliptical distributions. This provides significant speed-up compared to previous approaches where one uses  cross-validation to estimate the tuning parameters involved in the coupled RSCM. 
The coupled RSCM estimator was adapted and applied to a real data classification problem in the RDA
framework in   \cite[Sect.~V-B and VI-B]{raninen2021coupled} where the proposed method of estimating the MSE-optimal tuning parameters was
compared to different types of cross-validation based methods.  The proposed approach performed similarly to CV  in terms of classification accuracy but achieved the same performance with significant computational gain. 

 It should be emphasized that the main difference of \eqref{eq:RSCMstreamlined} to \eqref{eq:RSCMRDA} is that
the trace of  \eqref{eq:RSCMstreamlined}  depends on $\alpha$, while
in  \eqref{eq:RSCMRDA} it does not. However, when $\tr( \mathbf{I}_\mathbf{T})  \approx \tr(  \mathbf{I}_{\hat \M_k(\beta)}) $, the performance of
the two estimators is expected to be similar. Simulation
results in  \cite[Table I]{raninen2021coupled} illustrate that neither the coupled RSCM in  \eqref{eq:RSCMRDA}  nor the
streamlined estimator  \eqref{eq:RSCMstreamlined} was always better than the other. The codes to compute the coupled RSCM or streamlined RSCM with MSE-optimal estimated tuning parameters are available  in
Matlab, R, and Python programming languages at \url{https://github.com/EliasRaninen/CoupledRSCM}.

\input{results/AR1-definitionsforplotofclass4.tex}

\pgfplotstableread{results/AR1-estimatedtuningparametersofclass4.dat}\estimatedtuningparameters
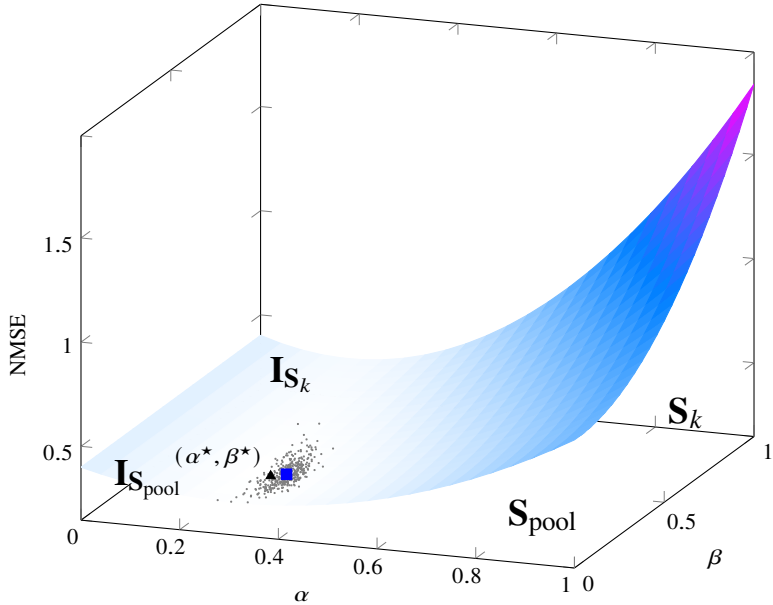
\begin{figure}[th]
    \centering
    \begin{tikzpicture}
	\begin{axis}[colormap/cool,
	    xlabel=$\alpha$,
	    ylabel=$\beta$,
	    zlabel={NMSE},
	    width=0.9\linewidth,
	    view={20}{20}]

	    \addplot3[surf,
	    shader=flat,
	    samples=25,
	    domain=0:1,
	    y domain=0:1,
	    ] {\Cxxyyfour*x^2*y^2 + \Cxxyfour*x^2*y +
	    \Cxxfour*x^2 + \Cyyfour*y^2 + \Cxyfour*x*y +
	    \Cxfour*x + \Cyfour*y + \Cfour};

	    \addplot3[only marks, mark size=0.2, mark=*, gray=0.5] table
	    [x=alpha, y=beta, z=nmse] {\estimatedtuningparameters};

	    \addplot3[only marks,mark size=2.0,blue,mark=square*, fill] coordinates
	    {(\almeanfour,\bemeanfour,\nmseofmeanfour)} node[anchor = north
	    west] {};

	    \addplot3[only marks,mark size=2.0,black,mark=triangle*]
	    coordinates {(\aloptfour,\beoptfour,\nmseofoptfour)} node[anchor =
	    south east]
	    {\footnotesize$(\alpha^\star,\beta^\star)$};

	    \addplot[] coordinates {(0.9,0.9)} node[] {{\Large $\displaystyle \S_k$}};
	    \addplot[] coordinates {(0.9,0.1)} node[] {{\Large $\displaystyle \S_{\textup{pool}}$}};
	    \addplot[] coordinates {(0.1,0.1)} node[] {{\Large $\displaystyle \mathbf{I}_{\S_{\textup{pool}}}$}}; 
	    \addplot[] coordinates {(0.1,0.9)} node[] {{\Large $\displaystyle \mathbf{I}_{\S_k}$}};
	\end{axis}
    \end{tikzpicture}
    \caption{NMSE of $\hat \M(\al,\be)$ for the AR$(1)$ covariance model in \eqref{eq:AR1cov} with
$\varrho_k = (0.2,0.3,0.4,0.5)$, $n_k = (25,50,75,100)$, dimension $p = 200$, and sampling
from MVT distributions with $\nu=8$ d.o.f.}
    \label{fig:abplane}
\end{figure}

\subsection{Linear pooling of  sample covariance matrices} 

In \cite{raninen2021linear} a method is proposed to estimate each class covariance matrix as a linear combination of
the SCM-s of the classes. For a vector of nonnegative
weights $\a \geq \mathbf{0}$, i.e., $\a = (a_i)$, $a_i \geq 0$, $i=1,\ldots,K$,
one defines
\begin{equation}
    \S(\a) = \sum_{i=1}^K a_{i} \S_i.
    \label{eq:linearcombination}
\end{equation}
Restricting the coefficients to be nonnegative ensures that
the estimator  is positive semidefinite. The goal is to find a $K \times K$ nonnegative coefficient matrix $\A^\star = 
    (\a_1^\star \ \cdots \ \a_K^\star)$ where
\begin{align} \label{eq:ak} 
    \a_k^\star &= \arg \min_{\a \geq \mathbf{0}}
    \expec \big[ \|\S(\a) - \M_k \|^2_\Fr \big], ~ k=1,\ldots,K.
\end{align}
 Let us define a diagonal matrix consisting of  scaled MSE-s of the
SCM-s as its diagonal elements as 
\begin{equation} \label{eq:Delta}
  \boldsymbol{\Delta} = \diag(\delta_1,\ldots,\delta_K),  \quad   \delta_k 
    =  p^{-1}\expec[ \| \S_k - \M_k \|_\Fr^2]
\end{equation} 
as well as the matrix of scaled inner products of the covariance matrices as
\begin{equation} \label{eq:cij}
 \C = \begin{pmatrix} \cv_1 \cdots \cv_K \end{pmatrix}  =(c_{ij}) = \big(p^{-1} \tr(\M_i \M_j) \big). 
\end{equation}
We can then state the following result. 

\begin{theorem}\label{prop:unconstrained}  \cite[Prop~1, Prop~2]{raninen2021linear} 
The scaled MSE     in \eqref{eq:ak} can be written as  
    \begin{equation}
	p^{-1}\expec[ \| \S_k - \M_k \|_\Fr^2]=    \a^\top (\boldsymbol{\Delta} + \C) \a - 2 \cv_k^\top \a  + c_{kk}. 
	\label{eq:thm:MSE}
    \end{equation} 
    where $\boldsymbol{\Delta}$ and $ \C$ are defined in \eqref{eq:Delta} and \eqref{eq:cij}, respectively. 
   Furthermore,  $\boldsymbol{\Delta} + \C$ is a positive definite symmetric matrix, and hence the MSE is a  strictly convex quadratic function in $\mathbf{a}$. 
   The unconstrained solution, which minimizes the MSE in \eqref{eq:thm:MSE}
    is
    \begin{align}
	\a_k^\star = (\boldsymbol{\Delta} + \C)^{-1} \cv_k 
	\Leftrightarrow
	\A^\star = (\boldsymbol{\Delta} + \C)^{-1} \C.
	\label{eq:Auncon_sol} 
    \end{align} 
\end{theorem}

It is important to notice that if the solution \eqref{eq:Auncon_sol} to the unconstrained problem  is also non-negative, i.e., verifies $\a_k^\star \geq 0$, then it is solution also to the constrained problem. If this is not the case, then the solution is found by solving the
    strictly convex quadratic programming (QP) problem
    \begin{equation} \label{eq:AQP}
	\begin{array}{ll}
	    {\text{minimize}}
	    & \frac{1}{2} \a^\top (\boldsymbol{\Delta} + \C) \a - \cv_k^\top \a \\
	    \text{subject to} & \a \geq \mathbf{0}
	\end{array}
    \end{equation}

It is often beneficial to incorporate regularization towards the identity
matrix. For example, if  $p > n = \sum_k n_k$, then all of the SCMs $\S_k$ are
singular. Regularization towards the identity can easily be added by using the
estimator
\begin{equation}\label{eq:LINPOOL}
    \tilde{\mathbf{S}}(\a) = \sum_{i=1}^K a_{i} \S_j + a_{I} \mathbf{I}, \quad a_i  \geq 0 ,  a_ I > \epsilon, 
\end{equation}
where the positive definiteness of the estimator is guaranteed due to the
constraint $a_I > \epsilon$, where $\epsilon$ is a small number (e.g., $\epsilon = 10^{-6}$). 
When using \eqref{eq:LINPOOL} one can simply replace $\boldsymbol{\Delta}$ and $\C$ with matrices
\begin{equation}\label{eq:tildeCD}
    \tilde{\boldsymbol{\Delta}} =  \bmat \boldsymbol{\Delta} & \mathbf{0} \\  \mathbf{0}^\top  & 0 \emat
    \quad\text{and}\quad
    \tilde{\C}  =
        \begin{pmatrix}
            \C & \boldsymbol{\eta} \\
            \boldsymbol{\eta}^\top & 1
        \end{pmatrix},
\end{equation}
where $\boldsymbol{\eta} = (\eta_1, \ldots, \eta_K)^\top$ is a vector consisting of scales $\eta_k = \tr(\M_k)/p$  of $\M_k$-s. 
The coefficient  vector $\a = (a_1, \ldots,  a_K, a_I)^\top$  that minimize the MSE  $ \expec \big[ \|   \tilde{\mathbf{S}}(\a) - \M_k \|^2_\Fr \big]$ under the stated constraints  in \eqref{eq:LINPOOL} can be found by solving the following strictly convex QP problem
\begin{equation} \label{eq:Linpool_QP}
      \begin{array}{ll}
            {\text{minimize}} &
            \frac{1}{2} \a^\top (\tilde{\boldsymbol{\Delta}} + \tilde{\C}) \a - \tilde \cv_k^\top
                \a \\
                \text{subject to} &
                a_j \geq 0, j=1, \ldots, K,  a_{I} \geq \epsilon. 
            \end{array}
\end{equation}
  The QP formulation of the problem makes it easy to incorporate additional
    constraints if needed. For example, in order to find a convex combination of
    the SCMs the equality constraint $\mathbf{1}^\top \a = 1$ can be added to the
    QP~\eqref{eq:Linpool_QP}. Such  constraint may be preferred in the case that the different population
covariance matrices have similar scales, so $\eta_j \approx \eta_k$. 

Linearly pooled estimator \eqref{eq:linearcombination}  offers more flexibility than the
partially pooled estimator \eqref{eq:partiallypooledSCM}  as it has individual weights for every
class SCM.  Same holds for their modifications  (i.e., \eqref{eq:LINPOOL} versus \eqref{eq:RSCMRDA}). 
Linearly pooled estimator requires estimation of more coefficients, and thus errors in these estimates may impact its performance. 
Another benefit of coupled estimator is that it has a similar form as
the popular estimator used in RDA and it can thus be easily be applied to
discriminant analysis classification problems without any modifications.  Codes for computing the linear pooled estimator are available at 
\url{https://github.com/EliasRaninen/LinearPoolingOfSampleCovarianceMatrices}. 

\section{Application to portfolio selection} \label{sec:portfolio}

Portfolio selection and optimization is one of the most important topics in
investment theory. It is a mathematical framework wherein one seeks
portfolio allocations which balance the return-risk tradeoff such that it
satisfies the investor's needs. Some historical key references are
\cite{markowitz1952portfolio, markowitz1959portfolio, tobin1958liquidity,
sharpe1964capital}, and~\cite{lintner1965valuation}.

We consider a portfolio $P$ that consists of $\pdim$  assets which can be stocks, bonds, currencies, exchange-traded funds (ETF-s), etc. We assume that assets are hold for a fixed investment period (e.g., 1 month, 1 year). The net return of the $i$th asset at time $t$ is 
\beq \label{eq:net_return}
r_{i,\time} = \frac{p_{i,\time} - p_{i,\time-1}}{p_{i,\time-1}} = \frac{p_{i, \time}}{p_{i,\time-1}} - 1  \in [-1,\infty). 
\eeq 
where $p_{i,t}$ denotes the price of $i$th asset at time $t$. 

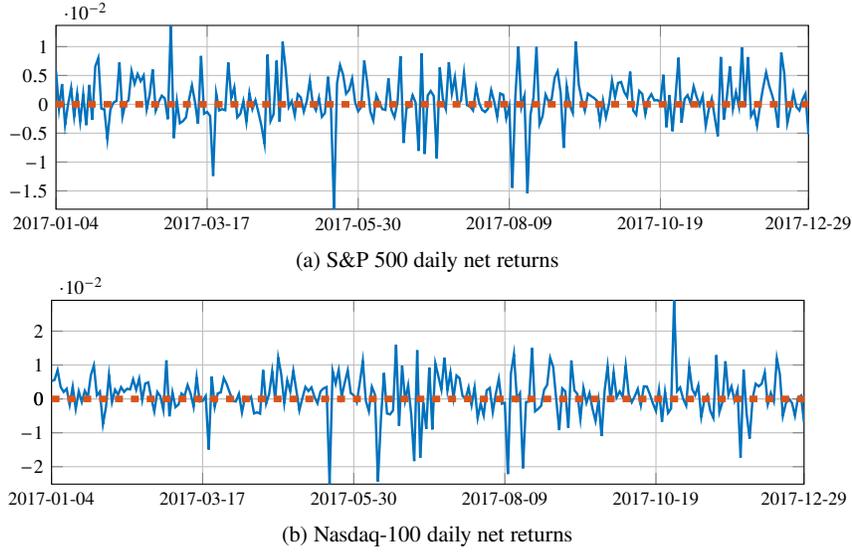
\begin{figure}[!t]
\centering
\setlength\fwidth{0.95\textwidth}
\subfloat[S\&P 500 daily net returns]{\hspace{2pt}\input{./tikz/SP500indx_r.tex}} \hspace{2pt}
\subfloat[Nasdaq-100 daily net returns]{\input{./tikz/nasdaq100indx_r.tex}}
\caption{Daily net returns of the stock indices for year 2017}\label{fig:indexes_ret}
\end{figure}

The original time series of stock prices  $p_{i,t}$ is not a stationary time series, but it can be
argued that a return time series $r_{i,t}$ is close to stationarity within a fixed sufficiently short time periods. 
This is illustrated in \autoref{fig:indexes_ret} which displays daily net returns of  Standard \& Poor’s 500
(S\&P 500)  and Nasdaq-100 stock indexes for year 2017.  Daily net returns are heavy-tailed and non-Gaussian distributed, having occasional large negative or positive returns. Overall the  returns are observed to fluctuate around zero which is displayed by the dotted red-line in the figure.

The objective in portfolio optimization is to find
optimal portfolio weights which determine the proportion of wealth that is
to be invested in each particular asset. That is, a fraction $w_i \in \R$ of
the total wealth is invested in the $i$th asset, $i=1,\ldots,\pdim$, and the
portfolio with $\pdim$ assets is described by the portfolio
\emph{weight} or \emph{allocation vector} $\w \in \R^\pdim$ which satisfies
the constraint $\bo 1^\top \w = 1$. The \emph{global mean variance
portfolio} (GMVP) aims at finding the weight vector that minimizes the
portfolio variance (risk or volatility), and hence does not require
specifying the mean vector.  The GMVP optimization problem is 
\beq \label{eq:GMVP_optim}
    \underset{\w \in \R^\pdim}{\mathrm{minimize}} \  \w^\top \M \w   \quad
    \mbox{subject to }  \quad \bo 1^\top \w = 1, 
\eeq 
where $\M$ is the
covariance matrix of $\bo r_t=(r_{1,t}, \ldots, r_{p,t})^\top$. The solution to \eqref{eq:GMVP_optim}
is \begin{equation} \label{eq:w_GMVP} 
    \w_{\textup{o}} = \frac{\M^{-1} \bo 1}{\bo 1^\top \M^{-1} \bo 1}.
\end{equation}
Naturally, the covariance matrix is unknown and needs to be estimated from the historical data.

\subsection{Are stock returns Gaussian?} 

Let us first investigate the hypothesis that the daily net returns of stocks are Gaussian. 

\begin{figure}[!t]
\centering
\setlength\fwidth{0.95\textwidth}
\subfloat[S\&P 500]{\input{tikz/sp500hist.tex}}
\subfloat[NASDAQ-100]{\input{tikz/nasdaq100hist.tex}}
\subfloat[Gaussian data]{\input{tikz/gaudatahist.tex}}
\caption{Empirical histograms of standardized daily net returns of S\&P 500 and Nasdaq-100 indexes for year 2017. Also plotted is synthetic Gaussian data of same length $n=100$ from  $\mathcal N(0,1)$ distribution.  The pdf of  $\mathcal N(0,1)$ distribution is plotted in red. }\label{fig:hist}
\end{figure}
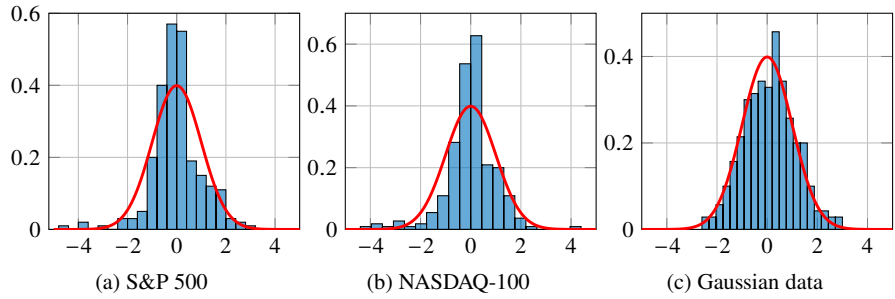

Let us start by plotting the histograms of historical daily net returns. These are shown in \autoref{fig:hist}a,b which display the histograms of standardized daily net returns of S\&P 500 and Nasdaq-100 indexes  For better comparison of Gaussianity assumption, \autoref{fig:hist}c displays histogram of one realisation from a standard Gaussian distribution $\mathcal N(0,1)$ of same length ($n=100$). Also shown is the p.d.f.  of  $\mathcal N(0,1)$ distribution plotted in red color.  As can be noted, the  histograms of daily net returns are not well matched with Gaussian distribution. Instead we observe that the empirical distribution is more peaked and heavier tailed. 
In fact,  when   Student’s t-distribution  is fitted to daily log-returns on stocks, it has been observed that  the number of degrees of freedom  typically lies between 3 and 7 (see e.g., \cite[p. 85]{mcneil2005quantitative}).

\begin{figure}[!t]
\centering
\setlength\fwidth{0.95\textwidth}
\input{tikz/scatterplot.tex}
\caption{Scatter plots of  daily net returns of Nasdaq-100 and S\&P 500 over year 2017 and the estimated 50\%, 95\% and 99\% tolerance ellipses based on the SCM.}\label{fig:scatterplot}
\end{figure}
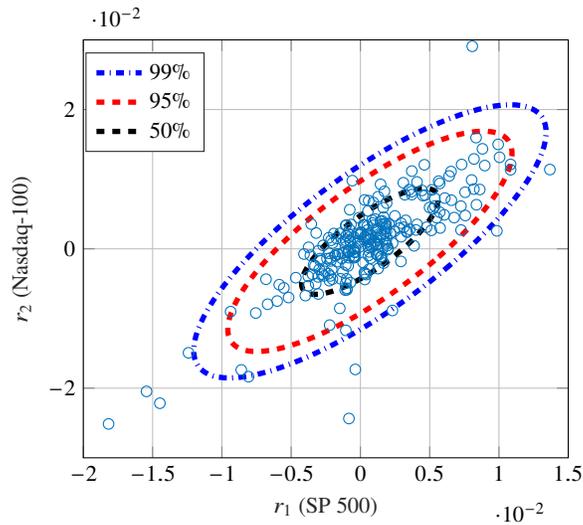

\autoref{fig:scatterplot} display the scatter plots of Nasdaq-100 and S\&P 500  historical daily net returns for  the whole year 2017 and the estimated 99\%,  95\% and 50\% tolerance ellipses computed using the SCM. Overall 95.6\%,  93.2\% and 65.6\% of observations lie inside the 99\%,  95\% and 50\% tolerance ellipses, respectively. The figure and the obtained numbers  further illustrate that the joint distribution of returns is more peaked (concentrated around the mean) and heavier tailed than bivariate Gaussian distribution as there are many observations that lie outside the 99\% tolerance ellipses. Hence, it is fair to say that the joint distribution is not well modelled by the MVN distribution. Instead,  an ES distribution that is more peaked and heavier tailed can provide a better fit.

Although many studies illustrate that for individual stocks or stock index, the value of $\nu$ is often very small, this may not be true when constructing a portfolio over a large set of stocks.  To inverstigate this, we considered  129 stocks in OMX Helsinki and their daily log returns for each year from 2015 to 2022.  This means that each year we have roughly 252 return values on 129 stocks. However,  for a given year we deleted stocks from our analysis that had missing values or several consecutive days  of 0 returns.  
 We fitted MVT distribution to the yearly log return data, where the d.o.f. $\nu$ was estimated using  OPP estimator \cite[Algorithm~1]{ollila2021shrinking} and TWE estimator\footnote{In the R package \texttt{fitHeavyTail} \cite{package_fitHeavyTail}, the function \texttt{fit\_Tyler} implements this method.}~\cite{ollila2023affine}. 
 As can be noted, the estimated values of $\nu$  based on TWE ranges from $5.3$ in year 2020 to $13.5$ in year 2021 while OPP obtains values from  5.7  in year 2020 to 15.5 in year 2021. Thus, only the year 2020 due to sudden fall of stock prices due to covid pandemic indicate a very heavy-tailed MVT distribution. However, the non-Gaussianity is clear from these estimated values. 
  
  \begin{svgraybox} The empirical data analysis thus testify that daily return data is not Gaussian but rather better modelled with a heavy-tailed ES distribution. Yet, since the data is not extremely heavy-tailed (as suggested by the obtained estimates of d.o.f. parameter $\nu$), we can anticipate that the SCM $\S$ can be an effective estimator of the covariance matrix for portfolio optimization problems.  However, it is important to take into account the fact that the data is non-Gaussian, but has higher peakedness and heavier tails. This is the case for linear shrinkage estimators that are reviewed in this chapter since they only assume ellipticity but do not specify the underlying ES distribution. 
  \end{svgraybox}

\begin{figure}[!t]
\centering
\setlength\fwidth{0.95\textwidth}
\input{tikz/OMXH_kurtosis.tex}
\caption{Estimated d.o.f. parameter $\nu$ of MVT distribution for OMX Helsinki stock data based on  historical daily net returns for each year. }\label{fig:OMXH_nu}
\end{figure}
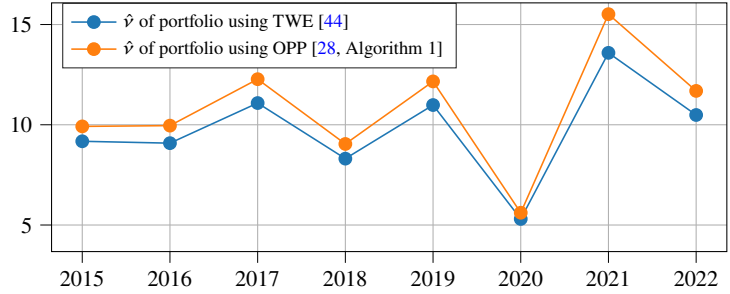

\subsection{Portfolio analysis} 

We now test the performance of RSCM estimators in portfolio optimization using GMVP portfolio selection and historical data. 
We investigate the out-of-sample portfolio performance of different
covariance matrix estimators for three different data sets.  The 1st and 2nd data sets consists of daily net returns of $p=45$ and $p=50$ stocks, respectively,  that are included in the Hang Seng Index (HSI) from Jan. 4, 2010 to Dec. 24, 2011 and from Jan. 1, 2016 to Dec. 27, 2017, both 
consisting of $T = 491$ trading days. The 3rd data set consists of  daily net returns of  $p=396$ stocks included in S\&P 500  from
Jan. 4, 2016 to Apr. 27, 2018 consisting of $T = 583$ trading days. 

At a particular day $t$, we used the previous $n$ days (i.e., from $t-n$ to
$t-1$) as the training window to estimate the covariance matrix, and the
portfolio weight vector. The estimated GMVP weight vector $\hat \w_{\textup{o}}$ was then used
to compute the portfolio returns for the following 20 days. (Note that $\hat \w_{\textup{o}}$  is computed as in  \eqref{eq:w_GMVP} but unkown $\M$ replaced by its estimate $\hat \M$).  Next, the window
was shifted 20 trading days forward, a new weight vector was computed, and the
portfolio returns for another 20 days were computed. Hence, this scenario
corresponds to the case that the portfolio manager holds the assets for
approximately a month (20 trading days), after which they are liquidated and
new weights are computed. In this manner, we obtained $T-n$ daily returns from
which the realized risk was computed as the sample standard deviation of the
obtained portfolio returns. To obtain the annualized realized risk, the
sample standard deviations of the daily returns were multiplied by
$\sqrt{250}$. In our tests, different training window lengths $n$ were
considered. 

In our analysis, we compare three different covariance matrix estimators: RSCM-Ell1 \cite{ollila2019optimal} described in Subsection~\ref{subsec:RegSCM} which is compared to  RSCM estimator by  Ledoit and Wolf (2004)~\cite{ledoit2004well}. These two estimators  both use RSCM in 
\eqref{eq:RSCM_final},  defined by  
\[
\hat \M_{\textup{RSCM}}= \hat  \be_{\textup{o}} \S + (1- \hat \beta_{\textup{o}}) [\tr(\S)/p ] \mathbf{I},
\]
while they  differ only in the approaches to compute $\hat \beta_{\textup{o}}$. The former utilize the ellipticity assumption while the latter builds upon random matrix theory.  We also
included in our study the robust GMVP weight estimator proposed in
\cite{yang2015robust} that uses a robust regularized Tyler's M-estimator
with a tuning parameter selection that is optimized for the GMVP problem.
The three estimators are denoted shortly as {\bf Ell1}, {\bf LW} and {\bf Rob} in the text and figure captions.

Figure~\ref{fig:risk} displays the annualized realized risks for HSI data set.   Overall we can notice that RSCM-Ell1 has the best
performance for all window lengths and for both periods.   For period 2016-2017, the differences
between the estimators were not as large as in the period 2010-2011.  
Also, note that the optimal training window length which yielded
the smallest realized risk was $n=90$ for the period 2010-2011, but much
larger ($n=230$) for the period 2016-2017. This could be explained by the
fact that the stock market were more turbulent in the first period.

\begin{figure}[t]
	\subfloat{%
        \begin{tikzpicture}[scale=1]
		 \begin{axis}[mystyle,
                xlabel = {$\ndim$},
                ylabel = {Realized risk},
				legend style={font=\footnotesize,
                  draw=none,					
                  overlay,at={(0.285,1.15)},	
                  anchor=north west},			
                ]
                \pgfplotstableread{data/SP500resultsRISK.dat}\loadedtable
                \addplot[blue,mark=o,thick] table [x=n, y=RiskEll1]{\loadedtable};
                \addplot[brown,mark=star,thick] table [x=n, y=RiskLWE]{\loadedtable};
				\legend{Ell1;,LW}
            \end{axis}
		\end{tikzpicture}
	\hfill
        \begin{tikzpicture}[scale=1]
		 \begin{axis}[mystyle,
                xlabel = {$\ndim$},
                ylabel = {$\hat\beta_{\textup{o}}$},
                ]
                \pgfplotstableread{data/SP500resultsBETA.dat}\loadedtable
                \addplot[blue,mark=o,thick] table [x=n, y=BETAEll1]{\loadedtable};
                \addplot[brown,mark=star,thick] table [x=n, y=BETALWE]{\loadedtable};
            \end{axis}
		\end{tikzpicture}
	}
\caption{Annualized realized portfolio risk achieved out-of-sample over 583
	trading days  for a portfolio consisting of  $p=396$ stocks in S\&P 500
	index  for Jan. 4, 2016 to Apr. 27, 2018. The portfolio allocations are
	obtained using  GMVP based on two RSCM estimators (Ell1 and LW) and
	different training window lengths $n$. The right panel shows the average
	$\hat \beta_{\textup{o}}$ of the RSCM estimators for different training window lengths.}
    \label{fig:risk2}
\end{figure}
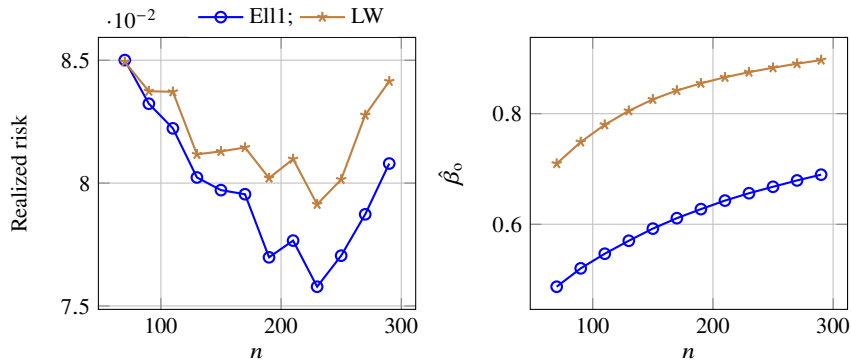

The left panel of Figure~\ref{fig:risk2} depicts the annualized realized risks of  RSCM-Ell1- and -LW estimators for S\&P 500 data. We have excluded the Rob estimator~\cite{yang2015robust} from this study as it is not well suited for very high-dimensional problems. 
With the S\&P 500 data, RSCM-Ell1 achieves the smallest realized risk and
outperformed RSCM-LW for all training window lengths $n$.  The optimal training window
length which produced the smallest realized risk was $n=230$ for both
methods.  Note that, the same result was achieved with HSI data for period 2016-2017. The right panel of Figure~\ref{fig:risk2} displays the estimated optimal shrinkage parameter $\hat \beta_{\textup{o}}$ used by the methods. As can be noted, RSCM-LW estimator uses much larger estimate of $\beta_{\textup{o}}$ and thus puts much more weight on the SCM $\S$ than RSCM-Ell1.

\section{Conclusions}  \label{sec:concl}

This chapter reviewed methods for linear shrinkage of the SCM(-s) under elliptical distributions in both the single and a multiple populations settings. Specifically, we considered approaches for choosing the shrinkage parameters that minimize the MSE. 

In the single population setting, we reviewed the RSCM estimator proposed in \cite{ollila2017optimal,ollila2019optimal} and its generalization called \tabasco{}~\cite{ollila2022regularized} that  imposes tapering/banding  templates to SCM, and thus allows  imposing structure to the covariance matrix estimator. In the multiple population setting, we reviewed the coupled RSCM estimator  \cite{raninen2021coupled}  and its genelization, the linearly pooled estimator proposed in \cite{raninen2021linear}. 

 It should be emphasized that only linear shrinkage of SCM was considered in this chapter.  Another popular approach is \emph{non-linear shrinkage} methods which perform nonlinear shrinkage to the eigenvalues $\hat \lambda_i$, $i=1,\ldots, p$,   of SCM $\S$. These estimators are written in the form
\[
\hat \M = \sum_{i=1}^p \phi_i(\hat \lambda_i) \mathbf{u}_i \mathbf{u}_i^\top
\]
where $\phi_i : \fR_{\geq 0} \to  \fR_{\geq 0} $ is a nonnegative function and 
$\mathbf{u}_1, \ldots \mathbf{u}_p$ are the eigenvector of $\S$. Such nonlinear  shrinkage approaches often rely upon random matrix theory in their design of the function $\phi_i$, see e.g. \cite{bun2017cleaning,ledoit2020analytical,donoho2018optimal_b}.

We also did not cover penalized SCM-s,  obtained by adding a penalty term on the covariance matrix  to the Gaussian negative log-likelihood function (see e.g., \cite{deng2013penalized,ollila2014regularized,yi2020shrinking,tyler2020lassoing}).  Also note that when a  penalty term $\tr(\M^{-1})$ is added to a (scaled) Gaussian negative log-likelihood, one recovers the regularized SCM  in \eqref{eq:regSCM} as the unique solution  \cite{ollila2014regularized}.

\begin{figure}[t]
	\centering
	\subfloat[HSI for Jan. 4, 2010 to Dec. 24, 2011.]{%
        \begin{tikzpicture}[scale=1]
		 \begin{axis}[mystyleportfolio,
                xlabel = {$\ndim$},
                ylabel = {Realized risk},
				xtick  = {50,100,150},
				ymax   = .13,
				scaled y ticks=base 10:1,
				legend style={font=\footnotesize,
                  overlay,at={(0.285,1.15)},	
                  anchor=north west},			
                ]
                	\legend{Ell1;,LW;,Rob}
                \pgfplotstableread{data/minvarportfolioresultsRISK2010.dat}\loadedtable
                \addplot[blue,mark=o,thick] table [x=n, y=RiskEll1]{\loadedtable};
                   \addplot[brown,mark=star,thick] table [x=n, y=RiskLWE]{\loadedtable};
                \addplot[black,mark=diamond,thick] table [x=n, y=RiskROB]{\loadedtable};              
            \end{axis}
		\end{tikzpicture}
	}
	\hfill
	\subfloat[HSI for Jan. 1, 2016 to Dec. 27, 2017)]{%
        \begin{tikzpicture}[scale=1]
		 \begin{axis}[mystyleportfolio,
                xlabel = {$\ndim$},
				legend style={font=\footnotesize,
                  draw=none,					
                  overlay,at={(0.315,1.15)},	
                  anchor=north west},			
                ]
                \pgfplotstableread{data/minvarportfolioresultsRISK2016.dat}\loadedtable
                \addplot[blue,mark=o,thick] table [x=n, y=RiskEll1]{\loadedtable};
                    \addplot[brown,mark=star,thick] table [x=n, y=RiskLWE]{\loadedtable};
                \addplot[black,mark=diamond,thick] table [x=n, y=RiskROB]{\loadedtable};
            \end{axis}
		\end{tikzpicture}
	}
\caption{Annualized realized portfolio risk achieved out-of-sample for the two HSI data sets. 
The portfolio allocations are obtained using GMVP based on the three different
covariance estimators (see text) and different training window lengths $n$. }
\label{fig:risk}
\end{figure}
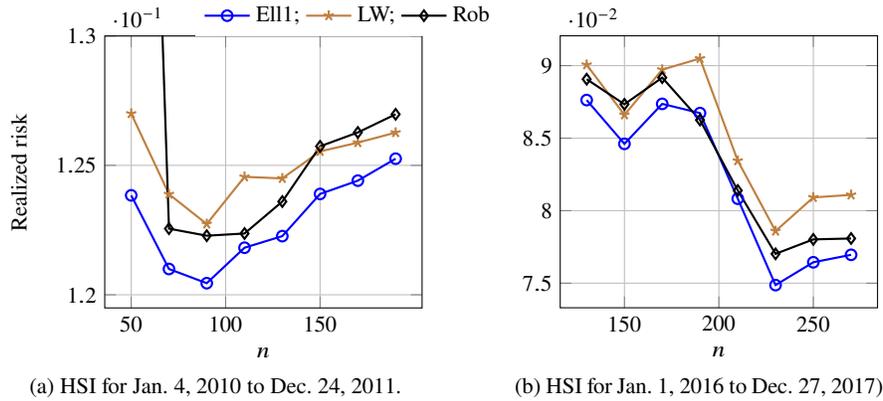

\begin{acknowledgement}
 I would like thank my former doctoral student Elias Raninen whose work has been essential in many of the results presented in this chapter. I would also like to thank  my collaborators David E. Tyler, Frederic Pascal, Daniel P. Palomar and Arnaud Breloy for many discussions regarding covariance matrix estimation and life.
\end{acknowledgement}

\section*{Appendix}
\addcontentsline{toc}{section}{Appendix}
%
%
\input{regularize_SCM.bbl}

\end{document}

%% file: tikz/bias_var.tex
%
%
\definecolor{mycolor1}{rgb}{0.49400,0.18400,0.55600}%
\definecolor{mycolor2}{rgb}{0.46600,0.67400,0.18800}%
\definecolor{mycolor3}{rgb}{0.30100,0.74500,0.93300}%
\begin{tikzpicture}

\begin{axis}[%
width=0.65\fwidth,
height=0.48\fwidth,
at={(1.011in,0.674in)},
scale only axis,
clip=false,
xmin=0,
xmax=1,
xlabel style={font=\color{white!15!black}},
xlabel={$\beta$},
ymin=0,
ymax=1,
axis background/.style={fill=white},
xmajorgrids,
ymajorgrids,
legend style={at={(0.2,0.717)}, anchor=south west, legend cell align=left, align=left, draw=white!15!black}
]
\addplot [color=blue, line width=2.0pt]
  table[row sep=crcr]{%
0	1\\
0.01	0.980182222222222\\
0.02	0.960728888888889\\
0.03	0.94164\\
0.04	0.922915555555555\\
0.05	0.904555555555556\\
0.06	0.88656\\
0.07	0.868928888888889\\
0.08	0.851662222222222\\
0.09	0.83476\\
0.1	0.818222222222222\\
0.11	0.802048888888889\\
0.12	0.78624\\
0.13	0.770795555555556\\
0.14	0.755715555555555\\
0.15	0.741\\
0.16	0.726648888888889\\
0.17	0.712662222222222\\
0.18	0.69904\\
0.19	0.685782222222222\\
0.2	0.672888888888889\\
0.21	0.66036\\
0.22	0.648195555555556\\
0.23	0.636395555555555\\
0.24	0.62496\\
0.25	0.613888888888889\\
0.26	0.603182222222222\\
0.27	0.59284\\
0.28	0.582862222222222\\
0.29	0.573248888888889\\
0.3	0.564\\
0.31	0.555115555555555\\
0.32	0.546595555555556\\
0.33	0.53844\\
0.34	0.530648888888889\\
0.35	0.523222222222222\\
0.36	0.51616\\
0.37	0.509462222222222\\
0.38	0.503128888888889\\
0.39	0.49716\\
0.4	0.491555555555556\\
0.41	0.486315555555556\\
0.42	0.48144\\
0.43	0.476928888888889\\
0.44	0.472782222222222\\
0.45	0.469\\
0.46	0.465582222222222\\
0.47	0.462528888888889\\
0.48	0.45984\\
0.49	0.457515555555556\\
0.5	0.455555555555556\\
0.51	0.45396\\
0.52	0.452728888888889\\
0.53	0.451862222222222\\
0.54	0.45136\\
0.55	0.451222222222222\\
0.56	0.451448888888889\\
0.57	0.45204\\
0.58	0.452995555555556\\
0.59	0.454315555555556\\
0.6	0.456\\
0.61	0.458048888888889\\
0.62	0.460462222222222\\
0.63	0.46324\\
0.64	0.466382222222222\\
0.65	0.469888888888889\\
0.66	0.47376\\
0.67	0.477995555555556\\
0.68	0.482595555555555\\
0.69	0.48756\\
0.7	0.492888888888889\\
0.71	0.498582222222222\\
0.72	0.50464\\
0.73	0.511062222222222\\
0.74	0.517848888888889\\
0.75	0.525\\
0.76	0.532515555555556\\
0.77	0.540395555555556\\
0.78	0.54864\\
0.79	0.557248888888889\\
0.8	0.566222222222222\\
0.81	0.57556\\
0.82	0.585262222222222\\
0.83	0.595328888888889\\
0.84	0.60576\\
0.85	0.616555555555555\\
0.86	0.627715555555555\\
0.87	0.63924\\
0.88	0.651128888888889\\
0.89	0.663382222222222\\
0.9	0.676\\
0.91	0.688982222222222\\
0.92	0.702328888888889\\
0.93	0.71604\\
0.94	0.730115555555556\\
0.95	0.744555555555556\\
0.96	0.75936\\
0.97	0.774528888888889\\
0.98	0.790062222222222\\
0.99	0.80596\\
1	0.822222222222222\\
};
\addlegendentry{MSE}

\addplot [color=red, line width=2.0pt]
  table[row sep=crcr]{%
0	1\\
0.01	0.9801\\
0.02	0.9604\\
0.03	0.9409\\
0.04	0.9216\\
0.05	0.9025\\
0.06	0.8836\\
0.07	0.8649\\
0.08	0.8464\\
0.09	0.8281\\
0.1	0.81\\
0.11	0.7921\\
0.12	0.7744\\
0.13	0.7569\\
0.14	0.7396\\
0.15	0.7225\\
0.16	0.7056\\
0.17	0.6889\\
0.18	0.6724\\
0.19	0.6561\\
0.2	0.64\\
0.21	0.6241\\
0.22	0.6084\\
0.23	0.5929\\
0.24	0.5776\\
0.25	0.5625\\
0.26	0.5476\\
0.27	0.5329\\
0.28	0.5184\\
0.29	0.5041\\
0.3	0.49\\
0.31	0.4761\\
0.32	0.4624\\
0.33	0.4489\\
0.34	0.4356\\
0.35	0.4225\\
0.36	0.4096\\
0.37	0.3969\\
0.38	0.3844\\
0.39	0.3721\\
0.4	0.36\\
0.41	0.3481\\
0.42	0.3364\\
0.43	0.3249\\
0.44	0.3136\\
0.45	0.3025\\
0.46	0.2916\\
0.47	0.2809\\
0.48	0.2704\\
0.49	0.2601\\
0.5	0.25\\
0.51	0.2401\\
0.52	0.2304\\
0.53	0.2209\\
0.54	0.2116\\
0.55	0.2025\\
0.56	0.1936\\
0.57	0.1849\\
0.58	0.1764\\
0.59	0.1681\\
0.6	0.16\\
0.61	0.1521\\
0.62	0.1444\\
0.63	0.1369\\
0.64	0.1296\\
0.65	0.1225\\
0.66	0.1156\\
0.67	0.1089\\
0.68	0.1024\\
0.69	0.0961\\
0.7	0.09\\
0.71	0.0841\\
0.72	0.0784\\
0.73	0.0729\\
0.74	0.0676\\
0.75	0.0625\\
0.76	0.0576\\
0.77	0.0529\\
0.78	0.0484\\
0.79	0.0441\\
0.8	0.04\\
0.81	0.0361\\
0.82	0.0324\\
0.83	0.0289\\
0.84	0.0256\\
0.85	0.0225\\
0.86	0.0196\\
0.87	0.0169\\
0.88	0.0144\\
0.89	0.0121\\
0.9	0.01\\
0.91	0.00809999999999999\\
0.92	0.00639999999999999\\
0.93	0.00490000000000001\\
0.94	0.00360000000000001\\
0.95	0.0025\\
0.96	0.0016\\
0.97	0.000900000000000002\\
0.98	0.000400000000000001\\
0.99	0.0001\\
1	0\\
};
\addlegendentry{Bias$^2$}

\addplot [color=black, line width=2.0pt]
  table[row sep=crcr]{%
0	0\\
0.01	8.22222222222058e-05\\
0.02	0.000328888888888934\\
0.03	0.000739999999999963\\
0.04	0.00131555555555551\\
0.05	0.00205555555555559\\
0.06	0.00295999999999996\\
0.07	0.00402888888888886\\
0.08	0.00526222222222217\\
0.09	0.00666\\
0.1	0.00822222222222224\\
0.11	0.0099488888888889\\
0.12	0.01184\\
0.13	0.0138955555555555\\
0.14	0.0161155555555555\\
0.15	0.0185\\
0.16	0.0210488888888889\\
0.17	0.0237622222222222\\
0.18	0.02664\\
0.19	0.0296822222222222\\
0.2	0.0328888888888889\\
0.21	0.03626\\
0.22	0.0397955555555556\\
0.23	0.0434955555555555\\
0.24	0.04736\\
0.25	0.0513888888888889\\
0.26	0.0555822222222222\\
0.27	0.05994\\
0.28	0.0644622222222222\\
0.29	0.0691488888888889\\
0.3	0.074\\
0.31	0.0790155555555556\\
0.32	0.0841955555555556\\
0.33	0.08954\\
0.34	0.0950488888888888\\
0.35	0.100722222222222\\
0.36	0.10656\\
0.37	0.112562222222222\\
0.38	0.118728888888889\\
0.39	0.12506\\
0.4	0.131555555555556\\
0.41	0.138215555555556\\
0.42	0.14504\\
0.43	0.152028888888889\\
0.44	0.159182222222222\\
0.45	0.1665\\
0.46	0.173982222222222\\
0.47	0.181628888888889\\
0.48	0.18944\\
0.49	0.197415555555556\\
0.5	0.205555555555556\\
0.51	0.21386\\
0.52	0.222328888888889\\
0.53	0.230962222222222\\
0.54	0.23976\\
0.55	0.248722222222222\\
0.56	0.257848888888889\\
0.57	0.26714\\
0.58	0.276595555555556\\
0.59	0.286215555555556\\
0.6	0.296\\
0.61	0.305948888888889\\
0.62	0.316062222222222\\
0.63	0.32634\\
0.64	0.336782222222222\\
0.65	0.347388888888889\\
0.66	0.35816\\
0.67	0.369095555555555\\
0.68	0.380195555555555\\
0.69	0.39146\\
0.7	0.402888888888889\\
0.71	0.414482222222222\\
0.72	0.42624\\
0.73	0.438162222222222\\
0.74	0.450248888888889\\
0.75	0.4625\\
0.76	0.474915555555556\\
0.77	0.487495555555556\\
0.78	0.50024\\
0.79	0.513148888888889\\
0.8	0.526222222222222\\
0.81	0.53946\\
0.82	0.552862222222222\\
0.83	0.566428888888889\\
0.84	0.58016\\
0.85	0.594055555555555\\
0.86	0.608115555555556\\
0.87	0.62234\\
0.88	0.636728888888889\\
0.89	0.651282222222222\\
0.9	0.666\\
0.91	0.680882222222222\\
0.92	0.695928888888889\\
0.93	0.71114\\
0.94	0.726515555555555\\
0.95	0.742055555555555\\
0.96	0.75776\\
0.97	0.773628888888889\\
0.98	0.789662222222222\\
0.99	0.80586\\
1	0.822222222222222\\
};
\addlegendentry{Variance}

\addplot [color=mycolor1, dashed,line width=2.0pt]
  table[row sep=crcr]{%
0	1\\
0.01	0.97997246857718\\
0.02	0.960310979996545\\
0.03	0.941015534258081\\
0.04	0.922086131361803\\
0.05	0.903522771307718\\
0.06	0.885325454095793\\
0.07	0.867494179726057\\
0.08	0.850028948198498\\
0.09	0.832929759513126\\
0.1	0.816196613669935\\
0.11	0.799829510668914\\
0.12	0.783828450510081\\
0.13	0.768193433193445\\
0.14	0.752924458718962\\
0.15	0.738021527086671\\
0.16	0.723484638296564\\
0.17	0.709313792348635\\
0.18	0.695508989242886\\
0.19	0.682070228979319\\
0.2	0.668997511557934\\
0.21	0.656290836978725\\
0.22	0.643950205241701\\
0.23	0.631975616346857\\
0.24	0.620367070294193\\
0.25	0.609124567083706\\
0.26	0.598248106715412\\
0.27	0.587737689189283\\
0.28	0.57759331450534\\
0.29	0.567814982663582\\
0.3	0.558402693664003\\
0.31	0.549356447506606\\
0.32	0.540676244191382\\
0.33	0.532362083718346\\
0.34	0.524413966087492\\
0.35	0.516831891298815\\
0.36	0.50961585935232\\
0.37	0.502765870248004\\
0.38	0.496281923985871\\
0.39	0.490164020565917\\
0.4	0.484412159988148\\
0.41	0.479026342252552\\
0.42	0.474006567359141\\
0.43	0.469352835307915\\
0.44	0.465065146098859\\
0.45	0.461143499731991\\
0.46	0.457587896207303\\
0.47	0.454398335524791\\
0.48	0.451574817684464\\
0.49	0.449117342686322\\
0.5	0.447025910530354\\
0.51	0.445300521216569\\
0.52	0.443941174744965\\
0.53	0.442947871115541\\
0.54	0.442320610328298\\
0.55	0.442059392383237\\
0.56	0.442164217280356\\
0.57	0.442635085019656\\
0.58	0.443471995601134\\
0.59	0.444674949024792\\
0.6	0.446243945290635\\
0.61	0.448178984398655\\
0.62	0.450480066348861\\
0.63	0.45314719114125\\
0.64	0.456180358775812\\
0.65	0.459579569252553\\
0.66	0.463344822571478\\
0.67	0.467476118732584\\
0.68	0.471973457735874\\
0.69	0.476836839581337\\
0.7	0.482066264268986\\
0.71	0.487661731798814\\
0.72	0.493623242170827\\
0.73	0.499950795385016\\
0.74	0.506644391441391\\
0.75	0.513704030339935\\
0.76	0.52112971208067\\
0.77	0.528921436663588\\
0.78	0.537079204088679\\
0.79	0.545603014355954\\
0.8	0.554492867465407\\
0.81	0.563748763417045\\
0.82	0.573370702210861\\
0.83	0.583358683846862\\
0.84	0.593712708325036\\
0.85	0.604432775645394\\
0.86	0.615518885807938\\
0.87	0.626971038812654\\
0.88	0.638789234659557\\
0.89	0.650973473348638\\
0.9	0.6635237548799\\
0.91	0.676440079253342\\
0.92	0.689722446468968\\
0.93	0.703370856526774\\
0.94	0.717385309426755\\
0.95	0.73176580516892\\
0.96	0.746512343753265\\
0.97	0.761624925179797\\
0.98	0.777103549448499\\
0.99	0.792948216559394\\
1	0.809158926512457\\
};
\addlegendentry{Emp. MSE}

\addplot [color=mycolor2, dotted, forget plot,line width=1.5pt]
  table[row sep=crcr]{%
0.548780487804878	0\\
0.548780487804878	1\\
};
\addplot [color=mycolor3, dotted, forget plot,line width=1.5pt]
  table[row sep=crcr]{%
0	0.451219512195122\\
1	0.451219512195122\\
};
\node[above right, align=left]
at (axis cs:0.529,1.01) {$\beta{}_\text{0}$};
\end{axis}

\end{tikzpicture}%

%% file: tikz/AR1_NMSE_large_p.tex
\begin{tikzpicture}
\begin{axis}[%
width=0.65\textwidth,
height=5cm,
scale only axis,
xmin=1,
xmax=15,
xlabel style={font=\color{white!15!black}},
xlabel={$\gamma_\text{0} = \lim_{p \to \infty} \gamma$ },
ymin=0,
ymax=20,
ylabel style={font=\color{white!15!black}},
ylabel={Limiting NMSE},
axis background/.style={fill=white},
xmajorgrids,
ymajorgrids,
legend style={legend cell align=left, align=left, fill=none, draw=none}
]
\addplot [color=blue, line width=1.5pt]
  table[row sep=crcr]{%
1	0.5\\
1.05	0.476190476190476\\
1.1	0.454545454545455\\
1.15	0.434782608695652\\
1.2	0.416666666666667\\
1.25	0.4\\
1.3	0.384615384615385\\
1.35	0.37037037037037\\
1.4	0.357142857142857\\
1.45	0.344827586206897\\
1.5	0.333333333333333\\
1.55	0.32258064516129\\
1.6	0.3125\\
1.65	0.303030303030303\\
1.7	0.294117647058823\\
1.75	0.285714285714286\\
1.8	0.277777777777778\\
1.85	0.27027027027027\\
1.9	0.263157894736842\\
1.95	0.256410256410256\\
2	0.25\\
2.05	0.24390243902439\\
2.1	0.238095238095238\\
2.15	0.232558139534884\\
2.2	0.227272727272727\\
2.25	0.222222222222222\\
2.3	0.217391304347826\\
2.35	0.212765957446809\\
2.4	0.208333333333333\\
2.45	0.204081632653061\\
2.5	0.2\\
2.55	0.196078431372549\\
2.6	0.192307692307692\\
2.65	0.188679245283019\\
2.7	0.185185185185185\\
2.75	0.181818181818182\\
2.8	0.178571428571429\\
2.85	0.175438596491228\\
2.9	0.172413793103448\\
2.95	0.169491525423729\\
3	0.166666666666667\\
3.05	0.163934426229508\\
3.1	0.161290322580645\\
3.15	0.158730158730159\\
3.2	0.15625\\
3.25	0.153846153846154\\
3.3	0.151515151515151\\
3.35	0.149253731343284\\
3.4	0.147058823529412\\
3.45	0.144927536231884\\
3.5	0.142857142857143\\
3.55	0.140845070422535\\
3.6	0.138888888888889\\
3.65	0.136986301369863\\
3.7	0.135135135135135\\
3.75	0.133333333333333\\
3.8	0.131578947368421\\
3.85	0.12987012987013\\
3.9	0.128205128205128\\
3.95	0.126582278481013\\
4	0.125\\
4.05	0.123456790123457\\
4.1	0.121951219512195\\
4.15	0.120481927710843\\
4.2	0.119047619047619\\
4.25	0.117647058823529\\
4.3	0.116279069767442\\
4.35	0.114942528735632\\
4.4	0.113636363636364\\
4.45	0.112359550561798\\
4.5	0.111111111111111\\
4.55	0.10989010989011\\
4.6	0.108695652173913\\
4.65	0.10752688172043\\
4.7	0.106382978723404\\
4.75	0.105263157894737\\
4.8	0.104166666666667\\
4.85	0.103092783505155\\
4.9	0.102040816326531\\
4.95	0.101010101010101\\
5	0.1\\
5.05	0.099009900990099\\
5.1	0.0980392156862745\\
5.15	0.0970873786407767\\
5.2	0.0961538461538461\\
5.25	0.0952380952380952\\
5.3	0.0943396226415094\\
5.35	0.0934579439252336\\
5.4	0.0925925925925926\\
5.45	0.091743119266055\\
5.5	0.0909090909090909\\
5.55	0.0900900900900901\\
5.6	0.0892857142857143\\
5.65	0.0884955752212389\\
5.7	0.087719298245614\\
5.75	0.0869565217391304\\
5.8	0.0862068965517241\\
5.85	0.0854700854700855\\
5.9	0.0847457627118644\\
5.95	0.0840336134453781\\
6	0.0833333333333333\\
6.05	0.0826446280991735\\
6.1	0.0819672131147541\\
6.15	0.0813008130081301\\
6.2	0.0806451612903226\\
6.25	0.08\\
6.3	0.0793650793650794\\
6.35	0.078740157480315\\
6.4	0.078125\\
6.45	0.0775193798449612\\
6.5	0.0769230769230769\\
6.55	0.0763358778625954\\
6.6	0.0757575757575757\\
6.65	0.075187969924812\\
6.7	0.0746268656716418\\
6.75	0.0740740740740741\\
6.8	0.0735294117647059\\
6.85	0.072992700729927\\
6.9	0.072463768115942\\
6.95	0.0719424460431655\\
7	0.0714285714285714\\
7.05	0.0709219858156028\\
7.1	0.0704225352112676\\
7.15	0.0699300699300699\\
7.2	0.0694444444444444\\
7.25	0.0689655172413793\\
7.3	0.0684931506849315\\
7.35	0.0680272108843537\\
7.4	0.0675675675675676\\
7.45	0.0671140939597315\\
7.5	0.0666666666666667\\
7.55	0.0662251655629139\\
7.6	0.0657894736842105\\
7.65	0.065359477124183\\
7.7	0.0649350649350649\\
7.75	0.0645161290322581\\
7.8	0.0641025641025641\\
7.85	0.0636942675159236\\
7.9	0.0632911392405063\\
7.95	0.0628930817610063\\
8	0.0625\\
8.05	0.062111801242236\\
8.1	0.0617283950617284\\
8.15	0.0613496932515338\\
8.2	0.0609756097560976\\
8.25	0.0606060606060606\\
8.3	0.0602409638554217\\
8.35	0.0598802395209581\\
8.4	0.0595238095238095\\
8.45	0.0591715976331361\\
8.5	0.0588235294117647\\
8.55	0.0584795321637427\\
8.6	0.0581395348837209\\
8.65	0.0578034682080925\\
8.7	0.0574712643678161\\
8.75	0.0571428571428571\\
8.8	0.0568181818181818\\
8.85	0.0564971751412429\\
8.9	0.0561797752808989\\
8.95	0.0558659217877095\\
9	0.0555555555555556\\
9.05	0.0552486187845304\\
9.1	0.0549450549450549\\
9.15	0.0546448087431694\\
9.2	0.0543478260869565\\
9.25	0.0540540540540541\\
9.3	0.053763440860215\\
9.35	0.053475935828877\\
9.4	0.0531914893617021\\
9.45	0.0529100529100529\\
9.5	0.0526315789473684\\
9.55	0.0523560209424084\\
9.6	0.0520833333333333\\
9.65	0.0518134715025907\\
9.7	0.0515463917525773\\
9.75	0.0512820512820513\\
9.8	0.0510204081632653\\
9.85	0.050761421319797\\
9.9	0.0505050505050505\\
9.95	0.050251256281407\\
10	0.05\\
10.05	0.0497512437810945\\
10.1	0.0495049504950495\\
10.15	0.0492610837438424\\
10.2	0.0490196078431373\\
10.25	0.0487804878048781\\
10.3	0.0485436893203883\\
10.35	0.0483091787439614\\
10.4	0.0480769230769231\\
10.45	0.0478468899521531\\
10.5	0.0476190476190476\\
10.55	0.0473933649289099\\
10.6	0.0471698113207547\\
10.65	0.0469483568075117\\
10.7	0.0467289719626168\\
10.75	0.0465116279069767\\
10.8	0.0462962962962963\\
10.85	0.0460829493087558\\
10.9	0.0458715596330275\\
10.95	0.045662100456621\\
11	0.0454545454545455\\
11.05	0.0452488687782805\\
11.1	0.045045045045045\\
11.15	0.0448430493273543\\
11.2	0.0446428571428571\\
11.25	0.0444444444444444\\
11.3	0.0442477876106195\\
11.35	0.0440528634361234\\
11.4	0.043859649122807\\
11.45	0.0436681222707424\\
11.5	0.0434782608695652\\
11.55	0.0432900432900433\\
11.6	0.0431034482758621\\
11.65	0.0429184549356223\\
11.7	0.0427350427350427\\
11.75	0.0425531914893617\\
11.8	0.0423728813559322\\
11.85	0.0421940928270042\\
11.9	0.0420168067226891\\
11.95	0.0418410041841004\\
12	0.0416666666666667\\
12.05	0.04149377593361\\
12.1	0.0413223140495868\\
12.15	0.0411522633744856\\
12.2	0.0409836065573771\\
12.25	0.0408163265306122\\
12.3	0.040650406504065\\
12.35	0.0404858299595142\\
12.4	0.0403225806451613\\
12.45	0.0401606425702811\\
12.5	0.04\\
12.55	0.0398406374501992\\
12.6	0.0396825396825397\\
12.65	0.0395256916996047\\
12.7	0.0393700787401575\\
12.75	0.0392156862745098\\
12.8	0.0390625\\
12.85	0.0389105058365759\\
12.9	0.0387596899224806\\
12.95	0.0386100386100386\\
13	0.0384615384615385\\
13.05	0.0383141762452107\\
13.1	0.0381679389312977\\
13.15	0.0380228136882129\\
13.2	0.0378787878787879\\
13.25	0.0377358490566038\\
13.3	0.037593984962406\\
13.35	0.0374531835205993\\
13.4	0.0373134328358209\\
13.45	0.0371747211895911\\
13.5	0.037037037037037\\
13.55	0.03690036900369\\
13.6	0.0367647058823529\\
13.65	0.0366300366300366\\
13.7	0.0364963503649635\\
13.75	0.0363636363636364\\
13.8	0.036231884057971\\
13.85	0.036101083032491\\
13.9	0.0359712230215827\\
13.95	0.03584229390681\\
14	0.0357142857142857\\
14.05	0.0355871886120996\\
14.1	0.0354609929078014\\
14.15	0.0353356890459364\\
14.2	0.0352112676056338\\
14.25	0.0350877192982456\\
14.3	0.034965034965035\\
14.35	0.0348432055749129\\
14.4	0.0347222222222222\\
14.45	0.0346020761245675\\
14.5	0.0344827586206897\\
14.55	0.0343642611683849\\
14.6	0.0342465753424658\\
14.65	0.0341296928327645\\
14.7	0.0340136054421769\\
14.75	0.0338983050847458\\
14.8	0.0337837837837838\\
14.85	0.0336700336700337\\
14.9	0.0335570469798658\\
14.95	0.0334448160535117\\
15	0.0333333333333333\\
};
\addlegendentry{$c_0 = 0.5$ ($\kappa = 0$)}

\addplot [color=blue, dashed, line width=1.5pt, forget plot]
  table[row sep=crcr]{%
1	1\\
1.05	0.952380952380952\\
1.1	0.909090909090909\\
1.15	0.869565217391304\\
1.2	0.833333333333333\\
1.25	0.8\\
1.3	0.769230769230769\\
1.35	0.740740740740741\\
1.4	0.714285714285714\\
1.45	0.689655172413793\\
1.5	0.666666666666667\\
1.55	0.645161290322581\\
1.6	0.625\\
1.65	0.606060606060606\\
1.7	0.588235294117647\\
1.75	0.571428571428571\\
1.8	0.555555555555556\\
1.85	0.54054054054054\\
1.9	0.526315789473684\\
1.95	0.512820512820513\\
2	0.5\\
2.05	0.487804878048781\\
2.1	0.476190476190476\\
2.15	0.465116279069767\\
2.2	0.454545454545455\\
2.25	0.444444444444444\\
2.3	0.434782608695652\\
2.35	0.425531914893617\\
2.4	0.416666666666667\\
2.45	0.408163265306122\\
2.5	0.4\\
2.55	0.392156862745098\\
2.6	0.384615384615385\\
2.65	0.377358490566038\\
2.7	0.37037037037037\\
2.75	0.363636363636364\\
2.8	0.357142857142857\\
2.85	0.350877192982456\\
2.9	0.344827586206897\\
2.95	0.338983050847458\\
3	0.333333333333333\\
3.05	0.327868852459016\\
3.1	0.32258064516129\\
3.15	0.317460317460317\\
3.2	0.3125\\
3.25	0.307692307692308\\
3.3	0.303030303030303\\
3.35	0.298507462686567\\
3.4	0.294117647058823\\
3.45	0.289855072463768\\
3.5	0.285714285714286\\
3.55	0.28169014084507\\
3.6	0.277777777777778\\
3.65	0.273972602739726\\
3.7	0.27027027027027\\
3.75	0.266666666666667\\
3.8	0.263157894736842\\
3.85	0.25974025974026\\
3.9	0.256410256410256\\
3.95	0.253164556962025\\
4	0.25\\
4.05	0.246913580246914\\
4.1	0.24390243902439\\
4.15	0.240963855421687\\
4.2	0.238095238095238\\
4.25	0.235294117647059\\
4.3	0.232558139534884\\
4.35	0.229885057471264\\
4.4	0.227272727272727\\
4.45	0.224719101123595\\
4.5	0.222222222222222\\
4.55	0.21978021978022\\
4.6	0.217391304347826\\
4.65	0.21505376344086\\
4.7	0.212765957446809\\
4.75	0.210526315789474\\
4.8	0.208333333333333\\
4.85	0.206185567010309\\
4.9	0.204081632653061\\
4.95	0.202020202020202\\
5	0.2\\
5.05	0.198019801980198\\
5.1	0.196078431372549\\
5.15	0.194174757281553\\
5.2	0.192307692307692\\
5.25	0.19047619047619\\
5.3	0.188679245283019\\
5.35	0.186915887850467\\
5.4	0.185185185185185\\
5.45	0.18348623853211\\
5.5	0.181818181818182\\
5.55	0.18018018018018\\
5.6	0.178571428571429\\
5.65	0.176991150442478\\
5.7	0.175438596491228\\
5.75	0.173913043478261\\
5.8	0.172413793103448\\
5.85	0.170940170940171\\
5.9	0.169491525423729\\
5.95	0.168067226890756\\
6	0.166666666666667\\
6.05	0.165289256198347\\
6.1	0.163934426229508\\
6.15	0.16260162601626\\
6.2	0.161290322580645\\
6.25	0.16\\
6.3	0.158730158730159\\
6.35	0.15748031496063\\
6.4	0.15625\\
6.45	0.155038759689922\\
6.5	0.153846153846154\\
6.55	0.152671755725191\\
6.6	0.151515151515151\\
6.65	0.150375939849624\\
6.7	0.149253731343284\\
6.75	0.148148148148148\\
6.8	0.147058823529412\\
6.85	0.145985401459854\\
6.9	0.144927536231884\\
6.95	0.143884892086331\\
7	0.142857142857143\\
7.05	0.141843971631206\\
7.1	0.140845070422535\\
7.15	0.13986013986014\\
7.2	0.138888888888889\\
7.25	0.137931034482759\\
7.3	0.136986301369863\\
7.35	0.136054421768707\\
7.4	0.135135135135135\\
7.45	0.134228187919463\\
7.5	0.133333333333333\\
7.55	0.132450331125828\\
7.6	0.131578947368421\\
7.65	0.130718954248366\\
7.7	0.12987012987013\\
7.75	0.129032258064516\\
7.8	0.128205128205128\\
7.85	0.127388535031847\\
7.9	0.126582278481013\\
7.95	0.125786163522013\\
8	0.125\\
8.05	0.124223602484472\\
8.1	0.123456790123457\\
8.15	0.122699386503068\\
8.2	0.121951219512195\\
8.25	0.121212121212121\\
8.3	0.120481927710843\\
8.35	0.119760479041916\\
8.4	0.119047619047619\\
8.45	0.118343195266272\\
8.5	0.117647058823529\\
8.55	0.116959064327485\\
8.6	0.116279069767442\\
8.65	0.115606936416185\\
8.7	0.114942528735632\\
8.75	0.114285714285714\\
8.8	0.113636363636364\\
8.85	0.112994350282486\\
8.9	0.112359550561798\\
8.95	0.111731843575419\\
9	0.111111111111111\\
9.05	0.110497237569061\\
9.1	0.10989010989011\\
9.15	0.109289617486339\\
9.2	0.108695652173913\\
9.25	0.108108108108108\\
9.3	0.10752688172043\\
9.35	0.106951871657754\\
9.4	0.106382978723404\\
9.45	0.105820105820106\\
9.5	0.105263157894737\\
9.55	0.104712041884817\\
9.6	0.104166666666667\\
9.65	0.103626943005181\\
9.7	0.103092783505155\\
9.75	0.102564102564103\\
9.8	0.102040816326531\\
9.85	0.101522842639594\\
9.9	0.101010101010101\\
9.95	0.100502512562814\\
10	0.1\\
10.05	0.099502487562189\\
10.1	0.099009900990099\\
10.15	0.0985221674876847\\
10.2	0.0980392156862745\\
10.25	0.0975609756097561\\
10.3	0.0970873786407767\\
10.35	0.0966183574879227\\
10.4	0.0961538461538462\\
10.45	0.0956937799043062\\
10.5	0.0952380952380952\\
10.55	0.0947867298578199\\
10.6	0.0943396226415094\\
10.65	0.0938967136150235\\
10.7	0.0934579439252337\\
10.75	0.0930232558139535\\
10.8	0.0925925925925926\\
10.85	0.0921658986175115\\
10.9	0.0917431192660551\\
10.95	0.091324200913242\\
11	0.0909090909090909\\
11.05	0.0904977375565611\\
11.1	0.0900900900900901\\
11.15	0.0896860986547085\\
11.2	0.0892857142857143\\
11.25	0.0888888888888889\\
11.3	0.0884955752212389\\
11.35	0.0881057268722467\\
11.4	0.087719298245614\\
11.45	0.0873362445414847\\
11.5	0.0869565217391304\\
11.55	0.0865800865800866\\
11.6	0.0862068965517241\\
11.65	0.0858369098712446\\
11.7	0.0854700854700855\\
11.75	0.0851063829787234\\
11.8	0.0847457627118644\\
11.85	0.0843881856540084\\
11.9	0.0840336134453781\\
11.95	0.0836820083682008\\
12	0.0833333333333333\\
12.05	0.0829875518672199\\
12.1	0.0826446280991736\\
12.15	0.0823045267489712\\
12.2	0.0819672131147541\\
12.25	0.0816326530612245\\
12.3	0.0813008130081301\\
12.35	0.0809716599190283\\
12.4	0.0806451612903226\\
12.45	0.0803212851405622\\
12.5	0.08\\
12.55	0.0796812749003984\\
12.6	0.0793650793650794\\
12.65	0.0790513833992095\\
12.7	0.078740157480315\\
12.75	0.0784313725490196\\
12.8	0.078125\\
12.85	0.0778210116731518\\
12.9	0.0775193798449612\\
12.95	0.0772200772200772\\
13	0.0769230769230769\\
13.05	0.0766283524904215\\
13.1	0.0763358778625954\\
13.15	0.0760456273764259\\
13.2	0.0757575757575758\\
13.25	0.0754716981132075\\
13.3	0.075187969924812\\
13.35	0.0749063670411985\\
13.4	0.0746268656716418\\
13.45	0.0743494423791822\\
13.5	0.0740740740740741\\
13.55	0.0738007380073801\\
13.6	0.0735294117647059\\
13.65	0.0732600732600733\\
13.7	0.072992700729927\\
13.75	0.0727272727272727\\
13.8	0.072463768115942\\
13.85	0.0722021660649819\\
13.9	0.0719424460431655\\
13.95	0.0716845878136201\\
14	0.0714285714285714\\
14.05	0.0711743772241993\\
14.1	0.0709219858156028\\
14.15	0.0706713780918728\\
14.2	0.0704225352112676\\
14.25	0.0701754385964912\\
14.3	0.0699300699300699\\
14.35	0.0696864111498258\\
14.4	0.0694444444444444\\
14.45	0.069204152249135\\
14.5	0.0689655172413793\\
14.55	0.0687285223367698\\
14.6	0.0684931506849315\\
14.65	0.068259385665529\\
14.7	0.0680272108843537\\
14.75	0.0677966101694915\\
14.8	0.0675675675675676\\
14.85	0.0673400673400673\\
14.9	0.0671140939597315\\
14.95	0.0668896321070234\\
15	0.0666666666666667\\
};
\addplot [color=red, line width=1.5pt]
  table[row sep=crcr]{%
1	2\\
1.05	1.9047619047619\\
1.1	1.81818181818182\\
1.15	1.73913043478261\\
1.2	1.66666666666667\\
1.25	1.6\\
1.3	1.53846153846154\\
1.35	1.48148148148148\\
1.4	1.42857142857143\\
1.45	1.37931034482759\\
1.5	1.33333333333333\\
1.55	1.29032258064516\\
1.6	1.25\\
1.65	1.21212121212121\\
1.7	1.17647058823529\\
1.75	1.14285714285714\\
1.8	1.11111111111111\\
1.85	1.08108108108108\\
1.9	1.05263157894737\\
1.95	1.02564102564103\\
2	1\\
2.05	0.975609756097561\\
2.1	0.952380952380952\\
2.15	0.930232558139535\\
2.2	0.909090909090909\\
2.25	0.888888888888889\\
2.3	0.869565217391304\\
2.35	0.851063829787234\\
2.4	0.833333333333333\\
2.45	0.816326530612245\\
2.5	0.8\\
2.55	0.784313725490196\\
2.6	0.769230769230769\\
2.65	0.754716981132075\\
2.7	0.740740740740741\\
2.75	0.727272727272727\\
2.8	0.714285714285714\\
2.85	0.701754385964912\\
2.9	0.689655172413793\\
2.95	0.677966101694915\\
3	0.666666666666667\\
3.05	0.655737704918033\\
3.1	0.645161290322581\\
3.15	0.634920634920635\\
3.2	0.625\\
3.25	0.615384615384615\\
3.3	0.606060606060606\\
3.35	0.597014925373134\\
3.4	0.588235294117647\\
3.45	0.579710144927536\\
3.5	0.571428571428571\\
3.55	0.563380281690141\\
3.6	0.555555555555556\\
3.65	0.547945205479452\\
3.7	0.54054054054054\\
3.75	0.533333333333333\\
3.8	0.526315789473684\\
3.85	0.519480519480519\\
3.9	0.512820512820513\\
3.95	0.506329113924051\\
4	0.5\\
4.05	0.493827160493827\\
4.1	0.487804878048781\\
4.15	0.481927710843373\\
4.2	0.476190476190476\\
4.25	0.470588235294118\\
4.3	0.465116279069767\\
4.35	0.459770114942529\\
4.4	0.454545454545455\\
4.45	0.449438202247191\\
4.5	0.444444444444444\\
4.55	0.439560439560439\\
4.6	0.434782608695652\\
4.65	0.43010752688172\\
4.7	0.425531914893617\\
4.75	0.421052631578947\\
4.8	0.416666666666667\\
4.85	0.412371134020619\\
4.9	0.408163265306122\\
4.95	0.404040404040404\\
5	0.4\\
5.05	0.396039603960396\\
5.1	0.392156862745098\\
5.15	0.388349514563107\\
5.2	0.384615384615385\\
5.25	0.380952380952381\\
5.3	0.377358490566038\\
5.35	0.373831775700935\\
5.4	0.37037037037037\\
5.45	0.36697247706422\\
5.5	0.363636363636364\\
5.55	0.36036036036036\\
5.6	0.357142857142857\\
5.65	0.353982300884956\\
5.7	0.350877192982456\\
5.75	0.347826086956522\\
5.8	0.344827586206897\\
5.85	0.341880341880342\\
5.9	0.338983050847458\\
5.95	0.336134453781513\\
6	0.333333333333333\\
6.05	0.330578512396694\\
6.1	0.327868852459016\\
6.15	0.32520325203252\\
6.2	0.32258064516129\\
6.25	0.32\\
6.3	0.317460317460317\\
6.35	0.31496062992126\\
6.4	0.3125\\
6.45	0.310077519379845\\
6.5	0.307692307692308\\
6.55	0.305343511450382\\
6.6	0.303030303030303\\
6.65	0.300751879699248\\
6.7	0.298507462686567\\
6.75	0.296296296296296\\
6.8	0.294117647058823\\
6.85	0.291970802919708\\
6.9	0.289855072463768\\
6.95	0.287769784172662\\
7	0.285714285714286\\
7.05	0.283687943262411\\
7.1	0.28169014084507\\
7.15	0.27972027972028\\
7.2	0.277777777777778\\
7.25	0.275862068965517\\
7.3	0.273972602739726\\
7.35	0.272108843537415\\
7.4	0.27027027027027\\
7.45	0.268456375838926\\
7.5	0.266666666666667\\
7.55	0.264900662251656\\
7.6	0.263157894736842\\
7.65	0.261437908496732\\
7.7	0.25974025974026\\
7.75	0.258064516129032\\
7.8	0.256410256410256\\
7.85	0.254777070063694\\
7.9	0.253164556962025\\
7.95	0.251572327044025\\
8	0.25\\
8.05	0.248447204968944\\
8.1	0.246913580246914\\
8.15	0.245398773006135\\
8.2	0.24390243902439\\
8.25	0.242424242424242\\
8.3	0.240963855421687\\
8.35	0.239520958083832\\
8.4	0.238095238095238\\
8.45	0.236686390532544\\
8.5	0.235294117647059\\
8.55	0.233918128654971\\
8.6	0.232558139534884\\
8.65	0.23121387283237\\
8.7	0.229885057471264\\
8.75	0.228571428571429\\
8.8	0.227272727272727\\
8.85	0.225988700564972\\
8.9	0.224719101123596\\
8.95	0.223463687150838\\
9	0.222222222222222\\
9.05	0.220994475138122\\
9.1	0.21978021978022\\
9.15	0.218579234972678\\
9.2	0.217391304347826\\
9.25	0.216216216216216\\
9.3	0.21505376344086\\
9.35	0.213903743315508\\
9.4	0.212765957446809\\
9.45	0.211640211640212\\
9.5	0.210526315789474\\
9.55	0.209424083769633\\
9.6	0.208333333333333\\
9.65	0.207253886010363\\
9.7	0.206185567010309\\
9.75	0.205128205128205\\
9.8	0.204081632653061\\
9.85	0.203045685279188\\
9.9	0.202020202020202\\
9.95	0.201005025125628\\
10	0.2\\
10.05	0.199004975124378\\
10.1	0.198019801980198\\
10.15	0.197044334975369\\
10.2	0.196078431372549\\
10.25	0.195121951219512\\
10.3	0.194174757281553\\
10.35	0.193236714975845\\
10.4	0.192307692307692\\
10.45	0.191387559808612\\
10.5	0.19047619047619\\
10.55	0.18957345971564\\
10.6	0.188679245283019\\
10.65	0.187793427230047\\
10.7	0.186915887850467\\
10.75	0.186046511627907\\
10.8	0.185185185185185\\
10.85	0.184331797235023\\
10.9	0.18348623853211\\
10.95	0.182648401826484\\
11	0.181818181818182\\
11.05	0.180995475113122\\
11.1	0.18018018018018\\
11.15	0.179372197309417\\
11.2	0.178571428571429\\
11.25	0.177777777777778\\
11.3	0.176991150442478\\
11.35	0.176211453744493\\
11.4	0.175438596491228\\
11.45	0.174672489082969\\
11.5	0.173913043478261\\
11.55	0.173160173160173\\
11.6	0.172413793103448\\
11.65	0.171673819742489\\
11.7	0.170940170940171\\
11.75	0.170212765957447\\
11.8	0.169491525423729\\
11.85	0.168776371308017\\
11.9	0.168067226890756\\
11.95	0.167364016736402\\
12	0.166666666666667\\
12.05	0.16597510373444\\
12.1	0.165289256198347\\
12.15	0.164609053497942\\
12.2	0.163934426229508\\
12.25	0.163265306122449\\
12.3	0.16260162601626\\
12.35	0.161943319838057\\
12.4	0.161290322580645\\
12.45	0.160642570281124\\
12.5	0.16\\
12.55	0.159362549800797\\
12.6	0.158730158730159\\
12.65	0.158102766798419\\
12.7	0.15748031496063\\
12.75	0.156862745098039\\
12.8	0.15625\\
12.85	0.155642023346304\\
12.9	0.155038759689922\\
12.95	0.154440154440154\\
13	0.153846153846154\\
13.05	0.153256704980843\\
13.1	0.152671755725191\\
13.15	0.152091254752852\\
13.2	0.151515151515152\\
13.25	0.150943396226415\\
13.3	0.150375939849624\\
13.35	0.149812734082397\\
13.4	0.149253731343284\\
13.45	0.148698884758364\\
13.5	0.148148148148148\\
13.55	0.14760147601476\\
13.6	0.147058823529412\\
13.65	0.146520146520147\\
13.7	0.145985401459854\\
13.75	0.145454545454545\\
13.8	0.144927536231884\\
13.85	0.144404332129964\\
13.9	0.143884892086331\\
13.95	0.14336917562724\\
14	0.142857142857143\\
14.05	0.142348754448399\\
14.1	0.141843971631206\\
14.15	0.141342756183746\\
14.2	0.140845070422535\\
14.25	0.140350877192982\\
14.3	0.13986013986014\\
14.35	0.139372822299652\\
14.4	0.138888888888889\\
14.45	0.13840830449827\\
14.5	0.137931034482759\\
14.55	0.13745704467354\\
14.6	0.136986301369863\\
14.65	0.136518771331058\\
14.7	0.136054421768707\\
14.75	0.135593220338983\\
14.8	0.135135135135135\\
14.85	0.134680134680135\\
14.9	0.134228187919463\\
14.95	0.133779264214047\\
15	0.133333333333333\\
};
\addlegendentry{$c_0 = 2.0$ ($\kappa =0$)}

\addplot [color=red, dashed, line width=1.5pt, forget plot]
  table[row sep=crcr]{%
1	4\\
1.05	3.80952380952381\\
1.1	3.63636363636364\\
1.15	3.47826086956522\\
1.2	3.33333333333333\\
1.25	3.2\\
1.3	3.07692307692308\\
1.35	2.96296296296296\\
1.4	2.85714285714286\\
1.45	2.75862068965517\\
1.5	2.66666666666667\\
1.55	2.58064516129032\\
1.6	2.5\\
1.65	2.42424242424242\\
1.7	2.35294117647059\\
1.75	2.28571428571429\\
1.8	2.22222222222222\\
1.85	2.16216216216216\\
1.9	2.10526315789474\\
1.95	2.05128205128205\\
2	2\\
2.05	1.95121951219512\\
2.1	1.9047619047619\\
2.15	1.86046511627907\\
2.2	1.81818181818182\\
2.25	1.77777777777778\\
2.3	1.73913043478261\\
2.35	1.70212765957447\\
2.4	1.66666666666667\\
2.45	1.63265306122449\\
2.5	1.6\\
2.55	1.56862745098039\\
2.6	1.53846153846154\\
2.65	1.50943396226415\\
2.7	1.48148148148148\\
2.75	1.45454545454545\\
2.8	1.42857142857143\\
2.85	1.40350877192982\\
2.9	1.37931034482759\\
2.95	1.35593220338983\\
3	1.33333333333333\\
3.05	1.31147540983607\\
3.1	1.29032258064516\\
3.15	1.26984126984127\\
3.2	1.25\\
3.25	1.23076923076923\\
3.3	1.21212121212121\\
3.35	1.19402985074627\\
3.4	1.17647058823529\\
3.45	1.15942028985507\\
3.5	1.14285714285714\\
3.55	1.12676056338028\\
3.6	1.11111111111111\\
3.65	1.0958904109589\\
3.7	1.08108108108108\\
3.75	1.06666666666667\\
3.8	1.05263157894737\\
3.85	1.03896103896104\\
3.9	1.02564102564103\\
3.95	1.0126582278481\\
4	1\\
4.05	0.987654320987654\\
4.1	0.975609756097561\\
4.15	0.963855421686747\\
4.2	0.952380952380952\\
4.25	0.941176470588235\\
4.3	0.930232558139535\\
4.35	0.919540229885058\\
4.4	0.909090909090909\\
4.45	0.898876404494382\\
4.5	0.888888888888889\\
4.55	0.879120879120879\\
4.6	0.869565217391304\\
4.65	0.860215053763441\\
4.7	0.851063829787234\\
4.75	0.842105263157895\\
4.8	0.833333333333333\\
4.85	0.824742268041237\\
4.9	0.816326530612245\\
4.95	0.808080808080808\\
5	0.8\\
5.05	0.792079207920792\\
5.1	0.784313725490196\\
5.15	0.776699029126214\\
5.2	0.769230769230769\\
5.25	0.761904761904762\\
5.3	0.754716981132076\\
5.35	0.747663551401869\\
5.4	0.740740740740741\\
5.45	0.73394495412844\\
5.5	0.727272727272727\\
5.55	0.720720720720721\\
5.6	0.714285714285714\\
5.65	0.707964601769911\\
5.7	0.701754385964912\\
5.75	0.695652173913043\\
5.8	0.689655172413793\\
5.85	0.683760683760684\\
5.9	0.677966101694915\\
5.95	0.672268907563025\\
6	0.666666666666667\\
6.05	0.661157024793388\\
6.1	0.655737704918033\\
6.15	0.650406504065041\\
6.2	0.645161290322581\\
6.25	0.64\\
6.3	0.634920634920635\\
6.35	0.62992125984252\\
6.4	0.625\\
6.45	0.62015503875969\\
6.5	0.615384615384615\\
6.55	0.610687022900763\\
6.6	0.606060606060606\\
6.65	0.601503759398496\\
6.7	0.597014925373134\\
6.75	0.592592592592593\\
6.8	0.588235294117647\\
6.85	0.583941605839416\\
6.9	0.579710144927536\\
6.95	0.575539568345324\\
7	0.571428571428571\\
7.05	0.567375886524823\\
7.1	0.563380281690141\\
7.15	0.559440559440559\\
7.2	0.555555555555556\\
7.25	0.551724137931034\\
7.3	0.547945205479452\\
7.35	0.54421768707483\\
7.4	0.54054054054054\\
7.45	0.536912751677852\\
7.5	0.533333333333333\\
7.55	0.529801324503311\\
7.6	0.526315789473684\\
7.65	0.522875816993464\\
7.7	0.519480519480519\\
7.75	0.516129032258065\\
7.8	0.512820512820513\\
7.85	0.509554140127389\\
7.9	0.506329113924051\\
7.95	0.50314465408805\\
8	0.5\\
8.05	0.496894409937888\\
8.1	0.493827160493827\\
8.15	0.49079754601227\\
8.2	0.487804878048781\\
8.25	0.484848484848485\\
8.3	0.481927710843373\\
8.35	0.479041916167665\\
8.4	0.476190476190476\\
8.45	0.473372781065089\\
8.5	0.470588235294118\\
8.55	0.467836257309941\\
8.6	0.465116279069767\\
8.65	0.46242774566474\\
8.7	0.459770114942529\\
8.75	0.457142857142857\\
8.8	0.454545454545455\\
8.85	0.451977401129944\\
8.9	0.449438202247191\\
8.95	0.446927374301676\\
9	0.444444444444444\\
9.05	0.441988950276243\\
9.1	0.43956043956044\\
9.15	0.437158469945355\\
9.2	0.434782608695652\\
9.25	0.432432432432432\\
9.3	0.43010752688172\\
9.35	0.427807486631016\\
9.4	0.425531914893617\\
9.45	0.423280423280423\\
9.5	0.421052631578947\\
9.55	0.418848167539267\\
9.6	0.416666666666667\\
9.65	0.414507772020725\\
9.7	0.412371134020619\\
9.75	0.41025641025641\\
9.8	0.408163265306122\\
9.85	0.406091370558376\\
9.9	0.404040404040404\\
9.95	0.402010050251256\\
10	0.4\\
10.05	0.398009950248756\\
10.1	0.396039603960396\\
10.15	0.394088669950739\\
10.2	0.392156862745098\\
10.25	0.390243902439024\\
10.3	0.388349514563107\\
10.35	0.386473429951691\\
10.4	0.384615384615385\\
10.45	0.382775119617225\\
10.5	0.380952380952381\\
10.55	0.37914691943128\\
10.6	0.377358490566038\\
10.65	0.375586854460094\\
10.7	0.373831775700935\\
10.75	0.372093023255814\\
10.8	0.37037037037037\\
10.85	0.368663594470046\\
10.9	0.36697247706422\\
10.95	0.365296803652968\\
11	0.363636363636364\\
11.05	0.361990950226244\\
11.1	0.36036036036036\\
11.15	0.358744394618834\\
11.2	0.357142857142857\\
11.25	0.355555555555556\\
11.3	0.353982300884956\\
11.35	0.352422907488987\\
11.4	0.350877192982456\\
11.45	0.349344978165939\\
11.5	0.347826086956522\\
11.55	0.346320346320346\\
11.6	0.344827586206897\\
11.65	0.343347639484979\\
11.7	0.341880341880342\\
11.75	0.340425531914894\\
11.8	0.338983050847458\\
11.85	0.337552742616034\\
11.9	0.336134453781513\\
11.95	0.334728033472803\\
12	0.333333333333333\\
12.05	0.33195020746888\\
12.1	0.330578512396694\\
12.15	0.329218106995885\\
12.2	0.327868852459016\\
12.25	0.326530612244898\\
12.3	0.32520325203252\\
12.35	0.323886639676113\\
12.4	0.32258064516129\\
12.45	0.321285140562249\\
12.5	0.32\\
12.55	0.318725099601594\\
12.6	0.317460317460317\\
12.65	0.316205533596838\\
12.7	0.31496062992126\\
12.75	0.313725490196078\\
12.8	0.3125\\
12.85	0.311284046692607\\
12.9	0.310077519379845\\
12.95	0.308880308880309\\
13	0.307692307692308\\
13.05	0.306513409961686\\
13.1	0.305343511450382\\
13.15	0.304182509505703\\
13.2	0.303030303030303\\
13.25	0.30188679245283\\
13.3	0.300751879699248\\
13.35	0.299625468164794\\
13.4	0.298507462686567\\
13.45	0.297397769516729\\
13.5	0.296296296296296\\
13.55	0.29520295202952\\
13.6	0.294117647058824\\
13.65	0.293040293040293\\
13.7	0.291970802919708\\
13.75	0.290909090909091\\
13.8	0.289855072463768\\
13.85	0.288808664259928\\
13.9	0.287769784172662\\
13.95	0.28673835125448\\
14	0.285714285714286\\
14.05	0.284697508896797\\
14.1	0.283687943262411\\
14.15	0.282685512367491\\
14.2	0.28169014084507\\
14.25	0.280701754385965\\
14.3	0.27972027972028\\
14.35	0.278745644599303\\
14.4	0.277777777777778\\
14.45	0.27681660899654\\
14.5	0.275862068965517\\
14.55	0.274914089347079\\
14.6	0.273972602739726\\
14.65	0.273037542662116\\
14.7	0.272108843537415\\
14.75	0.271186440677966\\
14.8	0.27027027027027\\
14.85	0.269360269360269\\
14.9	0.268456375838926\\
14.95	0.267558528428094\\
15	0.266666666666667\\
};
\addplot [color=black, line width=1.5pt]
  table[row sep=crcr]{%
1	10\\
1.05	9.52380952380952\\
1.1	9.09090909090909\\
1.15	8.69565217391304\\
1.2	8.33333333333333\\
1.25	8\\
1.3	7.69230769230769\\
1.35	7.40740740740741\\
1.4	7.14285714285714\\
1.45	6.89655172413793\\
1.5	6.66666666666667\\
1.55	6.45161290322581\\
1.6	6.25\\
1.65	6.06060606060606\\
1.7	5.88235294117647\\
1.75	5.71428571428571\\
1.8	5.55555555555556\\
1.85	5.4054054054054\\
1.9	5.26315789473684\\
1.95	5.12820512820513\\
2	5\\
2.05	4.87804878048781\\
2.1	4.76190476190476\\
2.15	4.65116279069767\\
2.2	4.54545454545454\\
2.25	4.44444444444444\\
2.3	4.34782608695652\\
2.35	4.25531914893617\\
2.4	4.16666666666667\\
2.45	4.08163265306122\\
2.5	4\\
2.55	3.92156862745098\\
2.6	3.84615384615385\\
2.65	3.77358490566038\\
2.7	3.7037037037037\\
2.75	3.63636363636364\\
2.8	3.57142857142857\\
2.85	3.50877192982456\\
2.9	3.44827586206897\\
2.95	3.38983050847458\\
3	3.33333333333333\\
3.05	3.27868852459016\\
3.1	3.2258064516129\\
3.15	3.17460317460317\\
3.2	3.125\\
3.25	3.07692307692308\\
3.3	3.03030303030303\\
3.35	2.98507462686567\\
3.4	2.94117647058823\\
3.45	2.89855072463768\\
3.5	2.85714285714286\\
3.55	2.8169014084507\\
3.6	2.77777777777778\\
3.65	2.73972602739726\\
3.7	2.7027027027027\\
3.75	2.66666666666667\\
3.8	2.63157894736842\\
3.85	2.5974025974026\\
3.9	2.56410256410256\\
3.95	2.53164556962025\\
4	2.5\\
4.05	2.46913580246914\\
4.1	2.4390243902439\\
4.15	2.40963855421687\\
4.2	2.38095238095238\\
4.25	2.35294117647059\\
4.3	2.32558139534884\\
4.35	2.29885057471264\\
4.4	2.27272727272727\\
4.45	2.24719101123596\\
4.5	2.22222222222222\\
4.55	2.1978021978022\\
4.6	2.17391304347826\\
4.65	2.1505376344086\\
4.7	2.12765957446809\\
4.75	2.10526315789474\\
4.8	2.08333333333333\\
4.85	2.06185567010309\\
4.9	2.04081632653061\\
4.95	2.02020202020202\\
5	2\\
5.05	1.98019801980198\\
5.1	1.96078431372549\\
5.15	1.94174757281553\\
5.2	1.92307692307692\\
5.25	1.9047619047619\\
5.3	1.88679245283019\\
5.35	1.86915887850467\\
5.4	1.85185185185185\\
5.45	1.8348623853211\\
5.5	1.81818181818182\\
5.55	1.8018018018018\\
5.6	1.78571428571429\\
5.65	1.76991150442478\\
5.7	1.75438596491228\\
5.75	1.73913043478261\\
5.8	1.72413793103448\\
5.85	1.70940170940171\\
5.9	1.69491525423729\\
5.95	1.68067226890756\\
6	1.66666666666667\\
6.05	1.65289256198347\\
6.1	1.63934426229508\\
6.15	1.6260162601626\\
6.2	1.61290322580645\\
6.25	1.6\\
6.3	1.58730158730159\\
6.35	1.5748031496063\\
6.4	1.5625\\
6.45	1.55038759689922\\
6.5	1.53846153846154\\
6.55	1.52671755725191\\
6.6	1.51515151515151\\
6.65	1.50375939849624\\
6.7	1.49253731343284\\
6.75	1.48148148148148\\
6.8	1.47058823529412\\
6.85	1.45985401459854\\
6.9	1.44927536231884\\
6.95	1.43884892086331\\
7	1.42857142857143\\
7.05	1.41843971631206\\
7.1	1.40845070422535\\
7.15	1.3986013986014\\
7.2	1.38888888888889\\
7.25	1.37931034482759\\
7.3	1.36986301369863\\
7.35	1.36054421768707\\
7.4	1.35135135135135\\
7.45	1.34228187919463\\
7.5	1.33333333333333\\
7.55	1.32450331125828\\
7.6	1.31578947368421\\
7.65	1.30718954248366\\
7.7	1.2987012987013\\
7.75	1.29032258064516\\
7.8	1.28205128205128\\
7.85	1.27388535031847\\
7.9	1.26582278481013\\
7.95	1.25786163522013\\
8	1.25\\
8.05	1.24223602484472\\
8.1	1.23456790123457\\
8.15	1.22699386503068\\
8.2	1.21951219512195\\
8.25	1.21212121212121\\
8.3	1.20481927710843\\
8.35	1.19760479041916\\
8.4	1.19047619047619\\
8.45	1.18343195266272\\
8.5	1.17647058823529\\
8.55	1.16959064327485\\
8.6	1.16279069767442\\
8.65	1.15606936416185\\
8.7	1.14942528735632\\
8.75	1.14285714285714\\
8.8	1.13636363636364\\
8.85	1.12994350282486\\
8.9	1.12359550561798\\
8.95	1.11731843575419\\
9	1.11111111111111\\
9.05	1.10497237569061\\
9.1	1.0989010989011\\
9.15	1.09289617486339\\
9.2	1.08695652173913\\
9.25	1.08108108108108\\
9.3	1.0752688172043\\
9.35	1.06951871657754\\
9.4	1.06382978723404\\
9.45	1.05820105820106\\
9.5	1.05263157894737\\
9.55	1.04712041884817\\
9.6	1.04166666666667\\
9.65	1.03626943005181\\
9.7	1.03092783505155\\
9.75	1.02564102564103\\
9.8	1.02040816326531\\
9.85	1.01522842639594\\
9.9	1.01010101010101\\
9.95	1.00502512562814\\
10	1\\
10.05	0.99502487562189\\
10.1	0.99009900990099\\
10.15	0.985221674876847\\
10.2	0.980392156862745\\
10.25	0.975609756097561\\
10.3	0.970873786407767\\
10.35	0.966183574879227\\
10.4	0.961538461538462\\
10.45	0.956937799043062\\
10.5	0.952380952380952\\
10.55	0.947867298578199\\
10.6	0.943396226415094\\
10.65	0.938967136150235\\
10.7	0.934579439252337\\
10.75	0.930232558139535\\
10.8	0.925925925925926\\
10.85	0.921658986175115\\
10.9	0.917431192660551\\
10.95	0.91324200913242\\
11	0.909090909090909\\
11.05	0.904977375565611\\
11.1	0.900900900900901\\
11.15	0.896860986547085\\
11.2	0.892857142857143\\
11.25	0.888888888888889\\
11.3	0.884955752212389\\
11.35	0.881057268722467\\
11.4	0.87719298245614\\
11.45	0.873362445414847\\
11.5	0.869565217391304\\
11.55	0.865800865800866\\
11.6	0.862068965517241\\
11.65	0.858369098712446\\
11.7	0.854700854700855\\
11.75	0.851063829787234\\
11.8	0.847457627118644\\
11.85	0.843881856540084\\
11.9	0.840336134453781\\
11.95	0.836820083682008\\
12	0.833333333333333\\
12.05	0.829875518672199\\
12.1	0.826446280991736\\
12.15	0.823045267489712\\
12.2	0.819672131147541\\
12.25	0.816326530612245\\
12.3	0.813008130081301\\
12.35	0.809716599190283\\
12.4	0.806451612903226\\
12.45	0.803212851405622\\
12.5	0.8\\
12.55	0.796812749003984\\
12.6	0.793650793650794\\
12.65	0.790513833992095\\
12.7	0.78740157480315\\
12.75	0.784313725490196\\
12.8	0.78125\\
12.85	0.778210116731518\\
12.9	0.775193798449612\\
12.95	0.772200772200772\\
13	0.769230769230769\\
13.05	0.766283524904214\\
13.1	0.763358778625954\\
13.15	0.760456273764259\\
13.2	0.757575757575758\\
13.25	0.754716981132075\\
13.3	0.75187969924812\\
13.35	0.749063670411985\\
13.4	0.746268656716418\\
13.45	0.743494423791822\\
13.5	0.740740740740741\\
13.55	0.738007380073801\\
13.6	0.735294117647059\\
13.65	0.732600732600733\\
13.7	0.72992700729927\\
13.75	0.727272727272727\\
13.8	0.72463768115942\\
13.85	0.72202166064982\\
13.9	0.719424460431655\\
13.95	0.716845878136201\\
14	0.714285714285714\\
14.05	0.711743772241993\\
14.1	0.709219858156028\\
14.15	0.706713780918728\\
14.2	0.704225352112676\\
14.25	0.701754385964912\\
14.3	0.699300699300699\\
14.35	0.696864111498258\\
14.4	0.694444444444444\\
14.45	0.69204152249135\\
14.5	0.689655172413793\\
14.55	0.687285223367697\\
14.6	0.684931506849315\\
14.65	0.68259385665529\\
14.7	0.680272108843538\\
14.75	0.677966101694915\\
14.8	0.675675675675676\\
14.85	0.673400673400673\\
14.9	0.671140939597315\\
14.95	0.668896321070234\\
15	0.666666666666667\\
};
\addlegendentry{$c_0 = 10$ \:($\kappa=0$)}

\addplot [color=black, dashed, line width=1.5pt, forget plot]
  table[row sep=crcr]{%
1	20\\
1.05	19.047619047619\\
1.1	18.1818181818182\\
1.15	17.3913043478261\\
1.2	16.6666666666667\\
1.25	16\\
1.3	15.3846153846154\\
1.35	14.8148148148148\\
1.4	14.2857142857143\\
1.45	13.7931034482759\\
1.5	13.3333333333333\\
1.55	12.9032258064516\\
1.6	12.5\\
1.65	12.1212121212121\\
1.7	11.7647058823529\\
1.75	11.4285714285714\\
1.8	11.1111111111111\\
1.85	10.8108108108108\\
1.9	10.5263157894737\\
1.95	10.2564102564103\\
2	10\\
2.05	9.75609756097561\\
2.1	9.52380952380952\\
2.15	9.30232558139535\\
2.2	9.09090909090909\\
2.25	8.88888888888889\\
2.3	8.69565217391304\\
2.35	8.51063829787234\\
2.4	8.33333333333333\\
2.45	8.16326530612245\\
2.5	8\\
2.55	7.84313725490196\\
2.6	7.69230769230769\\
2.65	7.54716981132075\\
2.7	7.40740740740741\\
2.75	7.27272727272727\\
2.8	7.14285714285714\\
2.85	7.01754385964912\\
2.9	6.89655172413793\\
2.95	6.77966101694915\\
3	6.66666666666667\\
3.05	6.55737704918033\\
3.1	6.45161290322581\\
3.15	6.34920634920635\\
3.2	6.25\\
3.25	6.15384615384615\\
3.3	6.06060606060606\\
3.35	5.97014925373134\\
3.4	5.88235294117647\\
3.45	5.79710144927536\\
3.5	5.71428571428571\\
3.55	5.63380281690141\\
3.6	5.55555555555556\\
3.65	5.47945205479452\\
3.7	5.4054054054054\\
3.75	5.33333333333333\\
3.8	5.26315789473684\\
3.85	5.19480519480519\\
3.9	5.12820512820513\\
3.95	5.06329113924051\\
4	5\\
4.05	4.93827160493827\\
4.1	4.87804878048781\\
4.15	4.81927710843373\\
4.2	4.76190476190476\\
4.25	4.70588235294118\\
4.3	4.65116279069767\\
4.35	4.59770114942529\\
4.4	4.54545454545454\\
4.45	4.49438202247191\\
4.5	4.44444444444444\\
4.55	4.3956043956044\\
4.6	4.34782608695652\\
4.65	4.3010752688172\\
4.7	4.25531914893617\\
4.75	4.21052631578947\\
4.8	4.16666666666667\\
4.85	4.12371134020619\\
4.9	4.08163265306122\\
4.95	4.04040404040404\\
5	4\\
5.05	3.96039603960396\\
5.1	3.92156862745098\\
5.15	3.88349514563107\\
5.2	3.84615384615385\\
5.25	3.80952380952381\\
5.3	3.77358490566038\\
5.35	3.73831775700935\\
5.4	3.7037037037037\\
5.45	3.6697247706422\\
5.5	3.63636363636364\\
5.55	3.6036036036036\\
5.6	3.57142857142857\\
5.65	3.53982300884956\\
5.7	3.50877192982456\\
5.75	3.47826086956522\\
5.8	3.44827586206897\\
5.85	3.41880341880342\\
5.9	3.38983050847458\\
5.95	3.36134453781513\\
6	3.33333333333333\\
6.05	3.30578512396694\\
6.1	3.27868852459016\\
6.15	3.2520325203252\\
6.2	3.2258064516129\\
6.25	3.2\\
6.3	3.17460317460317\\
6.35	3.1496062992126\\
6.4	3.125\\
6.45	3.10077519379845\\
6.5	3.07692307692308\\
6.55	3.05343511450382\\
6.6	3.03030303030303\\
6.65	3.00751879699248\\
6.7	2.98507462686567\\
6.75	2.96296296296296\\
6.8	2.94117647058823\\
6.85	2.91970802919708\\
6.9	2.89855072463768\\
6.95	2.87769784172662\\
7	2.85714285714286\\
7.05	2.83687943262411\\
7.1	2.8169014084507\\
7.15	2.7972027972028\\
7.2	2.77777777777778\\
7.25	2.75862068965517\\
7.3	2.73972602739726\\
7.35	2.72108843537415\\
7.4	2.7027027027027\\
7.45	2.68456375838926\\
7.5	2.66666666666667\\
7.55	2.64900662251656\\
7.6	2.63157894736842\\
7.65	2.61437908496732\\
7.7	2.5974025974026\\
7.75	2.58064516129032\\
7.8	2.56410256410256\\
7.85	2.54777070063694\\
7.9	2.53164556962025\\
7.95	2.51572327044025\\
8	2.5\\
8.05	2.48447204968944\\
8.1	2.46913580246914\\
8.15	2.45398773006135\\
8.2	2.4390243902439\\
8.25	2.42424242424242\\
8.3	2.40963855421687\\
8.35	2.39520958083832\\
8.4	2.38095238095238\\
8.45	2.36686390532544\\
8.5	2.35294117647059\\
8.55	2.33918128654971\\
8.6	2.32558139534884\\
8.65	2.3121387283237\\
8.7	2.29885057471264\\
8.75	2.28571428571429\\
8.8	2.27272727272727\\
8.85	2.25988700564972\\
8.9	2.24719101123596\\
8.95	2.23463687150838\\
9	2.22222222222222\\
9.05	2.20994475138122\\
9.1	2.1978021978022\\
9.15	2.18579234972678\\
9.2	2.17391304347826\\
9.25	2.16216216216216\\
9.3	2.1505376344086\\
9.35	2.13903743315508\\
9.4	2.12765957446809\\
9.45	2.11640211640212\\
9.5	2.10526315789474\\
9.55	2.09424083769633\\
9.6	2.08333333333333\\
9.65	2.07253886010363\\
9.7	2.06185567010309\\
9.75	2.05128205128205\\
9.8	2.04081632653061\\
9.85	2.03045685279188\\
9.9	2.02020202020202\\
9.95	2.01005025125628\\
10	2\\
10.05	1.99004975124378\\
10.1	1.98019801980198\\
10.15	1.97044334975369\\
10.2	1.96078431372549\\
10.25	1.95121951219512\\
10.3	1.94174757281553\\
10.35	1.93236714975845\\
10.4	1.92307692307692\\
10.45	1.91387559808612\\
10.5	1.9047619047619\\
10.55	1.8957345971564\\
10.6	1.88679245283019\\
10.65	1.87793427230047\\
10.7	1.86915887850467\\
10.75	1.86046511627907\\
10.8	1.85185185185185\\
10.85	1.84331797235023\\
10.9	1.8348623853211\\
10.95	1.82648401826484\\
11	1.81818181818182\\
11.05	1.80995475113122\\
11.1	1.8018018018018\\
11.15	1.79372197309417\\
11.2	1.78571428571429\\
11.25	1.77777777777778\\
11.3	1.76991150442478\\
11.35	1.76211453744493\\
11.4	1.75438596491228\\
11.45	1.74672489082969\\
11.5	1.73913043478261\\
11.55	1.73160173160173\\
11.6	1.72413793103448\\
11.65	1.71673819742489\\
11.7	1.70940170940171\\
11.75	1.70212765957447\\
11.8	1.69491525423729\\
11.85	1.68776371308017\\
11.9	1.68067226890756\\
11.95	1.67364016736402\\
12	1.66666666666667\\
12.05	1.6597510373444\\
12.1	1.65289256198347\\
12.15	1.64609053497942\\
12.2	1.63934426229508\\
12.25	1.63265306122449\\
12.3	1.6260162601626\\
12.35	1.61943319838057\\
12.4	1.61290322580645\\
12.45	1.60642570281124\\
12.5	1.6\\
12.55	1.59362549800797\\
12.6	1.58730158730159\\
12.65	1.58102766798419\\
12.7	1.5748031496063\\
12.75	1.56862745098039\\
12.8	1.5625\\
12.85	1.55642023346304\\
12.9	1.55038759689922\\
12.95	1.54440154440154\\
13	1.53846153846154\\
13.05	1.53256704980843\\
13.1	1.52671755725191\\
13.15	1.52091254752852\\
13.2	1.51515151515152\\
13.25	1.50943396226415\\
13.3	1.50375939849624\\
13.35	1.49812734082397\\
13.4	1.49253731343284\\
13.45	1.48698884758364\\
13.5	1.48148148148148\\
13.55	1.4760147601476\\
13.6	1.47058823529412\\
13.65	1.46520146520147\\
13.7	1.45985401459854\\
13.75	1.45454545454545\\
13.8	1.44927536231884\\
13.85	1.44404332129964\\
13.9	1.43884892086331\\
13.95	1.4336917562724\\
14	1.42857142857143\\
14.05	1.42348754448399\\
14.1	1.41843971631206\\
14.15	1.41342756183746\\
14.2	1.40845070422535\\
14.25	1.40350877192982\\
14.3	1.3986013986014\\
14.35	1.39372822299652\\
14.4	1.38888888888889\\
14.45	1.3840830449827\\
14.5	1.37931034482759\\
14.55	1.37457044673539\\
14.6	1.36986301369863\\
14.65	1.36518771331058\\
14.7	1.36054421768708\\
14.75	1.35593220338983\\
14.8	1.35135135135135\\
14.85	1.34680134680135\\
14.9	1.34228187919463\\
14.95	1.33779264214047\\
15	1.33333333333333\\
};
\end{axis}
\end{tikzpicture}%

%% file: tikz/taper_NMSE_versus_k_nuInf_alpha0dot1_p250_n100_ell1_uusi.tex
%
%
\begin{tikzpicture}

\begin{axis}[%
width=0.4\fwidth,
at={(0\fwidth,0\fwidth)},
scale only axis,
xmin=2,
xmax=30,
xlabel style={font=\color{white!15!black}},
xlabel={k},
ymin=0.0838069697070957,
ymax=0.2,
xlabel={bandwidth, $k$},
tick label style={font=\scriptsize} , 
ylabel style={font=\color{white!15!black}},
             yticklabel style={%
                 /pgf/number format/.cd,
                     fixed,
                     fixed zerofill,
                     precision=3,
                     },
ylabel={ $\expec[ \| \W \circ \S - \M \|_{\Fr}^2]$},
xtick={2,6,10,14,18,22,26,30},
axis background/.style={fill=white},
xmajorgrids,
ymajorgrids,
legend style={anchor=south west, legend cell align=left,  font = {\footnotesize}, align=right, draw=none, legend columns=2,at={(0.3,-0.47)}}
]

\addplot [color=blue, dashed, line width=0.7pt,  mark size=1.8pt,mark=o, mark options={solid, blue}]
  table[row sep=crcr]{%
2	0.189955993927895\\
4	0.104431584105162\\
6	0.0896832615565728\\
8	0.0899027360016629\\
10	0.0955617089271843\\
12	0.103743727075036\\
14	0.113269542816193\\
16	0.123575205819153\\
18	0.134358592780255\\
20	0.145443947722059\\
22	0.156722409378353\\
24	0.168123133584293\\
26	0.179598134726811\\
28	0.191113817206337\\
30	0.202645996719681\\
32	0.214176845827406\\
34	0.225692951088712\\
36	0.237184038237307\\
38	0.248642112793763\\
40	0.26006086683522\\
42	0.271435260810703\\
44	0.282761223186829\\
46	0.294035431073784\\
48	0.305255147554295\\
50	0.316418099391858\\
52	0.327522383937772\\
54	0.338566397449649\\
56	0.349548779313354\\
58	0.360468368217346\\
60	0.371324167408205\\
62	0.382115316915699\\
64	0.392841071176991\\
66	0.403500780880033\\
68	0.414093878130954\\
70	0.424619864260136\\
72	0.43507829973805\\
74	0.445468795789111\\
76	0.455791007380992\\
78	0.46604462733458\\
80	0.476229381352299\\
82	0.486345023802815\\
84	0.496391334132065\\
86	0.506368113795261\\
88	0.51627518362428\\
90	0.526112381560415\\
92	0.535879560695137\\
94	0.545576587571296\\
96	0.555203340705635\\
98	0.564759709299835\\
100	0.574245592112832\\
102	0.583660896471473\\
104	0.593005537400257\\
106	0.602279436853913\\
108	0.611482523038942\\
110	0.620614729812451\\
112	0.629675996148272\\
114	0.638666265661792\\
116	0.647585486186127\\
118	0.656433609393391\\
120	0.665210590455542\\
122	0.673916387740124\\
124	0.682550962536841\\
126	0.691114278811353\\
128	0.699606302983252\\
130	0.70802700372551\\
132	0.716376351783015\\
134	0.724654319808131\\
136	0.732860882211479\\
138	0.740996015026329\\
140	0.749059695785155\\
142	0.757051903407165\\
144	0.764972618095637\\
146	0.77282182124416\\
148	0.780599495350802\\
150	0.788305623939536\\
152	0.795940191488144\\
154	0.803503183362091\\
156	0.810994585753639\\
158	0.818414385625941\\
160	0.825762570661471\\
162	0.833039129214531\\
164	0.840244050267304\\
166	0.84737732338943\\
168	0.854438938700436\\
170	0.861428886835122\\
172	0.868347158911351\\
174	0.875193746500314\\
176	0.881968641598816\\
178	0.888671836603604\\
180	0.895303324287486\\
182	0.901863097777105\\
184	0.908351150532249\\
186	0.914767476326556\\
188	0.92111206922956\\
190	0.92738492358988\\
192	0.933586034019563\\
194	0.939715395379431\\
196	0.945773002765391\\
198	0.951758851495609\\
200	0.957672937098499\\
202	0.963515255301493\\
204	0.96928580202047\\
206	0.974984573349872\\
208	0.980611565553398\\
210	0.986166775055295\\
212	0.991650198432126\\
214	0.99706183240508\\
216	1.00240167383269\\
218	1.00766971970401\\
220	1.01286596713215\\
222	1.01799041334824\\
224	1.02304305569568\\
226	1.02802389162471\\
228	1.0329329186874\\
230	1.03777013453272\\
232	1.04253553690207\\
234	1.04722912362494\\
236	1.05185089261486\\
238	1.05640084186551\\
240	1.06087896944712\\
242	1.06528527350299\\
244	1.06961975224622\\
246	1.07388240395662\\
248	1.07807322697775\\
250	1.08219221971414\\
};

\end{axis}
\end{tikzpicture}%

%% file: tikz/taper_NMSE_versus_k_nu5_alpha0dot1_p250_n100_ell1.tex
%
%
\begin{tikzpicture}

\begin{axis}[%
width=0.4\fwidth,
at={(0\fwidth,0\fwidth)},
scale only axis,
xmin=2,
xmax=30,
xlabel style={font=\color{white!15!black}},
xlabel={bandwidth, $k$},
ymin=0.18,
ymax=0.6,
 tick label style={font=\scriptsize} , 
ylabel style={font=\color{white!15!black}},
xtick={2,6,10,14,18,22,26,30},
axis background/.style={fill=white},
xmajorgrids,
ymajorgrids,
legend style={anchor=south west, legend cell align=left,  font = {\footnotesize}, align=right, draw=none, legend columns=2,at={(-0.2,-0.47)}}
]

\addplot [color=blue, dashed, line width=0.7pt,  mark size=1.8pt, mark=o, mark options={solid, blue}]
  table[row sep=crcr]{%
2	0.252314442975713\\
4	0.194160501644641\\
6	0.205650972702826\\
8	0.23159831318618\\
10	0.262691115604645\\
12	0.296090732620476\\
14	0.33065121888011\\
16	0.36582490220224\\
18	0.401318437111787\\
20	0.436961168306168\\
22	0.472647377224081\\
24	0.508308249332154\\
26	0.543897161400158\\
28	0.579381462283172\\
30	0.614737640538039\\
32	0.649948359457984\\
34	0.685000570852554\\
36	0.719884277189121\\
38	0.754591696974678\\
40	0.789116688514135\\
42	0.82345434362127\\
44	0.857600695749555\\
46	0.891552506772088\\
48	0.925307108841565\\
50	0.958862285480451\\
52	0.992216181044058\\
54	1.02536723099308\\
56	1.05831410762518\\
58	1.0910556774269\\
60	1.12359096725624\\
62	1.15591913730345\\
64	1.18803945930375\\
66	1.21995129885501\\
68	1.25165410096975\\
70	1.28314737819527\\
72	1.31443070078742\\
74	1.34550368853742\\
76	1.37636600393804\\
78	1.40701734644106\\
80	1.43745744760936\\
82	1.46768606700572\\
84	1.49770298869202\\
86	1.52750801823598\\
88	1.55710098014235\\
90	1.58648171564013\\
92	1.61565008077025\\
94	1.64460594472701\\
96	1.67334918841546\\
98	1.70187970319264\\
100	1.73019738976616\\
102	1.75830215722769\\
104	1.78619392220269\\
106	1.81387260810052\\
108	1.84133814445136\\
110	1.86859046631854\\
112	1.89562951377666\\
114	1.92245523144695\\
116	1.94906756808279\\
118	1.97546647619934\\
120	2.00165191174171\\
122	2.02762383378732\\
124	2.0533822042784\\
126	2.07892698778098\\
128	2.1042581512676\\
130	2.12937566392103\\
132	2.15427949695655\\
134	2.17896962346098\\
136	2.20344601824651\\
138	2.22770865771792\\
140	2.25175751975164\\
142	2.27559258358553\\
144	2.29921382971832\\
146	2.32262123981772\\
148	2.34581479663631\\
150	2.3687944839345\\
152	2.39156028640981\\
154	2.41411218963215\\
156	2.43645017998396\\
158	2.4585742446055\\
160	2.48048437134421\\
162	2.50218054870817\\
164	2.52366276582287\\
166	2.54493101239169\\
168	2.56598527865888\\
170	2.58682555537574\\
172	2.60745183376885\\
174	2.627864105511\\
176	2.64806236269392\\
178	2.66804659780302\\
180	2.68781680369406\\
182	2.70737297357121\\
184	2.72671510096676\\
186	2.74584317972218\\
188	2.76475720397047\\
190	2.78345716811962\\
192	2.80194306683721\\
194	2.82021489503606\\
196	2.83827264786072\\
198	2.85611632067489\\
200	2.87374590904956\\
202	2.89116140875205\\
204	2.90836281573554\\
206	2.92535012612936\\
208	2.94212333622987\\
210	2.95868244249183\\
212	2.97502744152036\\
214	2.99115833006333\\
216	3.00707510500422\\
218	3.02277776335538\\
220	3.03826630225165\\
222	3.05354071894449\\
224	3.06860101079623\\
226	3.08344717527478\\
228	3.09807920994864\\
230	3.11249711248214\\
232	3.1267008806309\\
234	3.14069051223768\\
236	3.15446600522829\\
238	3.16802735760785\\
240	3.18137456745715\\
242	3.19450763292928\\
244	3.20742655224638\\
246	3.2201313236966\\
248	3.23262194563118\\
250	3.2448984164617\\
};

\end{axis}
\end{tikzpicture}%

%% file: tikz/AR1_NMSE_SCM.tex
\begin{tikzpicture}

\begin{axis}[%
width=0.44\textwidth,
height=5cm,
scale only axis,
xmin=7.93220338983051,
xmax=197,
tick label style={font=\footnotesize},
xtick={ 25,  50,  75, 100,  125, 150, 175, 200},
xlabel style={font=\color{white!15!black}},
xlabel={sample length, $n$},
ylabel = {NMSE},
ymin=0,
ymax=1,
axis background/.style={fill=white},
xmajorgrids,
ymajorgrids,
legend style={legend cell align=left, align=left, fill=none, draw=none}
]
\addplot [color=black, line width=1.5pt]
  table[row sep=crcr]{%
2	46.4545454545455\\
7	7.74242424242424\\
12	4.22314049586777\\
17	2.90340909090909\\
22	2.21212121212121\\
27	1.78671328671329\\
32	1.49853372434018\\
37	1.29040404040404\\
42	1.1330376940133\\
47	1.0098814229249\\
52	0.910873440285205\\
57	0.829545454545454\\
62	0.761549925484352\\
67	0.703856749311295\\
72	0.654289372599232\\
77	0.611244019138756\\
82	0.57351290684624\\
87	0.540169133192389\\
92	0.510489510489511\\
97	0.483901515151515\\
102	0.45994599459946\\
107	0.438250428816467\\
112	0.418509418509419\\
117	0.400470219435737\\
122	0.383921863260706\\
127	0.368686868686869\\
132	0.354614850798057\\
137	0.341577540106952\\
142	0.329464861379755\\
147	0.318181818181818\\
152	0.307645996387718\\
157	0.297785547785548\\
162	0.288537549407115\\
167	0.279846659364732\\
172	0.271664008506114\\
177	0.263946280991736\\
182	0.256654947262682\\
187	0.249755620723363\\
192	0.243217515468824\\
197	0.237012987012987\\
};
\addlegendentry{$\gamma = 1.1$}

\addplot [color=red, line width=1.5pt]
  table[row sep=crcr]{%
2	26\\
7	4.33333333333333\\
12	2.36363636363636\\
17	1.625\\
22	1.23809523809524\\
27	1\\
32	0.838709677419355\\
37	0.722222222222222\\
42	0.634146341463415\\
47	0.565217391304348\\
52	0.509803921568628\\
57	0.464285714285714\\
62	0.426229508196721\\
67	0.393939393939394\\
72	0.366197183098592\\
77	0.342105263157895\\
82	0.320987654320988\\
87	0.302325581395349\\
92	0.285714285714286\\
97	0.270833333333333\\
102	0.257425742574257\\
107	0.245283018867925\\
112	0.234234234234234\\
117	0.224137931034483\\
122	0.214876033057851\\
127	0.206349206349206\\
132	0.198473282442748\\
137	0.191176470588235\\
142	0.184397163120567\\
147	0.178082191780822\\
152	0.172185430463576\\
157	0.166666666666667\\
162	0.161490683229814\\
167	0.156626506024096\\
172	0.152046783625731\\
177	0.147727272727273\\
182	0.143646408839779\\
187	0.139784946236559\\
192	0.136125654450262\\
197	0.13265306122449\\
};
\addlegendentry{$\gamma = 2.0$}

\addplot [color=blue, line width=1.5pt]
  table[row sep=crcr]{%
2	6.55555555555555\\
7	1.09259259259259\\
12	0.595959595959596\\
17	0.409722222222222\\
22	0.312169312169312\\
27	0.252136752136752\\
32	0.211469534050179\\
37	0.182098765432099\\
42	0.159891598915989\\
47	0.142512077294686\\
52	0.128540305010893\\
57	0.117063492063492\\
62	0.107468123861566\\
67	0.0993265993265993\\
72	0.0923317683881064\\
77	0.0862573099415204\\
82	0.0809327846364883\\
87	0.0762273901808785\\
92	0.072039072039072\\
97	0.068287037037037\\
102	0.0649064906490649\\
107	0.0618448637316562\\
112	0.059059059059059\\
117	0.0565134099616858\\
122	0.054178145087236\\
127	0.0520282186948854\\
132	0.0500424088210348\\
137	0.048202614379085\\
142	0.0464933018124507\\
147	0.0449010654490106\\
152	0.0434142752023547\\
157	0.042022792022792\\
162	0.0407177363699103\\
167	0.0394912985274431\\
172	0.0383365821962313\\
177	0.0372474747474747\\
182	0.0362185389809699\\
187	0.0352449223416965\\
192	0.0343222803955788\\
197	0.0334467120181406\\
};
\addlegendentry{$\gamma = 9.0$}

\end{axis}
\end{tikzpicture}%

%% file: tikz/AR1_NMSE_LW.tex
\begin{tikzpicture}

\begin{axis}[%
width=0.44\textwidth,
height=5cm,
scale only axis,
xmin=2,
xmax=197,
tick label style={font=\footnotesize},
xtick={ 25,  50,  75, 100,  125, 150, 175, 200},
xlabel style={font=\color{white!15!black}},
xlabel={sample length, $n$},
ymin=0,
ymax=1,
axis background/.style={fill=white},
xmajorgrids,
ymajorgrids,
legend style={legend cell align=left, align=left, fill=none, draw=none}
]
\addplot [color=black, line width=1.5pt]
  table[row sep=crcr]{%
2	0.0907315340909092\\
7	0.0898540531035696\\
12	0.0889933820968304\\
17	0.0881490426082457\\
22	0.0873205741626795\\
27	0.0865075334349078\\
32	0.085709493458571\\
37	0.084926042878511\\
42	0.0841567852437418\\
47	0.0834013383385018\\
52	0.0826593335490133\\
57	0.0819304152637487\\
62	0.0812142403051495\\
67	0.0805104773908934\\
72	0.0798188066229304\\
77	0.0791389190026329\\
82	0.078470515970516\\
87	0.0778133089690879\\
92	0.0771670190274842\\
97	0.0765313763666318\\
102	0.0759061200237671\\
107	0.0752909974952115\\
112	0.0746857643963754\\
117	0.0740901841380311\\
122	0.0735040276179517\\
127	0.072927072927073\\
132	0.0723591050693855\\
137	0.0717999156948153\\
142	0.0712493028443949\\
147	0.0707070707070707\\
152	0.0701730293875309\\
157	0.069646994684476\\
162	0.0691287878787879\\
167	0.0686182355310864\\
172	0.0681151692881899\\
177	0.0676194256980284\\
182	0.0671308460325802\\
187	0.0666492761184297\\
192	0.0661745661745662\\
197	0.0657065706570657\\
};
\addlegendentry{$\gamma = 1.1$}

\addplot [color=red, line width=1.5pt]
  table[row sep=crcr]{%
2	0.490566037735849\\
7	0.448275862068966\\
12	0.412698412698413\\
17	0.382352941176471\\
22	0.356164383561644\\
27	0.333333333333333\\
32	0.313253012048193\\
37	0.295454545454545\\
42	0.279569892473118\\
47	0.26530612244898\\
52	0.252427184466019\\
57	0.240740740740741\\
62	0.230088495575221\\
67	0.220338983050847\\
72	0.211382113821138\\
77	0.203125\\
82	0.195488721804511\\
87	0.188405797101449\\
92	0.181818181818182\\
97	0.175675675675676\\
102	0.169934640522876\\
107	0.164556962025316\\
112	0.159509202453988\\
117	0.154761904761905\\
122	0.15028901734104\\
127	0.146067415730337\\
132	0.14207650273224\\
137	0.138297872340426\\
142	0.134715025906736\\
147	0.131313131313131\\
152	0.12807881773399\\
157	0.125\\
162	0.122065727699531\\
167	0.119266055045872\\
172	0.116591928251121\\
177	0.114035087719298\\
182	0.111587982832618\\
187	0.109243697478992\\
192	0.106995884773663\\
197	0.104838709677419\\
};
\addlegendentry{$\gamma = 2.0$}

\addplot [color=blue, line width=1.5pt]
  table[row sep=crcr]{%
2	0.782752902155887\\
7	0.490134994807892\\
12	0.3567649281935\\
17	0.280451574569222\\
22	0.231032794909447\\
27	0.196421140241365\\
32	0.170828809265291\\
37	0.151136727505604\\
42	0.135515360321562\\
47	0.122820712984647\\
52	0.112300737568403\\
57	0.103440718825334\\
62	0.0958764980702824\\
67	0.0893431762256294\\
72	0.0836434520645047\\
77	0.0786273529901716\\
82	0.0741788464560743\\
87	0.0702067529376766\\
92	0.0666384300437667\\
97	0.0634152895337902\\
102	0.0604895552992439\\
107	0.0578218792110743\\
112	0.0553795611873753\\
117	0.0531352020713723\\
122	0.0510656713188358\\
127	0.0491513068832656\\
132	0.0473752885677005\\
137	0.045723142497336\\
142	0.0441823457830197\\
147	0.0427420085121796\\
152	0.0413926159782513\\
157	0.0401258182436453\\
162	0.0389342571970634\\
167	0.0378114235360089\\
172	0.036751537802694\\
177	0.0357494508823753\\
182	0.0348005603480056\\
187	0.0339007397830926\\
192	0.0330462787929707\\
197	0.0322338318650549\\
};
\addlegendentry{$\gamma = 9.0$}

\end{axis}
\end{tikzpicture}%

%% file: results/AR1-definitionsforplotofclass4.tex
\def\aloptfour{3.097890e-01}
\def\beoptfour{2.048585e-01}
\def\nmseofoptfour{2.994326e-01}
\def\almeanfour{3.343960e-01}
\def\bemeanfour{2.257217e-01}
\def\nmseofmeanfour{3.002223e-01}
\def\Cxxyyfour{1.126043e+00}
\def\Cxxyfour{1.363044e-01}
\def\Cxxfour{9.699572e-01}
\def\Cyyfour{1.887015e-03}
\def\Cxyfour{-1.876994e-01}
\def\Cxfour{-6.090920e-01}
\def\Cyfour{1.680789e-05}
\def\Cfour{3.996509e-01}

%% file: tikz/sp500indx_r.tex
%
%
\definecolor{mycolor1}{rgb}{0.00000,0.44700,0.74100}%
\definecolor{mycolor2}{rgb}{0.85000,0.32500,0.09800}%
\begin{tikzpicture}

\begin{axis}[%
width=0.9\fwidth,
height=0.22\fwidth,
at={(0\fwidth,0\fwidth)},
scale only axis,
xmin=1,
xmax=250,
xtick={1,51,101,151,201,250},
ytick = {-0.015,-0.01,-0.005,0,0.005,0.01},
 tick label style={font=\scriptsize} , 
xticklabels={{2017-01-04},{2017-03-17},{2017-05-30},{2017-08-09},{2017-10-19},{2017-12-29}},
ymin=-0.0181782145892192,
ymax=0.0136738545053825,
axis background/.style={fill=white},
xmajorgrids,
ymajorgrids,
legend style={legend cell align=left, align=left, draw=white!15!black}
]
\addplot [color=mycolor1, line width=0.9pt]
  table[row sep=crcr]{%
1	0.00572227384420554\\
2	-0.000770670483320468\\
3	0.00351695901278104\\
4	-0.00354859422171994\\
5	0\\
6	0.00282963827286542\\
7	-0.00214480901769987\\
8	0.00184984060760929\\
9	-0.00296750268944657\\
10	0.00176375405717288\\
11	-0.00360930871925835\\
12	0.00336623751423915\\
13	-0.00269012501212185\\
14	0.00656459355538797\\
15	0.00802609062626392\\
16	-0.000735384169633257\\
17	-0.00086646422615233\\
18	-0.00600954349152305\\
19	-0.000889905338774533\\
20	0.000298363647374122\\
21	0.000570309478649111\\
22	0.00726475800164583\\
23	-0.00211535686335018\\
24	0.000226828953928004\\
25	0.000693322494601523\\
26	0.00575254631328193\\
27	0.00356605033332547\\
28	0.00524584494879643\\
29	0.00400733512294638\\
30	0.00499230897363945\\
31	-0.000864117909971318\\
32	0.00167855635546665\\
33	0.0060480662873772\\
34	-0.00108220037652196\\
35	0.000418987045779584\\
36	0.00149336406559386\\
37	0.001017983014868\\
38	-0.0025783762000211\\
39	0.0136738545053825\\
40	-0.00585988047735997\\
41	0.000503877140836995\\
42	-0.00327724059911494\\
43	-0.00291337376094525\\
44	-0.00228421554068836\\
45	0.000799895477743284\\
46	0.00326867042060042\\
47	0.000366632792746291\\
48	-0.00337902737257756\\
49	0.00837475296893309\\
50	-0.00162671028891304\\
51	-0.0013143148736342\\
52	-0.00200989340901925\\
53	-0.0124079728666598\\
54	0.00188988616232044\\
55	-0.00106026956160588\\
56	-0.000843996075344799\\
57	-0.00101958720654272\\
58	0.00725147415297744\\
59	0.00108532497496272\\
60	0.00293511002926894\\
61	-0.0022550475355152\\
62	-0.00164212562115773\\
63	0.000559522456275996\\
64	-0.00305486122501353\\
65	0.00192950937952174\\
66	-0.00082713012919311\\
67	0.000687686463902271\\
68	-0.00143387938289363\\
69	-0.00375995075621405\\
70	-0.00681469445288307\\
71	0.00861334911528977\\
72	-0.00290338013502112\\
73	-0.00171635055280095\\
74	0.00755726341090091\\
75	-0.00303507315136575\\
76	0.0108400689914652\\
77	0.00609068744472219\\
78	-0.00048570337896503\\
79	0.000552920072501184\\
80	-0.00191314733596659\\
81	0.0017322905313657\\
82	0.00118905005056003\\
83	-0.00127136050517784\\
84	0.000582102761619296\\
85	0.00408869518490174\\
86	3.74460771894736e-05\\
87	-0.0010252486558836\\
88	0.00113060139186416\\
89	-0.00216280937188174\\
90	-0.00147844134212094\\
91	0.00477651364260256\\
92	-0.00068689681361811\\
93	-0.0181782145892192\\
94	0.00368681853564956\\
95	0.00676749961798429\\
96	0.00516013154438255\\
97	0.00183787184870754\\
98	0.0024891266726228\\
99	0.00444194805139286\\
100	0.000310549996017651\\
101	-0.00120462448281966\\
102	-0.000459968685312506\\
103	0.00757111270794231\\
104	0.00370773099480837\\
105	-0.00121766489571795\\
106	-0.00277904015748698\\
107	0.00156825745274447\\
108	0.000267204529369902\\
109	-0.000829989016156052\\
110	-0.000978763197352017\\
111	0.00451150514439047\\
112	-0.000995830885900939\\
113	-0.00223959817167452\\
114	0.000283639201081209\\
115	0.00834722882601913\\
116	-0.0066966375083225\\
117	-0.000582644441432079\\
118	-0.00045578189908535\\
119	0.00156091558841642\\
120	0.000315801576723951\\
121	-0.00807282466310844\\
122	0.00880806612873708\\
123	-0.00860002315222397\\
124	0.00153323183664433\\
125	0.002310833991505\\
126	0.00145327890188485\\
127	-0.00936882379513426\\
128	0.00640312563543932\\
129	0.000927766212441172\\
130	-0.000782680881929565\\
131	0.00730560775918554\\
132	0.00187458426276477\\
133	0.00467350332149974\\
134	-5.29128558237613e-05\\
135	0.000597857000402824\\
136	0.00537263947766098\\
137	-0.000153659300766162\\
138	-0.000367871603640935\\
139	-0.00106373484696476\\
140	0.00292317179866441\\
141	0.000282663821871143\\
142	-0.00097268816832885\\
143	-0.00134111548933402\\
144	-0.000728145677214354\\
145	0.00244911503865652\\
146	0.00049264843488217\\
147	-0.00218365408505572\\
148	0.00188910352333216\\
149	0.00164719979631966\\
150	-0.00241443269303199\\
151	-0.000363608532139015\\
152	-0.0144744418842657\\
153	0.00127556980315369\\
154	0.0100437547380208\\
155	-0.000498808096269343\\
156	0.00142010291609984\\
157	-0.0154369518977056\\
158	-0.00183536733661438\\
159	0.00116265092165913\\
160	0.00994077996224996\\
161	-0.00345359283569247\\
162	-0.00207446192333016\\
163	0.00167286930487598\\
164	0.000487071888063406\\
165	0.000842821903098034\\
166	0.00461514890808878\\
167	0.00572097603609811\\
168	0.00198254089142424\\
169	-0.00755080681997555\\
170	0.00312872660796426\\
171	-0.000178435958468004\\
172	-0.00148885069737237\\
173	0.0108392989997979\\
174	0.00336394799267614\\
175	0.000757120832188596\\
176	-0.00110071761637232\\
177	0.00184718137532158\\
178	0.00145592086692758\\
179	0.00111019536561696\\
180	0.000634347859560069\\
181	-0.00304591746820848\\
182	0.000647793704117605\\
183	-0.00222205044498058\\
184	7.21668174081813e-05\\
185	0.00408514387806491\\
186	0.00120461578316267\\
187	0.00370510975092175\\
188	0.00387400355069611\\
189	0.00215883815217\\
190	0.00124672012828797\\
191	0.00564678732118651\\
192	-0.00107363431527852\\
193	-0.00180443405100716\\
194	0.00232241261212307\\
195	0.00180350703861598\\
196	-0.00168675271867513\\
197	0.000878107223526881\\
198	0.00175075343066022\\
199	0.000672578655309453\\
200	0.000742335162138286\\
201	0.000327997937233926\\
202	0.0051168426285273\\
203	-0.00397248424591656\\
204	0.00161790853431931\\
205	-0.00466304996071698\\
206	0.00127094621924906\\
207	0.00807302249303077\\
208	-0.00319247048042559\\
209	0.000944458796862779\\
210	0.00159210991669934\\
211	0.000189966107745132\\
212	0.00309707529371339\\
213	0.0012712512706079\\
214	-0.000189102832403254\\
215	0.00144365490939347\\
216	-0.00376188778827347\\
217	-0.00089764371357326\\
218	0.00098363433830384\\
219	-0.00230960941364045\\
220	-0.00552567572366813\\
221	0.00819605830144865\\
222	-0.00262596311976071\\
223	0.00127568291098101\\
224	0.00654113901643671\\
225	-0.000750261050562084\\
226	0.00205609524528505\\
227	-0.00038425774086126\\
228	0.00984851264624087\\
229	-0.000369225815214147\\
230	0.00819095052417307\\
231	-0.00202453064386598\\
232	-0.00105215690991378\\
233	-0.0037393815432909\\
234	-0.000114105345071946\\
235	0.00293235762829713\\
236	0.00550630649839068\\
237	0.00320195738261364\\
238	0.0015489219942515\\
239	-0.000472956803357905\\
240	-0.00407085926772277\\
241	0.00897434357723248\\
242	0.00536280703173797\\
243	-0.00323026930898673\\
244	-0.000827893291369564\\
245	0.00198565568722597\\
246	-0.000458166473157662\\
247	-0.00105841522388495\\
248	0.000790940869240808\\
249	0.00183399877188051\\
250	-0.00518315329180474\\
};

\addplot [color=mycolor2, dashed, line width=2.5pt]
  table[row sep=crcr]{%
1	0\\
250	0\\
};

\end{axis}

\end{tikzpicture}%

%% file: tikz/nasdaq100indx_r.tex
%
%
\definecolor{mycolor1}{rgb}{0.00000,0.44700,0.74100}%
\definecolor{mycolor2}{rgb}{0.85000,0.32500,0.09800}%
\begin{tikzpicture}

\begin{axis}[%
width=0.9\fwidth,
height=0.22\fwidth,
at={(0\fwidth,0\fwidth)},
scale only axis,
xmin=1,
xmax=250,
xtick={1,51,101,151,201,250},
ytick = {-0.02,-0.01,0,0, 0.01,0.02},
xticklabels={{2017-01-04},{2017-03-17},{2017-05-30},{2017-08-09},{2017-10-19},{2017-12-29}},
ymin=-0.0251361071970412,
ymax=0.0290831194771128,
 tick label style={font=\scriptsize} , 
axis background/.style={fill=white},
title style={font=\bfseries},
xmajorgrids,
ymajorgrids,
legend style={legend cell align=left, align=left, draw=white!15!black}
]
\addplot [color=mycolor1, line width=0.9pt]
  table[row sep=crcr]{%
1	0.00526942449173329\\
2	0.00561860528904501\\
3	0.00848545933903355\\
4	0.00355892530624713\\
5	0.00204382578763651\\
6	0.00298699730753582\\
7	-0.00173849900653655\\
8	0.00358620259903009\\
9	-0.00293701656628054\\
10	0.00222021274371476\\
11	-0.00092569516684271\\
12	0.00238168051872556\\
13	0.000493758868643779\\
14	0.00698025201627628\\
15	0.00988229023319565\\
16	0.00105789352797414\\
17	0.00216023075178562\\
18	-0.00749410427855879\\
19	-0.00244867415607952\\
20	0.00702003819980157\\
21	-0.000968376917131364\\
22	0.00270021611077897\\
23	0.00123214950388428\\
24	0.00346750403161655\\
25	0.00206333259570424\\
26	0.00299814065522797\\
27	0.0027876704792491\\
28	0.0057646203123034\\
29	0.00271076439313012\\
30	0.00594192716958397\\
31	-0.000330039841427099\\
32	0.00454286225392186\\
33	0.00488471956267844\\
34	0.00026162841429711\\
35	-0.00369194440941401\\
36	0.00205160965198625\\
37	0.000793468084985705\\
38	-0.00322385889400789\\
39	0.0113839859836191\\
40	-0.00514385424501584\\
41	0.0019055974250568\\
42	-0.00245091096440631\\
43	-0.00168465515998251\\
44	0.00159406241922011\\
45	0.000777997905541117\\
46	0.00408650332061833\\
47	0.00160974436171402\\
48	-0.00229858958258988\\
49	0.00633203308217656\\
50	-0.000769890976228926\\
51	-0.000613500161148139\\
52	0.000841282141722077\\
53	-0.0149225285674701\\
54	0.00656724320574975\\
55	-0.00231198003760291\\
56	0.00165445959831834\\
57	0.00191461968680096\\
58	0.00612919352347685\\
59	0.00426468717995476\\
60	0.00174396742061078\\
61	-0.000645298093107427\\
62	-0.000741283024232819\\
63	0.00151135096375055\\
64	-0.00402361869276679\\
65	0.000435517999617829\\
66	-0.0004629812971636\\
67	0.000610895698987868\\
68	-0.00429392240860205\\
69	-0.00395860206504584\\
70	-0.0043592730442723\\
71	0.00851958262195174\\
72	-0.00139651035851251\\
73	0.00148006008708101\\
74	0.00812825426999364\\
75	-0.000271878736491593\\
76	0.0121241044026057\\
77	0.00729120167602715\\
78	-0.00127971411856886\\
79	0.0054483760144568\\
80	0.00219877666761259\\
81	0.00825644346410526\\
82	0.00256498940429561\\
83	-0.00335036039412406\\
84	0.000206157330251866\\
85	0.00351384574969726\\
86	0.00229893259913205\\
87	0.00339989355112791\\
88	0.000593507040824282\\
89	-0.00131298502712474\\
90	0.00221877958961092\\
91	0.00310717622299261\\
92	0.00349899746689952\\
93	-0.0251361071970412\\
94	0.00819995441291477\\
95	0.0044878436728899\\
96	0.00846489155216812\\
97	0.000693089810843706\\
98	0.0047252415750263\\
99	0.0083887952874746\\
100	0.00172881726122198\\
101	0.00108321185074867\\
102	-0.00100611740831003\\
103	0.00478682316428802\\
104	0.0111665238455647\\
105	-0.000567859684864946\\
106	-0.00363213009857599\\
107	0.00355483038072246\\
108	0.00131175553324292\\
109	-0.0243589738416053\\
110	-0.00587950507091517\\
111	0.00764510696131904\\
112	-0.00430298596919332\\
113	-0.00457121840741148\\
114	-0.00340475899965587\\
115	0.0159712320239489\\
116	-0.00795363903142421\\
117	0.00979340577478149\\
118	-0.000435809404120802\\
119	0.00402080765303792\\
120	-0.00439764533198184\\
121	-0.0183449758224132\\
122	0.014357445093619\\
123	-0.0173838427294011\\
124	-0.00107908657291467\\
125	-0.0088472940452653\\
126	0.00926571984816116\\
127	-0.00901425847991433\\
128	0.0104629082379759\\
129	0.00666134277523112\\
130	0.00274841780939106\\
131	0.0121108256614262\\
132	0.00249347502812314\\
133	0.00771921925403096\\
134	0.000284366774319533\\
135	0.0069112430318421\\
136	0.00613255852774763\\
137	0.000855294459002787\\
138	5.22814536123573e-05\\
139	0.00335054162021753\\
140	-0.0018043338133954\\
141	0.00338581408982308\\
142	-0.00566320352515814\\
143	-0.00137059695399178\\
144	-0.00483842129820622\\
145	0.00252364132678884\\
146	0.00323316515930627\\
147	-0.00389396169541578\\
148	0.00147846970255605\\
149	0.00590175495546985\\
150	-0.00141200729068391\\
151	-0.0011744093556586\\
152	-0.0221644785971978\\
153	0.00748763334337177\\
154	0.0131423725549917\\
155	-7.44633288832786e-05\\
156	0.0016402140979368\\
157	-0.0204650167803319\\
158	-0.000933293566307558\\
159	-0.000754651148485164\\
160	0.0149986068384655\\
161	-0.00366917791334798\\
162	-0.00296317439088323\\
163	-0.00204135377524484\\
164	0.00267071076906489\\
165	0.00412122798566394\\
166	0.012070636891361\\
167	0.00938835930490289\\
168	-0.000116921482239851\\
169	-0.00921356784564364\\
170	0.00310142262702473\\
171	0.00221473506025327\\
172	-0.00854079373743033\\
173	0.0113572576502401\\
174	0.00252659087793528\\
175	0.0014576835500959\\
176	-0.00592235329757862\\
177	0.0032133950371358\\
178	-0.0011489450567802\\
179	0.001665233401966\\
180	-0.00291603763804682\\
181	-0.00647844567816558\\
182	-0.000436456817696107\\
183	-0.0109518245690592\\
184	0.00238433803443372\\
185	0.00959818621220965\\
186	-0.000794944746950854\\
187	0.00779191588357753\\
188	0.000438197963883669\\
189	0.0021966420766808\\
190	0.000630483258349468\\
191	0.00971859468098857\\
192	0.00122659981970963\\
193	-0.000995955059515885\\
194	0.000823670952704525\\
195	0.00292404081152853\\
196	-0.00185155453237407\\
197	0.00370016427278497\\
198	0.00362409035663847\\
199	0.0013214553341161\\
200	-0.00134905949992259\\
201	-0.00355393143207605\\
202	0.00265890646206524\\
203	-0.00670992878836629\\
204	0.00204193868989888\\
205	-0.00414132631872111\\
206	-0.00283564136478209\\
207	0.0290831194771128\\
208	0.00227242241636771\\
209	0.00336730830470544\\
210	1.43781925998621e-05\\
211	-0.00196198621978738\\
212	0.00949105808003092\\
213	0.00286387986120684\\
214	0.00113562956146818\\
215	0.00395999779321543\\
216	-0.00529484773222089\\
217	-0.000497470302699199\\
218	0.00112700480393357\\
219	-0.00356861874929504\\
220	-0.00560570246026459\\
221	0.012907578945337\\
222	-0.00388544352509879\\
223	-0.000934340624788921\\
224	0.0110991203324631\\
225	0.00117427004503945\\
226	0.00362816883733852\\
227	-0.000517970630100839\\
228	0.00258974728935724\\
229	-0.0173108814831872\\
230	0.0085845214524225\\
231	-0.00434996162841172\\
232	-0.0117026573013945\\
233	0.000225053555584331\\
234	0.00445960926639222\\
235	0.00369137075342119\\
236	0.00447890846557875\\
237	0.0077736260090373\\
238	-0.00160156567920089\\
239	0.00172628827852028\\
240	-0.00074433333667856\\
241	0.0119578626513634\\
242	0.00726072902019825\\
243	-0.0050051814065587\\
244	-0.00126374928804773\\
245	3.24390342880676e-05\\
246	-0.0011618073889752\\
247	-0.00495110977533253\\
248	0.000309295268849263\\
249	0.000974339385326761\\
250	-0.00698603732483072\\
};

\addplot [color=mycolor2, dashed, line width=2.5pt]
  table[row sep=crcr]{%
1	0\\
250	0\\
};

\end{axis}

\end{tikzpicture}%

%% file: tikz/sp500hist.tex
%
%
\definecolor{mycolor1}{rgb}{0.00000,0.44700,0.74100}%
\begin{tikzpicture}

\begin{axis}[%
width=0.3\fwidth,
at={(0\fwidth,0\fwidth)},
scale only axis,
xmin=-5.2,
xmax=5,
ymin=0,
ymax=0.6,
 tick label style={font=\footnotesize} , 
axis background/.style={fill=white},
title style={font=\bfseries},
xmajorgrids,
ymajorgrids,
legend style={legend cell align=left, align=left, draw=white!15!black}
]
\addplot[ybar interval, fill=mycolor1, fill opacity=0.6, draw=black, area legend] table[row sep=crcr] {%
x	y\\
-4.8	0.00999999999999999\\
-4.4	0\\
-4	0.02\\
-3.6	0\\
-3.2	0.01\\
-2.8	0\\
-2.4	0.03\\
-2	0.03\\
-1.6	0.05\\
-1.2	0.2\\
-0.800000000000001	0.4\\
-0.4	0.57\\
0	0.550000000000001\\
0.399999999999999	0.19\\
0.8	0.15\\
1.2	0.12\\
1.6	0.11\\
2	0.03\\
2.4	0.02\\
2.8	0.01\\
3.2	0.01\\
};

\addplot [color=red,line width=1.1pt]
  table[row sep=crcr]{%
-5	1.4867195147343e-06\\
-4.99	1.56286710894929e-06\\
-4.98	1.64275058804507e-06\\
-4.97	1.72654452167707e-06\\
-4.96	1.81443119018203e-06\\
-4.95	1.90660090312281e-06\\
-4.94	2.00325232994849e-06\\
-4.93	2.10459284318313e-06\\
-4.92	2.21083887456842e-06\\
-4.91	2.322216284598e-06\\
-4.9	2.43896074589335e-06\\
-4.89	2.56131814088454e-06\\
-4.88	2.68954497427152e-06\\
-4.87	2.82390880075582e-06\\
-4.86	2.96468866854527e-06\\
-4.85	3.11217557914894e-06\\
-4.84	3.26667296399328e-06\\
-4.83	3.42849717840504e-06\\
-4.82	3.59797801352125e-06\\
-4.81	3.77545922670135e-06\\
-4.8	3.96129909103208e-06\\
-4.79	4.1558709645312e-06\\
-4.78	4.35956387967164e-06\\
-4.77	4.57278315386414e-06\\
-4.76	4.79595102155252e-06\\
-4.75	5.02950728859245e-06\\
-4.74	5.2739100096013e-06\\
-4.73	5.52963618898405e-06\\
-4.72	5.79718250635729e-06\\
-4.71	6.07706606711112e-06\\
-4.7	6.36982517886709e-06\\
-4.69	6.67602015460746e-06\\
-4.68	6.99623414327041e-06\\
-4.67	7.33107398862395e-06\\
-4.66	7.68117111725046e-06\\
-4.65	8.04718245649229e-06\\
-4.64	8.42979138322877e-06\\
-4.63	8.8297087043741e-06\\
-4.62	9.24767367000562e-06\\
-4.61	9.68445502005144e-06\\
-4.6	1.01408520654868e-05\\
-4.59	1.06176958050084e-05\\
-4.58	1.11158500781778e-05\\
-4.57	1.16362127560427e-05\\
-4.56	1.21797169702687e-05\\
-4.55	1.27473323818335e-05\\
-4.54	1.33400664903558e-05\\
-4.53	1.39589659851548e-05\\
-4.52	1.46051181391529e-05\\
-4.51	1.52796522467616e-05\\
-4.5	1.59837411069055e-05\\
-4.49	1.67186025523651e-05\\
-4.48	1.74855010266391e-05\\
-4.47	1.82857492095474e-05\\
-4.46	1.91207096928177e-05\\
-4.45	1.99917967069228e-05\\
-4.44	2.09004779004505e-05\\
-4.43	2.18482761733165e-05\\
-4.42	2.28367715651469e-05\\
-4.41	2.38676032001796e-05\\
-4.4	2.49424712900535e-05\\
-4.39	2.60631391958783e-05\\
-4.38	2.72314355509926e-05\\
-4.37	2.84492564458443e-05\\
-4.36	2.97185676764422e-05\\
-4.35	3.10414070578503e-05\\
-4.34	3.24198868042138e-05\\
-4.33	3.38561959768279e-05\\
-4.32	3.53526030017731e-05\\
-4.31	3.69114582586662e-05\\
-4.3	3.85351967420871e-05\\
-4.29	4.0226340797265e-05\\
-4.28	4.19875029316173e-05\\
-4.27	4.38213887037581e-05\\
-4.26	4.57307996916013e-05\\
-4.25	4.7718636541205e-05\\
-4.24	4.97879020980121e-05\\
-4.23	5.19417046221598e-05\\
-4.22	5.41832610895402e-05\\
-4.21	5.65159005803074e-05\\
-4.2	5.89430677565399e-05\\
-4.19	6.14683264307694e-05\\
-4.18	6.40953632271061e-05\\
-4.17	6.68279913366906e-05\\
-4.16	6.96701543692143e-05\\
-4.15	7.26259303022523e-05\\
-4.14	7.56995355301612e-05\\
-4.13	7.88953290142931e-05\\
-4.12	8.2217816536286e-05\\
-4.11	8.56716550561819e-05\\
-4.1	8.92616571771329e-05\\
-4.09	9.29927957184459e-05\\
-4.08	9.68702083987193e-05\\
-4.07	0.000100899202630814\\
-4.06	0.0001050852604304\\
-4.05	0.000109434043439801\\
-4.04	0.000113951398068865\\
-4.03	0.000118643360754566\\
-4.02	0.000123516163341024\\
-4.01	0.000128576238581621\\
-4	0.000133830225764885\\
-3.99	0.00013928497646576\\
-3.98	0.000144947560423891\\
-3.97	0.000150825271550518\\
-3.96	0.000156925634065532\\
-3.95	0.000163256408766242\\
-3.94	0.000169825599429344\\
-3.93	0.000176641459347571\\
-3.92	0.000183712498002457\\
-3.91	0.000191047487874598\\
-3.9	0.000198655471392773\\
-3.89	0.000206545768023226\\
-3.88	0.000214727981500367\\
-3.87	0.000223212007200102\\
-3.86	0.000232008039656942\\
-3.85	0.000241126580225994\\
-3.84	0.000250578444890861\\
-3.83	0.000260374772218443\\
-3.82	0.000270527031461521\\
-3.81	0.000281047030809986\\
-3.8	0.00029194692579146\\
-3.79	0.000303239227822004\\
-3.78	0.000314936812907522\\
-3.77	0.000327052930496375\\
-3.76	0.000339601212483655\\
-3.75	0.000352595682367445\\
-3.74	0.000366050764557335\\
-3.73	0.000379981293835321\\
-3.72	0.000394402524969157\\
-3.71	0.000409330142478079\\
-3.7	0.000424780270550751\\
-3.69	0.000440769483115133\\
-3.68	0.000457314814059858\\
-3.67	0.000474433767606621\\
-3.66	0.000492144328832893\\
-3.65	0.000510464974344186\\
-3.64	0.000529414683094936\\
-3.63	0.000549012947356959\\
-3.62	0.000569279783834253\\
-3.61	0.000590235744922786\\
-3.6	0.000611901930113773\\
-3.59	0.000634299997538757\\
-3.58	0.000657452175654677\\
-3.57	0.000681381275066892\\
-3.56	0.000706110700488036\\
-3.55	0.000731664462830311\\
-3.54	0.00075806719142871\\
-3.53	0.000785344146392469\\
-3.52	0.000813521231081809\\
-3.51	0.000842625004706903\\
-3.5	0.00087268269504576\\
-3.49	0.000903722211277524\\
-3.48	0.00093577215692748\\
-3.47	0.000968861842919847\\
-3.46	0.00100302130073424\\
-3.45	0.00103828129566141\\
-3.44	0.00107467334015374\\
-3.43	0.00111222970726557\\
-3.42	0.00115098344417848\\
-3.41	0.00119096838580612\\
-3.4	0.00123221916847302\\
-3.39	0.00127477124366184\\
-3.38	0.00131866089182274\\
-3.37	0.0013639252362389\\
-3.36	0.00141060225694139\\
-3.35	0.00145873080466675\\
-3.34	0.00150835061485031\\
-3.33	0.00155950232164769\\
-3.32	0.00161222747197712\\
-3.31	0.00166656853957458\\
-3.3	0.00172256893905368\\
-3.29	0.00178027303996188\\
-3.28	0.00183972618082428\\
-3.27	0.00190097468316608\\
-3.26	0.00196406586550438\\
-3.25	0.00202904805729977\\
-3.24	0.00209597061285794\\
-3.23	0.00216488392517106\\
-3.22	0.00223583943968854\\
-3.21	0.0023088896680065\\
-3.2	0.00238408820146484\\
-3.19	0.0024614897246407\\
-3.18	0.00254115002872653\\
-3.17	0.00262312602478102\\
-3.16	0.0027074757568407\\
-3.15	0.00279425841487945\\
-3.14	0.00288353434760344\\
-3.13	0.00297536507506825\\
-3.12	0.00306981330110474\\
-3.11	0.00316694292554008\\
-3.1	0.00326681905619992\\
-3.09	0.00336950802067748\\
-3.08	0.00347507737785494\\
-3.07	0.00358359592916236\\
-3.06	0.00369513372955903\\
-3.05	0.00380976209822181\\
-3.04	0.00392755362892478\\
-3.03	0.00404858220009443\\
-3.02	0.00417292298452396\\
-3.01	0.00430065245873045\\
-3	0.00443184841193801\\
-2.99	0.00456658995467015\\
-2.98	0.00470495752693398\\
-2.97	0.00484703290597895\\
-2.96	0.00499289921361238\\
-2.95	0.00514264092305394\\
-2.94	0.00529634386531102\\
-2.93	0.00545409523505655\\
-2.92	0.00561598359599097\\
-2.91	0.00578209888566947\\
-2.9	0.00595253241977585\\
-2.89	0.00612737689582369\\
-2.88	0.00630672639626593\\
-2.87	0.00649067639099336\\
-2.86	0.00667932373920262\\
-2.85	0.00687276669061397\\
-2.84	0.00707110488601945\\
-2.83	0.00727443935714122\\
-2.82	0.00748287252578056\\
-2.81	0.00769650820223732\\
-2.8	0.00791545158297997\\
-2.79	0.00813980924754602\\
-2.78	0.00836968915465303\\
-2.77	0.00860520063749967\\
-2.76	0.00884645439823723\\
-2.75	0.00909356250159105\\
-2.74	0.00934663836761229\\
-2.73	0.00960579676353959\\
-2.72	0.00987115379475115\\
-2.71	0.0101428268947871\\
-2.7	0.0104209348144226\\
-2.69	0.0107055976097722\\
-2.68	0.0109969366294056\\
-2.67	0.0112950745004561\\
-2.66	0.0116001351137026\\
-2.65	0.0119122436076052\\
-2.64	0.012231526351278\\
-2.63	0.0125581109263782\\
-2.62	0.0128921261078953\\
-2.61	0.0132337018438214\\
-2.6	0.0135829692336856\\
-2.59	0.0139400605059358\\
-2.58	0.0143051089941497\\
-2.57	0.01467824911206\\
-2.56	0.0150596163273774\\
-2.55	0.0154493471343952\\
-2.54	0.0158475790253608\\
-2.53	0.0162544504606005\\
-2.52	0.0166701008373811\\
-2.51	0.017094670457497\\
-2.5	0.0175283004935685\\
-2.49	0.0179711329540397\\
-2.48	0.018423310646862\\
-2.47	0.0188849771418562\\
-2.46	0.019356276731737\\
-2.45	0.0198373543917953\\
-2.44	0.0203283557382258\\
-2.43	0.0208294269850922\\
-2.42	0.0213407148999228\\
-2.41	0.0218623667579294\\
-2.4	0.0223945302948429\\
-2.39	0.0229373536583607\\
-2.38	0.0234909853582014\\
-2.37	0.024055574214763\\
-2.36	0.0246312693063825\\
-2.35	0.0252182199151944\\
-2.34	0.0258165754715877\\
-2.33	0.0264264854972617\\
-2.32	0.0270480995468818\\
-2.31	0.0276815671483366\\
-2.3	0.0283270377416012\\
-2.29	0.0289846606162094\\
-2.28	0.0296545848473413\\
-2.27	0.0303369592305316\\
-2.26	0.0310319322150083\\
-2.25	0.0317396518356674\\
-2.24	0.0324602656436975\\
-2.23	0.0331939206358611\\
-2.22	0.0339407631824492\\
-2.21	0.0347009389539188\\
-2.2	0.0354745928462315\\
-2.19	0.0362618689049062\\
-2.18	0.0370629102478065\\
-2.17	0.0378778589866775\\
-2.16	0.0387068561474556\\
-2.15	0.0395500415893702\\
-2.14	0.0404075539228603\\
-2.13	0.0412795304263304\\
-2.12	0.0421661069617703\\
-2.11	0.0430674178892657\\
-2.1	0.0439835959804272\\
-2.09	0.0449147723307671\\
-2.08	0.0458610762710549\\
-2.07	0.0468226352776832\\
-2.06	0.047799574882077\\
-2.05	0.0487920185791828\\
-2.04	0.0498000877350708\\
-2.03	0.0508239014936912\\
-2.02	0.0518635766828206\\
-2.01	0.0529192277192403\\
-2	0.0539909665131881\\
-1.99	0.0550789023721258\\
-1.98	0.056183141903868\\
-1.97	0.0573037889191172\\
-1.96	0.0584409443334515\\
-1.95	0.0595947060688161\\
-1.94	0.0607651689545648\\
-1.93	0.0619524246281052\\
-1.92	0.0631565614351987\\
-1.91	0.0643776643299693\\
-1.9	0.0656158147746766\\
-1.89	0.0668710906393071\\
-1.88	0.0681435661010446\\
-1.87	0.0694333115436742\\
-1.86	0.0707403934569834\\
-1.85	0.072064874336218\\
-1.84	0.0734068125816569\\
-1.83	0.0747662623983676\\
-1.82	0.0761432736962074\\
-1.81	0.077537891990134\\
-1.8	0.0789501583008942\\
-1.79	0.0803801090561542\\
-1.78	0.0818277759921428\\
-1.77	0.0832931860558745\\
-1.76	0.0847763613080223\\
-1.75	0.0862773188265115\\
-1.74	0.0877960706109057\\
-1.73	0.089332623487655\\
-1.72	0.0908869790162829\\
-1.71	0.0924591333965807\\
-1.7	0.094049077376887\\
-1.69	0.095656796163524\\
-1.68	0.0972822693314675\\
-1.67	0.0989254707363237\\
-1.66	0.100586368427691\\
-1.65	0.102264924563978\\
-1.64	0.103961095328764\\
-1.63	0.105674830848764\\
-1.62	0.107406075113484\\
-1.61	0.109154765896647\\
-1.6	0.110920834679456\\
-1.59	0.112704206575771\\
-1.58	0.114504800259292\\
-1.57	0.116322527892807\\
-1.56	0.118157295059582\\
-1.55	0.120009000696986\\
-1.54	0.121877537032402\\
-1.53	0.123762789521523\\
-1.52	0.125664636789088\\
-1.51	0.127582950572142\\
-1.5	0.129517595665892\\
-1.49	0.131468429872231\\
-1.48	0.133435303951002\\
-1.47	0.135418061574071\\
-1.46	0.137416539282282\\
-1.45	0.13943056644536\\
-1.44	0.141459965224839\\
-1.43	0.143504550540062\\
-1.42	0.145564130037348\\
-1.41	0.147638504062356\\
-1.4	0.149727465635745\\
-1.39	0.151830800432162\\
-1.38	0.153948286762634\\
-1.37	0.156079695560421\\
-1.36	0.158224790370383\\
-1.35	0.16038332734192\\
-1.34	0.162555055225534\\
-1.33	0.164739715373077\\
-1.32	0.166937041741714\\
-1.31	0.169146760901672\\
-1.3	0.171368592047807\\
-1.29	0.173602247015033\\
-1.28	0.175847430297662\\
-1.27	0.178103839072694\\
-1.26	0.18037116322708\\
-1.25	0.182649085389022\\
-1.24	0.184937280963305\\
-1.23	0.18723541817073\\
-1.22	0.18954315809164\\
-1.21	0.191860154713599\\
-1.2	0.194186054983213\\
-1.19	0.196520498862137\\
-1.18	0.198863119387276\\
-1.17	0.201213542735197\\
-1.16	0.203571388290759\\
-1.15	0.205936268719975\\
-1.14	0.208307790047108\\
-1.13	0.210685551736015\\
-1.12	0.213069146775718\\
-1.11	0.21545816177022\\
-1.1	0.217852177032551\\
-1.09	0.220250766683033\\
-1.08	0.222653498751761\\
-1.07	0.22505993528527\\
-1.06	0.227469632457386\\
-1.05	0.229882140684233\\
-1.04	0.232297004743366\\
-1.03	0.234713763897012\\
-1.02	0.23713195201938\\
-1.01	0.239551097728013\\
-1	0.241970724519143\\
-0.99	0.244390350907\\
-0.98	0.246809490567043\\
-0.97	0.249227652483066\\
-0.96	0.251644341098117\\
-0.95	0.254059056469189\\
-0.94	0.25647129442562\\
-0.93	0.258880546731149\\
-0.92	0.261286301249553\\
-0.91	0.263688042113818\\
-0.9	0.266085249898755\\
-0.89	0.268477401797002\\
-0.88	0.270863971798338\\
-0.87	0.273244430872216\\
-0.86	0.275618247153457\\
-0.85	0.277984886130997\\
-0.84	0.280343810839621\\
-0.83	0.28269448205458\\
-0.82	0.285036358489007\\
-0.81	0.287368896994028\\
-0.8	0.289691552761483\\
-0.79	0.292003779529141\\
-0.78	0.294305029788325\\
-0.77	0.296594754993816\\
-0.76	0.298872405775953\\
-0.75	0.301137432154804\\
-0.74	0.3033892837563\\
-0.73	0.30562741003021\\
-0.72	0.307851260469853\\
-0.71	0.310060284833416\\
-0.7	0.312253933366761\\
-0.69	0.314431657027597\\
-0.68	0.316592907710893\\
-0.67	0.318737138475402\\
-0.66	0.320863803771172\\
-0.649999999999999	0.322972359667914\\
-0.64	0.325062264084082\\
-0.63	0.327132977016555\\
-0.62	0.329183960770765\\
-0.61	0.331214680191153\\
-0.6	0.3332246028918\\
-0.59	0.335213199487106\\
-0.58	0.337179943822381\\
-0.57	0.339124313204192\\
-0.56	0.341045788630353\\
-0.55	0.342943855019384\\
-0.54	0.344818001439333\\
-0.53	0.346667721335792\\
-0.52	0.348492512758975\\
-0.51	0.350291878589726\\
-0.5	0.3520653267643\\
-0.49	0.35381237049778\\
-0.48	0.355532528505997\\
-0.47	0.357225325225801\\
-0.46	0.358890291033545\\
-0.45	0.360526962461648\\
-0.44	0.362134882413092\\
-0.43	0.363713600373713\\
-0.42	0.365262672622154\\
-0.41	0.366781662437336\\
-0.399999999999999	0.368270140303323\\
-0.39	0.369727684111432\\
-0.38	0.371153879359466\\
-0.37	0.372548319347933\\
-0.36	0.373910605373128\\
-0.35	0.375240346916938\\
-0.34	0.376537161833254\\
-0.33	0.377800676530865\\
-0.32	0.379030526152702\\
-0.31	0.380226354751325\\
-0.3	0.381387815460524\\
-0.29	0.382514570662924\\
-0.28	0.383606292153479\\
-0.27	0.384662661298743\\
-0.26	0.385683369191816\\
-0.25	0.386668116802849\\
-0.24	0.387616615125014\\
-0.23	0.388528585315836\\
-0.22	0.38940375883379\\
-0.21	0.390241877570074\\
-0.2	0.391042693975456\\
-0.19	0.391805971182121\\
-0.18	0.392531483120429\\
-0.17	0.393219014630497\\
-0.16	0.393868361568541\\
-0.149999999999999	0.394479330907889\\
-0.14	0.395051740834611\\
-0.13	0.395585420837687\\
-0.12	0.396080211793656\\
-0.11	0.396535966045686\\
-0.0999999999999996	0.396952547477012\\
-0.0899999999999999	0.397329831578688\\
-0.0800000000000001	0.397667705511609\\
-0.0700000000000003	0.397966068162751\\
-0.0599999999999996	0.398224830195607\\
-0.0499999999999998	0.398443914094764\\
-0.04	0.398623254204605\\
-0.0300000000000002	0.3987627967621\\
-0.0199999999999996	0.398862499923666\\
-0.00999999999999979	0.398922333786082\\
0	0.398942280401433\\
0.00999999999999979	0.398922333786082\\
0.0199999999999996	0.398862499923666\\
0.0300000000000002	0.3987627967621\\
0.04	0.398623254204605\\
0.0499999999999998	0.398443914094764\\
0.0599999999999996	0.398224830195607\\
0.0700000000000003	0.397966068162751\\
0.0800000000000001	0.397667705511609\\
0.0899999999999999	0.397329831578688\\
0.0999999999999996	0.396952547477012\\
0.11	0.396535966045686\\
0.12	0.396080211793656\\
0.13	0.395585420837687\\
0.14	0.395051740834611\\
0.149999999999999	0.394479330907889\\
0.16	0.393868361568541\\
0.17	0.393219014630497\\
0.18	0.392531483120429\\
0.19	0.391805971182121\\
0.2	0.391042693975456\\
0.21	0.390241877570074\\
0.22	0.38940375883379\\
0.23	0.388528585315836\\
0.24	0.387616615125014\\
0.25	0.386668116802849\\
0.26	0.385683369191816\\
0.27	0.384662661298743\\
0.28	0.383606292153479\\
0.29	0.382514570662924\\
0.3	0.381387815460524\\
0.31	0.380226354751325\\
0.32	0.379030526152702\\
0.33	0.377800676530865\\
0.34	0.376537161833254\\
0.35	0.375240346916938\\
0.36	0.373910605373128\\
0.37	0.372548319347933\\
0.38	0.371153879359466\\
0.39	0.369727684111432\\
0.399999999999999	0.368270140303323\\
0.41	0.366781662437336\\
0.42	0.365262672622154\\
0.43	0.363713600373713\\
0.44	0.362134882413092\\
0.45	0.360526962461648\\
0.46	0.358890291033545\\
0.47	0.357225325225801\\
0.48	0.355532528505997\\
0.49	0.35381237049778\\
0.5	0.3520653267643\\
0.51	0.350291878589726\\
0.52	0.348492512758975\\
0.53	0.346667721335792\\
0.54	0.344818001439333\\
0.55	0.342943855019384\\
0.56	0.341045788630353\\
0.57	0.339124313204192\\
0.58	0.337179943822381\\
0.59	0.335213199487106\\
0.6	0.3332246028918\\
0.61	0.331214680191153\\
0.62	0.329183960770765\\
0.63	0.327132977016555\\
0.64	0.325062264084082\\
0.649999999999999	0.322972359667914\\
0.66	0.320863803771172\\
0.67	0.318737138475402\\
0.68	0.316592907710893\\
0.69	0.314431657027597\\
0.7	0.312253933366761\\
0.71	0.310060284833416\\
0.72	0.307851260469853\\
0.73	0.30562741003021\\
0.74	0.3033892837563\\
0.75	0.301137432154804\\
0.76	0.298872405775953\\
0.77	0.296594754993816\\
0.78	0.294305029788325\\
0.79	0.292003779529141\\
0.8	0.289691552761483\\
0.81	0.287368896994028\\
0.82	0.285036358489007\\
0.83	0.28269448205458\\
0.84	0.280343810839621\\
0.85	0.277984886130997\\
0.86	0.275618247153457\\
0.87	0.273244430872216\\
0.88	0.270863971798338\\
0.89	0.268477401797002\\
0.9	0.266085249898755\\
0.91	0.263688042113818\\
0.92	0.261286301249553\\
0.93	0.258880546731149\\
0.94	0.25647129442562\\
0.95	0.254059056469189\\
0.96	0.251644341098117\\
0.97	0.249227652483066\\
0.98	0.246809490567043\\
0.99	0.244390350907\\
1	0.241970724519143\\
1.01	0.239551097728013\\
1.02	0.23713195201938\\
1.03	0.234713763897012\\
1.04	0.232297004743366\\
1.05	0.229882140684233\\
1.06	0.227469632457386\\
1.07	0.22505993528527\\
1.08	0.222653498751761\\
1.09	0.220250766683033\\
1.1	0.217852177032551\\
1.11	0.21545816177022\\
1.12	0.213069146775718\\
1.13	0.210685551736015\\
1.14	0.208307790047108\\
1.15	0.205936268719975\\
1.16	0.203571388290759\\
1.17	0.201213542735197\\
1.18	0.198863119387276\\
1.19	0.196520498862137\\
1.2	0.194186054983213\\
1.21	0.191860154713599\\
1.22	0.18954315809164\\
1.23	0.18723541817073\\
1.24	0.184937280963305\\
1.25	0.182649085389022\\
1.26	0.18037116322708\\
1.27	0.178103839072694\\
1.28	0.175847430297662\\
1.29	0.173602247015033\\
1.3	0.171368592047807\\
1.31	0.169146760901672\\
1.32	0.166937041741714\\
1.33	0.164739715373077\\
1.34	0.162555055225534\\
1.35	0.16038332734192\\
1.36	0.158224790370383\\
1.37	0.156079695560421\\
1.38	0.153948286762634\\
1.39	0.151830800432162\\
1.4	0.149727465635745\\
1.41	0.147638504062356\\
1.42	0.145564130037348\\
1.43	0.143504550540062\\
1.44	0.141459965224839\\
1.45	0.13943056644536\\
1.46	0.137416539282282\\
1.47	0.135418061574071\\
1.48	0.133435303951002\\
1.49	0.131468429872231\\
1.5	0.129517595665892\\
1.51	0.127582950572142\\
1.52	0.125664636789088\\
1.53	0.123762789521523\\
1.54	0.121877537032402\\
1.55	0.120009000696986\\
1.56	0.118157295059582\\
1.57	0.116322527892807\\
1.58	0.114504800259292\\
1.59	0.112704206575771\\
1.6	0.110920834679456\\
1.61	0.109154765896647\\
1.62	0.107406075113484\\
1.63	0.105674830848764\\
1.64	0.103961095328764\\
1.65	0.102264924563978\\
1.66	0.100586368427691\\
1.67	0.0989254707363237\\
1.68	0.0972822693314675\\
1.69	0.095656796163524\\
1.7	0.094049077376887\\
1.71	0.0924591333965807\\
1.72	0.0908869790162829\\
1.73	0.089332623487655\\
1.74	0.0877960706109057\\
1.75	0.0862773188265115\\
1.76	0.0847763613080223\\
1.77	0.0832931860558745\\
1.78	0.0818277759921428\\
1.79	0.0803801090561542\\
1.8	0.0789501583008942\\
1.81	0.077537891990134\\
1.82	0.0761432736962074\\
1.83	0.0747662623983676\\
1.84	0.0734068125816569\\
1.85	0.072064874336218\\
1.86	0.0707403934569834\\
1.87	0.0694333115436742\\
1.88	0.0681435661010446\\
1.89	0.0668710906393071\\
1.9	0.0656158147746766\\
1.91	0.0643776643299693\\
1.92	0.0631565614351987\\
1.93	0.0619524246281052\\
1.94	0.0607651689545648\\
1.95	0.0595947060688161\\
1.96	0.0584409443334515\\
1.97	0.0573037889191172\\
1.98	0.056183141903868\\
1.99	0.0550789023721258\\
2	0.0539909665131881\\
2.01	0.0529192277192403\\
2.02	0.0518635766828206\\
2.03	0.0508239014936912\\
2.04	0.0498000877350708\\
2.05	0.0487920185791828\\
2.06	0.047799574882077\\
2.07	0.0468226352776832\\
2.08	0.0458610762710549\\
2.09	0.0449147723307671\\
2.1	0.0439835959804272\\
2.11	0.0430674178892657\\
2.12	0.0421661069617703\\
2.13	0.0412795304263304\\
2.14	0.0404075539228603\\
2.15	0.0395500415893702\\
2.16	0.0387068561474556\\
2.17	0.0378778589866775\\
2.18	0.0370629102478065\\
2.19	0.0362618689049062\\
2.2	0.0354745928462315\\
2.21	0.0347009389539188\\
2.22	0.0339407631824492\\
2.23	0.0331939206358611\\
2.24	0.0324602656436975\\
2.25	0.0317396518356674\\
2.26	0.0310319322150083\\
2.27	0.0303369592305316\\
2.28	0.0296545848473413\\
2.29	0.0289846606162094\\
2.3	0.0283270377416012\\
2.31	0.0276815671483366\\
2.32	0.0270480995468818\\
2.33	0.0264264854972617\\
2.34	0.0258165754715877\\
2.35	0.0252182199151944\\
2.36	0.0246312693063825\\
2.37	0.024055574214763\\
2.38	0.0234909853582014\\
2.39	0.0229373536583607\\
2.4	0.0223945302948429\\
2.41	0.0218623667579294\\
2.42	0.0213407148999228\\
2.43	0.0208294269850922\\
2.44	0.0203283557382258\\
2.45	0.0198373543917953\\
2.46	0.019356276731737\\
2.47	0.0188849771418562\\
2.48	0.018423310646862\\
2.49	0.0179711329540397\\
2.5	0.0175283004935685\\
2.51	0.017094670457497\\
2.52	0.0166701008373811\\
2.53	0.0162544504606005\\
2.54	0.0158475790253608\\
2.55	0.0154493471343952\\
2.56	0.0150596163273774\\
2.57	0.01467824911206\\
2.58	0.0143051089941497\\
2.59	0.0139400605059358\\
2.6	0.0135829692336856\\
2.61	0.0132337018438214\\
2.62	0.0128921261078953\\
2.63	0.0125581109263782\\
2.64	0.012231526351278\\
2.65	0.0119122436076052\\
2.66	0.0116001351137026\\
2.67	0.0112950745004561\\
2.68	0.0109969366294056\\
2.69	0.0107055976097722\\
2.7	0.0104209348144226\\
2.71	0.0101428268947871\\
2.72	0.00987115379475115\\
2.73	0.00960579676353959\\
2.74	0.00934663836761229\\
2.75	0.00909356250159105\\
2.76	0.00884645439823723\\
2.77	0.00860520063749967\\
2.78	0.00836968915465303\\
2.79	0.00813980924754602\\
2.8	0.00791545158297997\\
2.81	0.00769650820223732\\
2.82	0.00748287252578056\\
2.83	0.00727443935714122\\
2.84	0.00707110488601945\\
2.85	0.00687276669061397\\
2.86	0.00667932373920262\\
2.87	0.00649067639099336\\
2.88	0.00630672639626593\\
2.89	0.00612737689582369\\
2.9	0.00595253241977585\\
2.91	0.00578209888566947\\
2.92	0.00561598359599097\\
2.93	0.00545409523505655\\
2.94	0.00529634386531102\\
2.95	0.00514264092305394\\
2.96	0.00499289921361238\\
2.97	0.00484703290597895\\
2.98	0.00470495752693398\\
2.99	0.00456658995467015\\
3	0.00443184841193801\\
3.01	0.00430065245873045\\
3.02	0.00417292298452396\\
3.03	0.00404858220009443\\
3.04	0.00392755362892478\\
3.05	0.00380976209822181\\
3.06	0.00369513372955903\\
3.07	0.00358359592916236\\
3.08	0.00347507737785494\\
3.09	0.00336950802067748\\
3.1	0.00326681905619992\\
3.11	0.00316694292554008\\
3.12	0.00306981330110474\\
3.13	0.00297536507506825\\
3.14	0.00288353434760344\\
3.15	0.00279425841487945\\
3.16	0.0027074757568407\\
3.17	0.00262312602478102\\
3.18	0.00254115002872653\\
3.19	0.0024614897246407\\
3.2	0.00238408820146484\\
3.21	0.0023088896680065\\
3.22	0.00223583943968854\\
3.23	0.00216488392517106\\
3.24	0.00209597061285794\\
3.25	0.00202904805729977\\
3.26	0.00196406586550438\\
3.27	0.00190097468316608\\
3.28	0.00183972618082428\\
3.29	0.00178027303996188\\
3.3	0.00172256893905368\\
3.31	0.00166656853957458\\
3.32	0.00161222747197712\\
3.33	0.00155950232164769\\
3.34	0.00150835061485031\\
3.35	0.00145873080466675\\
3.36	0.00141060225694139\\
3.37	0.0013639252362389\\
3.38	0.00131866089182274\\
3.39	0.00127477124366184\\
3.4	0.00123221916847302\\
3.41	0.00119096838580612\\
3.42	0.00115098344417848\\
3.43	0.00111222970726557\\
3.44	0.00107467334015374\\
3.45	0.00103828129566141\\
3.46	0.00100302130073424\\
3.47	0.000968861842919847\\
3.48	0.00093577215692748\\
3.49	0.000903722211277524\\
3.5	0.00087268269504576\\
3.51	0.000842625004706903\\
3.52	0.000813521231081809\\
3.53	0.000785344146392469\\
3.54	0.00075806719142871\\
3.55	0.000731664462830311\\
3.56	0.000706110700488036\\
3.57	0.000681381275066892\\
3.58	0.000657452175654677\\
3.59	0.000634299997538757\\
3.6	0.000611901930113773\\
3.61	0.000590235744922786\\
3.62	0.000569279783834253\\
3.63	0.000549012947356959\\
3.64	0.000529414683094936\\
3.65	0.000510464974344186\\
3.66	0.000492144328832893\\
3.67	0.000474433767606621\\
3.68	0.000457314814059858\\
3.69	0.000440769483115133\\
3.7	0.000424780270550751\\
3.71	0.000409330142478079\\
3.72	0.000394402524969157\\
3.73	0.000379981293835321\\
3.74	0.000366050764557335\\
3.75	0.000352595682367445\\
3.76	0.000339601212483655\\
3.77	0.000327052930496375\\
3.78	0.000314936812907522\\
3.79	0.000303239227822004\\
3.8	0.00029194692579146\\
3.81	0.000281047030809986\\
3.82	0.000270527031461521\\
3.83	0.000260374772218443\\
3.84	0.000250578444890861\\
3.85	0.000241126580225994\\
3.86	0.000232008039656942\\
3.87	0.000223212007200102\\
3.88	0.000214727981500367\\
3.89	0.000206545768023226\\
3.9	0.000198655471392773\\
3.91	0.000191047487874598\\
3.92	0.000183712498002457\\
3.93	0.000176641459347571\\
3.94	0.000169825599429344\\
3.95	0.000163256408766242\\
3.96	0.000156925634065532\\
3.97	0.000150825271550518\\
3.98	0.000144947560423891\\
3.99	0.00013928497646576\\
4	0.000133830225764885\\
4.01	0.000128576238581621\\
4.02	0.000123516163341024\\
4.03	0.000118643360754566\\
4.04	0.000113951398068865\\
4.05	0.000109434043439801\\
4.06	0.0001050852604304\\
4.07	0.000100899202630814\\
4.08	9.68702083987193e-05\\
4.09	9.29927957184459e-05\\
4.1	8.92616571771329e-05\\
4.11	8.56716550561819e-05\\
4.12	8.2217816536286e-05\\
4.13	7.88953290142931e-05\\
4.14	7.56995355301612e-05\\
4.15	7.26259303022523e-05\\
4.16	6.96701543692143e-05\\
4.17	6.68279913366906e-05\\
4.18	6.40953632271061e-05\\
4.19	6.14683264307694e-05\\
4.2	5.89430677565399e-05\\
4.21	5.65159005803074e-05\\
4.22	5.41832610895402e-05\\
4.23	5.19417046221598e-05\\
4.24	4.97879020980121e-05\\
4.25	4.7718636541205e-05\\
4.26	4.57307996916013e-05\\
4.27	4.38213887037581e-05\\
4.28	4.19875029316173e-05\\
4.29	4.0226340797265e-05\\
4.3	3.85351967420871e-05\\
4.31	3.69114582586662e-05\\
4.32	3.53526030017731e-05\\
4.33	3.38561959768279e-05\\
4.34	3.24198868042138e-05\\
4.35	3.10414070578503e-05\\
4.36	2.97185676764422e-05\\
4.37	2.84492564458443e-05\\
4.38	2.72314355509926e-05\\
4.39	2.60631391958783e-05\\
4.4	2.49424712900535e-05\\
4.41	2.38676032001796e-05\\
4.42	2.28367715651469e-05\\
4.43	2.18482761733165e-05\\
4.44	2.09004779004505e-05\\
4.45	1.99917967069228e-05\\
4.46	1.91207096928177e-05\\
4.47	1.82857492095474e-05\\
4.48	1.74855010266391e-05\\
4.49	1.67186025523651e-05\\
4.5	1.59837411069055e-05\\
4.51	1.52796522467616e-05\\
4.52	1.46051181391529e-05\\
4.53	1.39589659851548e-05\\
4.54	1.33400664903558e-05\\
4.55	1.27473323818335e-05\\
4.56	1.21797169702687e-05\\
4.57	1.16362127560427e-05\\
4.58	1.11158500781778e-05\\
4.59	1.06176958050084e-05\\
4.6	1.01408520654868e-05\\
4.61	9.68445502005144e-06\\
4.62	9.24767367000562e-06\\
4.63	8.8297087043741e-06\\
4.64	8.42979138322877e-06\\
4.65	8.04718245649229e-06\\
4.66	7.68117111725046e-06\\
4.67	7.33107398862395e-06\\
4.68	6.99623414327041e-06\\
4.69	6.67602015460746e-06\\
4.7	6.36982517886709e-06\\
4.71	6.07706606711112e-06\\
4.72	5.79718250635729e-06\\
4.73	5.52963618898405e-06\\
4.74	5.2739100096013e-06\\
4.75	5.02950728859245e-06\\
4.76	4.79595102155252e-06\\
4.77	4.57278315386414e-06\\
4.78	4.35956387967164e-06\\
4.79	4.1558709645312e-06\\
4.8	3.96129909103208e-06\\
4.81	3.77545922670135e-06\\
4.82	3.59797801352125e-06\\
4.83	3.42849717840504e-06\\
4.84	3.26667296399328e-06\\
4.85	3.11217557914894e-06\\
4.86	2.96468866854527e-06\\
4.87	2.82390880075582e-06\\
4.88	2.68954497427152e-06\\
4.89	2.56131814088454e-06\\
4.9	2.43896074589335e-06\\
4.91	2.322216284598e-06\\
4.92	2.21083887456842e-06\\
4.93	2.10459284318313e-06\\
4.94	2.00325232994849e-06\\
4.95	1.90660090312281e-06\\
4.96	1.81443119018203e-06\\
4.97	1.72654452167707e-06\\
4.98	1.64275058804507e-06\\
4.99	1.56286710894929e-06\\
5	1.4867195147343e-06\\
};

\end{axis}

\end{tikzpicture}%

%% file: tikz/nasdaq100hist.tex
%
%
\definecolor{mycolor1}{rgb}{0.00000,0.44700,0.74100}%
\begin{tikzpicture}

\begin{axis}[%
width=0.3\fwidth,
at={(0\fwidth,0\fwidth)},
scale only axis,
xmin=-5,
xmax=5,
ymin=0,
ymax=0.7,
 tick label style={font=\footnotesize} , 
axis background/.style={fill=white},
title style={font=\bfseries},
xmajorgrids,
ymajorgrids,
legend style={legend cell align=left, align=left, draw=white!15!black}
]
\addplot[ybar interval, fill=mycolor1, fill opacity=0.6, draw=black, area legend] table[row sep=crcr] {%
x	y\\
-4.4	0.00909090909090909\\
-3.96	0.0181818181818182\\
-3.52	0.00909090909090908\\
-3.08	0.0272727272727273\\
-2.64	0.00909090909090908\\
-2.2	0.0181818181818182\\
-1.76	0.0545454545454545\\
-1.32	0.109090909090909\\
-0.88	0.281818181818182\\
-0.44	0.536363636363636\\
0	0.627272727272728\\
0.44	0.209090909090909\\
0.88	0.2\\
1.32	0.109090909090909\\
1.76	0.0363636363636364\\
2.2	0.00909090909090908\\
2.64	0\\
3.08	0\\
3.52	0\\
3.96	0.00909090909090906\\
4.4	0.00909090909090906\\
};

\addplot [color=red,line width=1.1pt]
  table[row sep=crcr]{%
-5	1.4867195147343e-06\\
-4.95	1.90660090312281e-06\\
-4.9	2.43896074589335e-06\\
-4.85	3.11217557914894e-06\\
-4.8	3.96129909103208e-06\\
-4.75	5.02950728859245e-06\\
-4.7	6.36982517886709e-06\\
-4.65	8.04718245649229e-06\\
-4.6	1.01408520654868e-05\\
-4.55	1.27473323818335e-05\\
-4.5	1.59837411069055e-05\\
-4.45	1.99917967069228e-05\\
-4.4	2.49424712900535e-05\\
-4.35	3.10414070578503e-05\\
-4.3	3.85351967420871e-05\\
-4.25	4.7718636541205e-05\\
-4.2	5.89430677565399e-05\\
-4.15	7.26259303022523e-05\\
-4.1	8.92616571771329e-05\\
-4.05	0.000109434043439801\\
-4	0.000133830225764885\\
-3.95	0.000163256408766242\\
-3.9	0.000198655471392773\\
-3.85	0.000241126580225994\\
-3.8	0.00029194692579146\\
-3.75	0.000352595682367445\\
-3.7	0.000424780270550751\\
-3.65	0.000510464974344186\\
-3.6	0.000611901930113773\\
-3.55	0.000731664462830311\\
-3.5	0.00087268269504576\\
-3.45	0.00103828129566141\\
-3.4	0.00123221916847302\\
-3.35	0.00145873080466675\\
-3.3	0.00172256893905368\\
-3.25	0.00202904805729977\\
-3.2	0.00238408820146484\\
-3.15	0.00279425841487945\\
-3.1	0.00326681905619992\\
-3.05	0.00380976209822181\\
-3	0.00443184841193801\\
-2.95	0.00514264092305394\\
-2.9	0.00595253241977585\\
-2.85	0.00687276669061397\\
-2.8	0.00791545158297997\\
-2.75	0.00909356250159105\\
-2.7	0.0104209348144226\\
-2.65	0.0119122436076052\\
-2.6	0.0135829692336856\\
-2.55	0.0154493471343952\\
-2.5	0.0175283004935685\\
-2.45	0.0198373543917953\\
-2.4	0.0223945302948429\\
-2.35	0.0252182199151944\\
-2.3	0.0283270377416012\\
-2.25	0.0317396518356674\\
-2.2	0.0354745928462315\\
-2.15	0.0395500415893702\\
-2.1	0.0439835959804272\\
-2.05	0.0487920185791828\\
-2	0.0539909665131881\\
-1.95	0.0595947060688161\\
-1.9	0.0656158147746766\\
-1.85	0.0720648743362181\\
-1.8	0.0789501583008942\\
-1.75	0.0862773188265115\\
-1.7	0.094049077376887\\
-1.65	0.102264924563978\\
-1.6	0.110920834679456\\
-1.55	0.120009000696986\\
-1.5	0.129517595665892\\
-1.45	0.13943056644536\\
-1.4	0.149727465635745\\
-1.35	0.16038332734192\\
-1.3	0.171368592047807\\
-1.25	0.182649085389022\\
-1.2	0.194186054983213\\
-1.15	0.205936268719975\\
-1.1	0.217852177032551\\
-1.05	0.229882140684233\\
-1	0.241970724519143\\
-0.95	0.254059056469189\\
-0.899999999999999	0.266085249898755\\
-0.85	0.277984886130997\\
-0.8	0.289691552761483\\
-0.75	0.301137432154804\\
-0.7	0.312253933366761\\
-0.649999999999999	0.322972359667914\\
-0.6	0.3332246028918\\
-0.55	0.342943855019384\\
-0.5	0.3520653267643\\
-0.45	0.360526962461648\\
-0.399999999999999	0.368270140303323\\
-0.35	0.375240346916938\\
-0.3	0.381387815460524\\
-0.25	0.386668116802849\\
-0.199999999999999	0.391042693975456\\
-0.149999999999999	0.394479330907889\\
-0.0999999999999996	0.396952547477012\\
-0.0499999999999998	0.398443914094764\\
0	0.398942280401433\\
0.0499999999999998	0.398443914094764\\
0.0999999999999996	0.396952547477012\\
0.149999999999999	0.394479330907889\\
0.199999999999999	0.391042693975456\\
0.25	0.386668116802849\\
0.3	0.381387815460524\\
0.35	0.375240346916938\\
0.399999999999999	0.368270140303323\\
0.45	0.360526962461648\\
0.5	0.3520653267643\\
0.55	0.342943855019384\\
0.6	0.3332246028918\\
0.649999999999999	0.322972359667914\\
0.7	0.312253933366761\\
0.75	0.301137432154804\\
0.8	0.289691552761483\\
0.85	0.277984886130997\\
0.899999999999999	0.266085249898755\\
0.95	0.254059056469189\\
1	0.241970724519143\\
1.05	0.229882140684233\\
1.1	0.217852177032551\\
1.15	0.205936268719975\\
1.2	0.194186054983213\\
1.25	0.182649085389022\\
1.3	0.171368592047807\\
1.35	0.16038332734192\\
1.4	0.149727465635745\\
1.45	0.13943056644536\\
1.5	0.129517595665892\\
1.55	0.120009000696986\\
1.6	0.110920834679456\\
1.65	0.102264924563978\\
1.7	0.094049077376887\\
1.75	0.0862773188265115\\
1.8	0.0789501583008942\\
1.85	0.0720648743362181\\
1.9	0.0656158147746766\\
1.95	0.0595947060688161\\
2	0.0539909665131881\\
2.05	0.0487920185791828\\
2.1	0.0439835959804272\\
2.15	0.0395500415893702\\
2.2	0.0354745928462315\\
2.25	0.0317396518356674\\
2.3	0.0283270377416012\\
2.35	0.0252182199151944\\
2.4	0.0223945302948429\\
2.45	0.0198373543917953\\
2.5	0.0175283004935685\\
2.55	0.0154493471343952\\
2.6	0.0135829692336856\\
2.65	0.0119122436076052\\
2.7	0.0104209348144226\\
2.75	0.00909356250159105\\
2.8	0.00791545158297997\\
2.85	0.00687276669061397\\
2.9	0.00595253241977585\\
2.95	0.00514264092305394\\
3	0.00443184841193801\\
3.05	0.00380976209822181\\
3.1	0.00326681905619992\\
3.15	0.00279425841487945\\
3.2	0.00238408820146484\\
3.25	0.00202904805729977\\
3.3	0.00172256893905368\\
3.35	0.00145873080466675\\
3.4	0.00123221916847302\\
3.45	0.00103828129566141\\
3.5	0.00087268269504576\\
3.55	0.000731664462830311\\
3.6	0.000611901930113773\\
3.65	0.000510464974344186\\
3.7	0.000424780270550751\\
3.75	0.000352595682367445\\
3.8	0.00029194692579146\\
3.85	0.000241126580225994\\
3.9	0.000198655471392773\\
3.95	0.000163256408766242\\
4	0.000133830225764885\\
4.05	0.000109434043439801\\
4.1	8.92616571771329e-05\\
4.15	7.26259303022523e-05\\
4.2	5.89430677565399e-05\\
4.25	4.7718636541205e-05\\
4.3	3.85351967420871e-05\\
4.35	3.10414070578503e-05\\
4.4	2.49424712900535e-05\\
4.45	1.99917967069228e-05\\
4.5	1.59837411069055e-05\\
4.55	1.27473323818335e-05\\
4.6	1.01408520654868e-05\\
4.65	8.04718245649229e-06\\
4.7	6.36982517886709e-06\\
4.75	5.02950728859245e-06\\
4.8	3.96129909103208e-06\\
4.85	3.11217557914894e-06\\
4.9	2.43896074589335e-06\\
4.95	1.90660090312281e-06\\
5	1.4867195147343e-06\\
};

\end{axis}

\end{tikzpicture}%

%% file: tikz/gaudatahist.tex
%
%
\definecolor{mycolor1}{rgb}{0.00000,0.44700,0.74100}%
\begin{tikzpicture}

\begin{axis}[%
width=0.3\fwidth,
at={(0\fwidth,0\fwidth)},
scale only axis,
xmin=-5,
xmax=5,
ymin=0,
ymax=0.5,
 tick label style={font=\footnotesize} , 
axis background/.style={fill=white},
title style={font=\bfseries},
xmajorgrids,
ymajorgrids,
legend style={legend cell align=left, align=left, draw=white!15!black}
]
\addplot[ybar interval, fill=mycolor1, fill opacity=0.6, draw=black, area legend] table[row sep=crcr] {%
x	y\\
-2.6	0.0285714285714286\\
-2.32	0.0285714285714285\\
-2.04	0.0571428571428571\\
-1.76	0.1\\
-1.48	0.157142857142857\\
-1.2	0.214285714285714\\
-0.92	0.3\\
-0.64	0.314285714285714\\
-0.36	0.342857142857143\\
-0.0799999999999996	0.328571428571429\\
0.2	0.457142857142857\\
0.48	0.342857142857143\\
0.76	0.257142857142857\\
1.04	0.2\\
1.32	0.2\\
1.6	0.0999999999999999\\
1.88	0.0428571428571428\\
2.16	0.0428571428571428\\
2.44	0.0285714285714286\\
2.72	0.0285714285714285\\
3	0.0285714285714285\\
};

\addplot [color=red,line width=1.1pt]
  table[row sep=crcr]{%
-5	1.4867195147343e-06\\
-4.95	1.90660090312281e-06\\
-4.9	2.43896074589335e-06\\
-4.85	3.11217557914894e-06\\
-4.8	3.96129909103208e-06\\
-4.75	5.02950728859245e-06\\
-4.7	6.36982517886709e-06\\
-4.65	8.04718245649229e-06\\
-4.6	1.01408520654868e-05\\
-4.55	1.27473323818335e-05\\
-4.5	1.59837411069055e-05\\
-4.45	1.99917967069228e-05\\
-4.4	2.49424712900535e-05\\
-4.35	3.10414070578503e-05\\
-4.3	3.85351967420871e-05\\
-4.25	4.7718636541205e-05\\
-4.2	5.89430677565399e-05\\
-4.15	7.26259303022523e-05\\
-4.1	8.92616571771329e-05\\
-4.05	0.000109434043439801\\
-4	0.000133830225764885\\
-3.95	0.000163256408766242\\
-3.9	0.000198655471392773\\
-3.85	0.000241126580225994\\
-3.8	0.00029194692579146\\
-3.75	0.000352595682367445\\
-3.7	0.000424780270550751\\
-3.65	0.000510464974344186\\
-3.6	0.000611901930113773\\
-3.55	0.000731664462830311\\
-3.5	0.00087268269504576\\
-3.45	0.00103828129566141\\
-3.4	0.00123221916847302\\
-3.35	0.00145873080466675\\
-3.3	0.00172256893905368\\
-3.25	0.00202904805729977\\
-3.2	0.00238408820146484\\
-3.15	0.00279425841487945\\
-3.1	0.00326681905619992\\
-3.05	0.00380976209822181\\
-3	0.00443184841193801\\
-2.95	0.00514264092305394\\
-2.9	0.00595253241977585\\
-2.85	0.00687276669061397\\
-2.8	0.00791545158297997\\
-2.75	0.00909356250159105\\
-2.7	0.0104209348144226\\
-2.65	0.0119122436076052\\
-2.6	0.0135829692336856\\
-2.55	0.0154493471343952\\
-2.5	0.0175283004935685\\
-2.45	0.0198373543917953\\
-2.4	0.0223945302948429\\
-2.35	0.0252182199151944\\
-2.3	0.0283270377416012\\
-2.25	0.0317396518356674\\
-2.2	0.0354745928462315\\
-2.15	0.0395500415893702\\
-2.1	0.0439835959804272\\
-2.05	0.0487920185791828\\
-2	0.0539909665131881\\
-1.95	0.0595947060688161\\
-1.9	0.0656158147746766\\
-1.85	0.0720648743362181\\
-1.8	0.0789501583008942\\
-1.75	0.0862773188265115\\
-1.7	0.094049077376887\\
-1.65	0.102264924563978\\
-1.6	0.110920834679456\\
-1.55	0.120009000696986\\
-1.5	0.129517595665892\\
-1.45	0.13943056644536\\
-1.4	0.149727465635745\\
-1.35	0.16038332734192\\
-1.3	0.171368592047807\\
-1.25	0.182649085389022\\
-1.2	0.194186054983213\\
-1.15	0.205936268719975\\
-1.1	0.217852177032551\\
-1.05	0.229882140684233\\
-1	0.241970724519143\\
-0.95	0.254059056469189\\
-0.899999999999999	0.266085249898755\\
-0.85	0.277984886130997\\
-0.8	0.289691552761483\\
-0.75	0.301137432154804\\
-0.7	0.312253933366761\\
-0.649999999999999	0.322972359667914\\
-0.6	0.3332246028918\\
-0.55	0.342943855019384\\
-0.5	0.3520653267643\\
-0.45	0.360526962461648\\
-0.399999999999999	0.368270140303323\\
-0.35	0.375240346916938\\
-0.3	0.381387815460524\\
-0.25	0.386668116802849\\
-0.199999999999999	0.391042693975456\\
-0.149999999999999	0.394479330907889\\
-0.0999999999999996	0.396952547477012\\
-0.0499999999999998	0.398443914094764\\
0	0.398942280401433\\
0.0499999999999998	0.398443914094764\\
0.0999999999999996	0.396952547477012\\
0.149999999999999	0.394479330907889\\
0.199999999999999	0.391042693975456\\
0.25	0.386668116802849\\
0.3	0.381387815460524\\
0.35	0.375240346916938\\
0.399999999999999	0.368270140303323\\
0.45	0.360526962461648\\
0.5	0.3520653267643\\
0.55	0.342943855019384\\
0.6	0.3332246028918\\
0.649999999999999	0.322972359667914\\
0.7	0.312253933366761\\
0.75	0.301137432154804\\
0.8	0.289691552761483\\
0.85	0.277984886130997\\
0.899999999999999	0.266085249898755\\
0.95	0.254059056469189\\
1	0.241970724519143\\
1.05	0.229882140684233\\
1.1	0.217852177032551\\
1.15	0.205936268719975\\
1.2	0.194186054983213\\
1.25	0.182649085389022\\
1.3	0.171368592047807\\
1.35	0.16038332734192\\
1.4	0.149727465635745\\
1.45	0.13943056644536\\
1.5	0.129517595665892\\
1.55	0.120009000696986\\
1.6	0.110920834679456\\
1.65	0.102264924563978\\
1.7	0.094049077376887\\
1.75	0.0862773188265115\\
1.8	0.0789501583008942\\
1.85	0.0720648743362181\\
1.9	0.0656158147746766\\
1.95	0.0595947060688161\\
2	0.0539909665131881\\
2.05	0.0487920185791828\\
2.1	0.0439835959804272\\
2.15	0.0395500415893702\\
2.2	0.0354745928462315\\
2.25	0.0317396518356674\\
2.3	0.0283270377416012\\
2.35	0.0252182199151944\\
2.4	0.0223945302948429\\
2.45	0.0198373543917953\\
2.5	0.0175283004935685\\
2.55	0.0154493471343952\\
2.6	0.0135829692336856\\
2.65	0.0119122436076052\\
2.7	0.0104209348144226\\
2.75	0.00909356250159105\\
2.8	0.00791545158297997\\
2.85	0.00687276669061397\\
2.9	0.00595253241977585\\
2.95	0.00514264092305394\\
3	0.00443184841193801\\
3.05	0.00380976209822181\\
3.1	0.00326681905619992\\
3.15	0.00279425841487945\\
3.2	0.00238408820146484\\
3.25	0.00202904805729977\\
3.3	0.00172256893905368\\
3.35	0.00145873080466675\\
3.4	0.00123221916847302\\
3.45	0.00103828129566141\\
3.5	0.00087268269504576\\
3.55	0.000731664462830311\\
3.6	0.000611901930113773\\
3.65	0.000510464974344186\\
3.7	0.000424780270550751\\
3.75	0.000352595682367445\\
3.8	0.00029194692579146\\
3.85	0.000241126580225994\\
3.9	0.000198655471392773\\
3.95	0.000163256408766242\\
4	0.000133830225764885\\
4.05	0.000109434043439801\\
4.1	8.92616571771329e-05\\
4.15	7.26259303022523e-05\\
4.2	5.89430677565399e-05\\
4.25	4.7718636541205e-05\\
4.3	3.85351967420871e-05\\
4.35	3.10414070578503e-05\\
4.4	2.49424712900535e-05\\
4.45	1.99917967069228e-05\\
4.5	1.59837411069055e-05\\
4.55	1.27473323818335e-05\\
4.6	1.01408520654868e-05\\
4.65	8.04718245649229e-06\\
4.7	6.36982517886709e-06\\
4.75	5.02950728859245e-06\\
4.8	3.96129909103208e-06\\
4.85	3.11217557914894e-06\\
4.9	2.43896074589335e-06\\
4.95	1.90660090312281e-06\\
5	1.4867195147343e-06\\
};

\end{axis}

\end{tikzpicture}%

%% file: tikz/scatterplot.tex
%
%
\definecolor{mycolor1}{rgb}{0.00000,0.44700,0.74100}%
\begin{tikzpicture}

\begin{axis}[%
width=0.58\fwidth,
at={(0\fwidth,0\fwidth)},
scale only axis,
xmin=-0.02,
xmax=0.015,
xlabel style={font=\color{white!15!black}},
xlabel={$r_1$ (SP 500)},
ymin=-0.03,
ymax=0.03,
ylabel style={font=\color{white!15!black}},
ylabel={$r_2$ (Nasdaq-100)},
axis background/.style={fill=white},
xmajorgrids,
ymajorgrids,
legend style={anchor=north east, legend cell align=left, align=left, draw=white!15!black,at={(0.25,0.97)}}
]

\addplot [color=blue, dashdotted, line width=2.0pt]
  table[row sep=crcr]{%
0.00647515544459925	-0.00231637622910503\\
0.00575262701445428	-0.00352260618422464\\
0.00501009930425805	-0.00471067898825941\\
0.00425050273165406	-0.00587590586063067\\
0.00347683507728771	-0.0070136881832147\\
0.00269214965393566	-0.00811953564898753\\
0.0018995432564752	-0.00918908398321544\\
0.0011021439402497	-0.0102181121672534\\
0.0003030986760632	-0.0112025590969772\\
-0.000494439069475613	-0.0121385396101062\\
-0.00128732177924543	-0.0130223598191641\\
-0.00207242030742445	-0.0138505316895646\\
-0.00284663622879935	-0.0146197868052911\\
-0.00360691406683406	-0.015327089267841\\
-0.00435025335223978	-0.0159696476775304\\
-0.00507372046445642	-0.0165449261498732\\
-0.00577446020931274	-0.0170506543235589\\
-0.00644970708717334	-0.0174848363205311\\
-0.00709679620710225	-0.0178457586228066\\
-0.00771317380397002	-0.0181319968349482\\
-0.00829640731699795	-0.0183424213055023\\
-0.00884419498996415	-0.0184762015852173\\
-0.00935437495518381	-0.0185328097044477\\
-0.00982493376541328	-0.0185120222568087\\
-0.0102540143400064	-0.0184139212808591\\
-0.0106399232939636	-0.0182388939363331\\
-0.0109811376209486	-0.017987630976197\\
-0.0112763107038988	-0.0176611240205634\\
-0.0115242776295076	-0.0172606616432193\\
-0.0117240597856048	-0.0167878242862143\\
-0.0118748687232922	-0.0162444780225776\\
-0.0119761092685907	-0.0156327671917805\\
-0.0120273818713204	-0.014955105937008\\
-0.0120284841819427	-0.0142141686776388\\
-0.011979411850141	-0.0134128795545335\\
-0.0118803585419901	-0.012554400889787\\
-0.0117317151756439	-0.0116421207064872\\
-0.0115340683785608	-0.0106796393577366\\
-0.0112881981723527	-0.00967075531770271\\
-0.0109950748943959	-0.00861945019077635\\
-0.0106558553683534	-0.00752987299799784\\
-0.0102718783387198	-0.00640632380276661\\
-0.00984465918740854	-0.00525323674045543\\
-0.00937588395323207	-0.00407516251890388\\
-0.00886740267787662	-0.00287675045885351\\
-0.00832122210463304	-0.00166273014520294\\
-0.00773949775869774	-0.000437892761497031\\
-0.00712452544029955	0.00079292781868571\\
-0.00647873216422507	0.00202487410882373\\
-0.00580466658150007	0.00325308417973259\\
-0.00510498892102831	0.0044727108473782\\
-0.00438246049088334	0.00567894080249779\\
-0.00363993278068711	0.00686701360653257\\
-0.00288033620808312	0.00803224047890383\\
-0.00210666855371677	0.00917002280148786\\
-0.00132198313036472	0.0102758702672607\\
-0.000529376732904261	0.0113454186014886\\
0.000268022583321242	0.0123744467855265\\
0.00106706784750774	0.0133588937152503\\
0.00186460559304656	0.0142948742283794\\
0.00265748830281638	0.0151786944374373\\
0.00344258683099539	0.0160068663078378\\
0.00421680275237029	0.0167761214235643\\
0.00497708059040501	0.0174834238861142\\
0.00572041987581072	0.0181259822958036\\
0.00644388698802736	0.0187012607681464\\
0.00714462673288368	0.019206988941832\\
0.00781987361074428	0.0196411709388042\\
0.0084669627306732	0.0200020932410798\\
0.00908334032754097	0.0202883314532214\\
0.00966657384056889	0.0204987559237754\\
0.0102143615135351	0.0206325362034905\\
0.0107245414787548	0.0206891443227209\\
0.0111951002889842	0.0206683568750819\\
0.0116241808635774	0.0205702558991323\\
0.0120100898175346	0.0203952285546063\\
0.0123513041445196	0.0201439655944702\\
0.0126464772274698	0.0198174586388366\\
0.0128944441530785	0.0194169962614925\\
0.0130942263091757	0.0189441589044875\\
0.0132450352468631	0.0184008126408508\\
0.0133462757921616	0.0177891018100537\\
0.0133975483948914	0.0171114405552812\\
0.0133986507055136	0.016370503295912\\
0.013349578373712	0.0155692141728067\\
0.013250525065561	0.0147107355080601\\
0.0131018816992149	0.0137984553247604\\
0.0129042349021318	0.0128359739760097\\
0.0126583646959236	0.0118270899359759\\
0.0123652414179669	0.0107757848090495\\
0.0120260218919244	0.00968620761627101\\
0.0116420448622907	0.00856265842103978\\
0.0112148257109795	0.0074095713587286\\
0.010746050476803	0.00623149713717706\\
0.0102375692014476	0.00503308507712668\\
0.00969138862820399	0.00381906476347611\\
0.00910966428226869	0.0025942273797702\\
0.0084946919638705	0.00136340679958746\\
0.00784889868779602	0.000131460509449442\\
0.00717483310507102	-0.00109674956145941\\
0.00647515544459926	-0.00231637622910503\\
};
\addlegendentry{99\%}

\addplot [color=red, dashed, line width=2.0pt]
  table[row sep=crcr]{%
0.00535503760767827	-0.00165968514731242\\
0.00477228584295901	-0.00263256400671196\\
0.00417340375691711	-0.00359079830487972\\
0.00356075486354297	-0.00453060632881645\\
0.00293675700812607	-0.00544827908560911\\
0.00230387282512425	-0.00634019494013146\\
0.00166460001927054	-0.00720283390798327\\
0.00102146150827367	-0.00803279154726071\\
0.000376995466014525	-0.0088267923943331\\
-0.000266254694466647	-0.00958170289060156\\
-0.000905750358659459	-0.0102945437492231\\
-0.00153896772933512	-0.0109625017129948\\
-0.00216340778682108	-0.0115829406569945\\
-0.00277660615148996	-0.0121534119921619\\
-0.00337614280954011	-0.0126716643287607\\
-0.00395965166368443	-0.0131356523615852\\
-0.00452482987105547	-0.013543544941846\\
-0.00506944693147409	-0.013893732303877\\
-0.00559135349021468	-0.0141848324181457\\
-0.00608848982052633	-0.0144156964454922\\
-0.00655889395243327	-0.0145854132710722\\
-0.00700070941573409	-0.0146933131001113\\
-0.0074121925666414	-0.0147389701012792\\
-0.00779171946914736	-0.0147222040872509\\
-0.00813779230395712	-0.0146430812258241\\
-0.00844904527969732	-0.0145019137787851\\
-0.00872425002307069	-0.0142992588695541\\
-0.0089623204266842	-0.0140359162844735\\
-0.00916231693541882	-0.0137129253164169\\
-0.00932345025442429	-0.0133315606631747\\
-0.0094450844641053	-0.0128933273968043\\
-0.00952673952980563	-0.0123999550237983\\
-0.00956809319628561	-0.0118533906595126\\
-0.00956898225951634	-0.0112557913437916\\
-0.00952940321077142	-0.0106095155281171\\
-0.00944951225047429	-0.00991711376787835\\
-0.00932962467174655	-0.00918131865649474\\
-0.00917021361609	-0.00840503404111807\\
-0.00897190820611334	-0.00759132356247454\\
-0.00873549106267262	-0.00674339856407439\\
-0.00846189521622429	-0.00586460541850623\\
-0.00815220042458021	-0.0049584123208332\\
-0.00780762891159693	-0.00402839560121158\\
-0.00742954054361644	-0.0030782256107495\\
-0.007019427462695	-0.00211165223630793\\
-0.00657890819780018	-0.00113249010141042\\
-0.0061097212772164	-0.000144603511666791\\
-0.00561371836736809	0.000848108795875705\\
-0.00509285696513825	0.00184172903926631\\
-0.00454919267252253	0.00283233585334626\\
-0.00398487108410732	0.00381601976558558\\
-0.00340211931938807	0.00478889862498511\\
-0.00280323723334617	0.00574713292315288\\
-0.00219058833997202	0.00668694094708961\\
-0.00156659048455513	0.00760461370388228\\
-0.000933706301553311	0.00849652955840462\\
-0.000294433495699599	0.00935916852625643\\
0.000348705015297268	0.0101891261655339\\
0.000993171057556421	0.0109831270126063\\
0.00163642121803759	0.0117380375088747\\
0.0022759168822304	0.0124508783674963\\
0.00290913425290606	0.0131188363312679\\
0.00353357431039203	0.0137392752752676\\
0.00414677267506091	0.0143097466104351\\
0.00474630933311105	0.0148279989470338\\
0.00532981818725538	0.0152919869798584\\
0.00589499639462641	0.0156998795601191\\
0.00643961345504503	0.0160500669221501\\
0.00696152001378563	0.0163411670364189\\
0.00745865634409727	0.0165720310637654\\
0.00792906047600422	0.0167417478893453\\
0.00837087593930503	0.0168496477183845\\
0.00878235909021234	0.0168953047195524\\
0.0091618859927183	0.0168785387055241\\
0.00950795882752806	0.0167994158440973\\
0.00981921180326827	0.0166582483970583\\
0.0100944165466416	0.0164555934878272\\
0.0103324869502551	0.0161922509027466\\
0.0105324834589898	0.0158692599346901\\
0.0106936167779952	0.0154878952814479\\
0.0108152509876763	0.0150496620150775\\
0.0108969060533766	0.0145562896420715\\
0.0109382597198566	0.0140097252777858\\
0.0109391487830873	0.0134121259620647\\
0.0108995697343424	0.0127658501463902\\
0.0108196787740452	0.0120734483861515\\
0.0106997911953175	0.0113376532747679\\
0.0105403801396609	0.0105613686593912\\
0.0103420747296843	0.00974765818074771\\
0.0101056575862436	0.00889973318234756\\
0.00983206173979523	0.00802094003677939\\
0.00952236694815116	0.00711474693910637\\
0.00917779543516788	0.00618473021948475\\
0.00879970706718739	0.00523456022902267\\
0.00838959398626595	0.0042679868545811\\
0.00794907472137112	0.00328882471968359\\
0.00747988780078734	0.00230093812993996\\
0.00698388489093903	0.00130822582239746\\
0.0064630234887092	0.000314605579006868\\
0.00591935919609349	-0.000676001235073083\\
0.00535503760767827	-0.00165968514731242\\
};
\addlegendentry{95\%}

\addplot [color=black, dashed, line width=2.0pt]
  table[row sep=crcr]{%
0.00293141486055402	-0.000238788710165079\\
0.00265110083468387	-0.000706760849028736\\
0.00236302783564149	-0.00116768869190822\\
0.00206833275594605	-0.00161975316718559\\
0.00176817862244375	-0.00206117018290554\\
0.00146375000637213	-0.00249019766777127\\
0.00115624834839862	-0.00290514244630404\\
0.0008468872170831	-0.00330436692103321\\
0.000536887519477221	-0.00368629553534533\\
0.000227472682762035	-0.00404942099148631\\
-8.01361740601323e-05	-0.004392310199177\\
-0.000384725059364523	-0.00471360993136582\\
-0.000685091899974265	-0.0050120521647977\\
-0.000980051285194029	-0.00528645908432258\\
-0.00126843914508448	-0.00553574773119369\\
-0.00154911734451451	-0.00575893427701098\\
-0.00182097817486064	-0.00595513790644234\\
-0.00208294872562688	-0.00612358429339931\\
-0.00233399511873234	-0.00626360865694842\\
-0.00257312658875561	-0.00637465838489778\\
-0.00279939939303328	-0.00645629521470497\\
-0.00301192053618108	-0.00650819696309906\\
-0.00320985129433868	-0.00653015879759079\\
-0.0033924105252296	-0.00652209404485283\\
-0.00355887775097289	-0.00648403453277981\\
-0.00370859600148017	-0.00641613046487813\\
-0.00384097440721647	-0.00631864982748139\\
-0.00395549053109227	-0.00619197733213073\\
-0.00405169243028409	-0.00603661289729413\\
-0.00412920043984628	-0.00585316967541664\\
-0.00418770867107513	-0.00564237163308778\\
-0.00422698621871187	-0.00540505069387614\\
-0.00424687807222016	-0.00514214345510718\\
-0.00424730572754194	-0.00485468749154142\\
-0.00422826749691702	-0.00454381726054092\\
-0.00418983851554395	-0.00421075962488434\\
-0.00413217044505568	-0.00385682901090005\\
-0.00405549087498037	-0.00348342222102594\\
-0.00396010242454957	-0.0030920129212683\\
-0.00384638154839827	-0.00268414582531538\\
-0.00371477705087053	-0.00226143059825815\\
-0.00356580831479371	-0.00182553550397759\\
-0.00340006325171176	-0.00137818082126929\\
-0.00321819598166695	-0.000921132054688996\\
-0.00302092425168689	-0.000456192966912679\\
-0.00280902660316494	1.48015398906715e-05\\
-0.00258333929931299	0.000489992665579624\\
-0.00234475302481261	0.000967505047875264\\
-0.00209420937068957	0.00144545416355932\\
-0.00183269711828417	0.00192195376582638\\
-0.00156124833698307	0.00239512332843824\\
-0.00128093431111293	0.0028630954673019\\
-0.000992861312070543	0.00332402331018138\\
-0.000698166232375107	0.00377608778545875\\
-0.000398012098872801	0.0042175048011787\\
-9.35834828011815e-05	0.00464653228604444\\
0.000213918175172326	0.0050614770645772\\
0.000523279306487844	0.00546070153930637\\
0.000833279004093725	0.0058426301536185\\
0.00114269384080891	0.00620575560975948\\
0.00145030269763108	0.00654864481745017\\
0.00175489158293547	0.00686994454963898\\
0.00205525842354521	0.00716838678307086\\
0.00235021780876498	0.00744279370259575\\
0.00263860566865543	0.00769208234946685\\
0.00291928386808545	0.00791526889528414\\
0.00319114469843158	0.0081114725247155\\
0.00345311524919783	0.00827991891167248\\
0.00370416164230329	0.00841994327522158\\
0.00394329311232656	0.00853099300317095\\
0.00416956591660423	0.00861262983297813\\
0.00438208705975203	0.00866453158137222\\
0.00458001781790963	0.00868649341586395\\
0.00476257704880055	0.008678428663126\\
0.00492904427454383	0.00864036915105297\\
0.00507876252505112	0.00857246508315129\\
0.00521114093078741	0.00847498444575456\\
0.00532565705466322	0.0083483119504039\\
0.00542185895385503	0.0081929475155673\\
0.00549936696341722	0.00800950429368981\\
0.00555787519464608	0.00779870625136094\\
0.00559715274228281	0.00756138531214931\\
0.0056170445957911	0.00729847807338035\\
0.00561747225111288	0.00701102210981459\\
0.00559843402048797	0.00670015187881409\\
0.0055600050391149	0.00636709424315751\\
0.00550233696862662	0.00601316362917322\\
0.00542565739855132	0.00563975683929911\\
0.00533026894812051	0.00524834753954147\\
0.00521654807196922	0.00484048044358855\\
0.00508494357444148	0.00441776521653132\\
0.00493597483836465	0.00398187012225075\\
0.0047702297752827	0.00353451543954246\\
0.00458836250523789	0.00307746667296217\\
0.00439109077525784	0.00261252758518585\\
0.00417919312673589	0.0021415330783825\\
0.00395350582288394	0.00166634195269354\\
0.00371491954838356	0.0011888295703979\\
0.00346437589426052	0.000710880454713845\\
0.00320286364185511	0.000234380852446789\\
0.00293141486055402	-0.000238788710165077\\
};
\addlegendentry{50\%}

\addplot [color=mycolor1, draw=none, mark=o, mark options={solid, mycolor1}]
  table[row sep=crcr]{%
0.00572227384420554	0.00526942449173329\\
-0.000770670483320468	0.00561860528904501\\
0.00351695901278104	0.00848545933903355\\
-0.00354859422171994	0.00355892530624713\\
0	0.00204382578763651\\
0.00282963827286542	0.00298699730753582\\
-0.00214480901769987	-0.00173849900653655\\
0.00184984060760929	0.00358620259903009\\
-0.00296750268944657	-0.00293701656628054\\
0.00176375405717288	0.00222021274371476\\
-0.00360930871925835	-0.00092569516684271\\
0.00336623751423915	0.00238168051872556\\
-0.00269012501212185	0.000493758868643779\\
0.00656459355538797	0.00698025201627628\\
0.00802609062626392	0.00988229023319565\\
-0.000735384169633257	0.00105789352797414\\
-0.00086646422615233	0.00216023075178562\\
-0.00600954349152305	-0.00749410427855879\\
-0.000889905338774533	-0.00244867415607952\\
0.000298363647374122	0.00702003819980157\\
0.000570309478649111	-0.000968376917131364\\
0.00726475800164583	0.00270021611077897\\
-0.00211535686335018	0.00123214950388428\\
0.000226828953928004	0.00346750403161655\\
0.000693322494601523	0.00206333259570424\\
0.00575254631328193	0.00299814065522797\\
0.00356605033332547	0.0027876704792491\\
0.00524584494879643	0.0057646203123034\\
0.00400733512294638	0.00271076439313012\\
0.00499230897363945	0.00594192716958397\\
-0.000864117909971318	-0.000330039841427099\\
0.00167855635546665	0.00454286225392186\\
0.0060480662873772	0.00488471956267844\\
-0.00108220037652196	0.00026162841429711\\
0.000418987045779584	-0.00369194440941401\\
0.00149336406559386	0.00205160965198625\\
0.001017983014868	0.000793468084985705\\
-0.0025783762000211	-0.00322385889400789\\
0.0136738545053825	0.0113839859836191\\
-0.00585988047735997	-0.00514385424501584\\
0.000503877140836995	0.0019055974250568\\
-0.00327724059911494	-0.00245091096440631\\
-0.00291337376094525	-0.00168465515998251\\
-0.00228421554068836	0.00159406241922011\\
0.000799895477743284	0.000777997905541117\\
0.00326867042060042	0.00408650332061833\\
0.000366632792746291	0.00160974436171402\\
-0.00337902737257756	-0.00229858958258988\\
0.00837475296893309	0.00633203308217656\\
-0.00162671028891304	-0.000769890976228926\\
-0.0013143148736342	-0.000613500161148139\\
-0.00200989340901925	0.000841282141722077\\
-0.0124079728666598	-0.0149225285674701\\
0.00188988616232044	0.00656724320574975\\
-0.00106026956160588	-0.00231198003760291\\
-0.000843996075344799	0.00165445959831834\\
-0.00101958720654272	0.00191461968680096\\
0.00725147415297744	0.00612919352347685\\
0.00108532497496272	0.00426468717995476\\
0.00293511002926894	0.00174396742061078\\
-0.0022550475355152	-0.000645298093107427\\
-0.00164212562115773	-0.000741283024232819\\
0.000559522456275996	0.00151135096375055\\
-0.00305486122501353	-0.00402361869276679\\
0.00192950937952174	0.000435517999617829\\
-0.00082713012919311	-0.0004629812971636\\
0.000687686463902271	0.000610895698987868\\
-0.00143387938289363	-0.00429392240860205\\
-0.00375995075621405	-0.00395860206504584\\
-0.00681469445288307	-0.0043592730442723\\
0.00861334911528977	0.00851958262195174\\
-0.00290338013502112	-0.00139651035851251\\
-0.00171635055280095	0.00148006008708101\\
0.00755726341090091	0.00812825426999364\\
-0.00303507315136575	-0.000271878736491593\\
0.0108400689914652	0.0121241044026057\\
0.00609068744472219	0.00729120167602715\\
-0.00048570337896503	-0.00127971411856886\\
0.000552920072501184	0.0054483760144568\\
-0.00191314733596659	0.00219877666761259\\
0.0017322905313657	0.00825644346410526\\
0.00118905005056003	0.00256498940429561\\
-0.00127136050517784	-0.00335036039412406\\
0.000582102761619296	0.000206157330251866\\
0.00408869518490174	0.00351384574969726\\
3.74460771894736e-05	0.00229893259913205\\
-0.0010252486558836	0.00339989355112791\\
0.00113060139186416	0.000593507040824282\\
-0.00216280937188174	-0.00131298502712474\\
-0.00147844134212094	0.00221877958961092\\
0.00477651364260256	0.00310717622299261\\
-0.00068689681361811	0.00349899746689952\\
-0.0181782145892192	-0.0251361071970412\\
0.00368681853564956	0.00819995441291477\\
0.00676749961798429	0.0044878436728899\\
0.00516013154438255	0.00846489155216812\\
0.00183787184870754	0.000693089810843706\\
0.0024891266726228	0.0047252415750263\\
0.00444194805139286	0.0083887952874746\\
0.000310549996017651	0.00172881726122198\\
-0.00120462448281966	0.00108321185074867\\
-0.000459968685312506	-0.00100611740831003\\
0.00757111270794231	0.00478682316428802\\
0.00370773099480837	0.0111665238455647\\
-0.00121766489571795	-0.000567859684864946\\
-0.00277904015748698	-0.00363213009857599\\
0.00156825745274447	0.00355483038072246\\
0.000267204529369902	0.00131175553324292\\
-0.000829989016156052	-0.0243589738416053\\
-0.000978763197352017	-0.00587950507091517\\
0.00451150514439047	0.00764510696131904\\
-0.000995830885900939	-0.00430298596919332\\
-0.00223959817167452	-0.00457121840741148\\
0.000283639201081209	-0.00340475899965587\\
0.00834722882601913	0.0159712320239489\\
-0.0066966375083225	-0.00795363903142421\\
-0.000582644441432079	0.00979340577478149\\
-0.00045578189908535	-0.000435809404120802\\
0.00156091558841642	0.00402080765303792\\
0.000315801576723951	-0.00439764533198184\\
-0.00807282466310844	-0.0183449758224132\\
0.00880806612873708	0.014357445093619\\
-0.00860002315222397	-0.0173838427294011\\
0.00153323183664433	-0.00107908657291467\\
0.002310833991505	-0.0088472940452653\\
0.00145327890188485	0.00926571984816116\\
-0.00936882379513426	-0.00901425847991433\\
0.00640312563543932	0.0104629082379759\\
0.000927766212441172	0.00666134277523112\\
-0.000782680881929565	0.00274841780939106\\
0.00730560775918554	0.0121108256614262\\
0.00187458426276477	0.00249347502812314\\
0.00467350332149974	0.00771921925403096\\
-5.29128558237613e-05	0.000284366774319533\\
0.000597857000402824	0.0069112430318421\\
0.00537263947766098	0.00613255852774763\\
-0.000153659300766162	0.000855294459002787\\
-0.000367871603640935	5.22814536123573e-05\\
-0.00106373484696476	0.00335054162021753\\
0.00292317179866441	-0.0018043338133954\\
0.000282663821871143	0.00338581408982308\\
-0.00097268816832885	-0.00566320352515814\\
-0.00134111548933402	-0.00137059695399178\\
-0.000728145677214354	-0.00483842129820622\\
0.00244911503865652	0.00252364132678884\\
0.00049264843488217	0.00323316515930627\\
-0.00218365408505572	-0.00389396169541578\\
0.00188910352333216	0.00147846970255605\\
0.00164719979631966	0.00590175495546985\\
-0.00241443269303199	-0.00141200729068391\\
-0.000363608532139015	-0.0011744093556586\\
-0.0144744418842657	-0.0221644785971978\\
0.00127556980315369	0.00748763334337177\\
0.0100437547380208	0.0131423725549917\\
-0.000498808096269343	-7.44633288832786e-05\\
0.00142010291609984	0.0016402140979368\\
-0.0154369518977056	-0.0204650167803319\\
-0.00183536733661438	-0.000933293566307558\\
0.00116265092165913	-0.000754651148485164\\
0.00994077996224996	0.0149986068384655\\
-0.00345359283569247	-0.00366917791334798\\
-0.00207446192333016	-0.00296317439088323\\
0.00167286930487598	-0.00204135377524484\\
0.000487071888063406	0.00267071076906489\\
0.000842821903098034	0.00412122798566394\\
0.00461514890808878	0.012070636891361\\
0.00572097603609811	0.00938835930490289\\
0.00198254089142424	-0.000116921482239851\\
-0.00755080681997555	-0.00921356784564364\\
0.00312872660796426	0.00310142262702473\\
-0.000178435958468004	0.00221473506025327\\
-0.00148885069737237	-0.00854079373743033\\
0.0108392989997979	0.0113572576502401\\
0.00336394799267614	0.00252659087793528\\
0.000757120832188596	0.0014576835500959\\
-0.00110071761637232	-0.00592235329757862\\
0.00184718137532158	0.0032133950371358\\
0.00145592086692758	-0.0011489450567802\\
0.00111019536561696	0.001665233401966\\
0.000634347859560069	-0.00291603763804682\\
-0.00304591746820848	-0.00647844567816558\\
0.000647793704117605	-0.000436456817696107\\
-0.00222205044498058	-0.0109518245690592\\
7.21668174081813e-05	0.00238433803443372\\
0.00408514387806491	0.00959818621220965\\
0.00120461578316267	-0.000794944746950854\\
0.00370510975092175	0.00779191588357753\\
0.00387400355069611	0.000438197963883669\\
0.00215883815217	0.0021966420766808\\
0.00124672012828797	0.000630483258349468\\
0.00564678732118651	0.00971859468098857\\
-0.00107363431527852	0.00122659981970963\\
-0.00180443405100716	-0.000995955059515885\\
0.00232241261212307	0.000823670952704525\\
0.00180350703861598	0.00292404081152853\\
-0.00168675271867513	-0.00185155453237407\\
0.000878107223526881	0.00370016427278497\\
0.00175075343066022	0.00362409035663847\\
0.000672578655309453	0.0013214553341161\\
0.000742335162138286	-0.00134905949992259\\
0.000327997937233926	-0.00355393143207605\\
0.0051168426285273	0.00265890646206524\\
-0.00397248424591656	-0.00670992878836629\\
0.00161790853431931	0.00204193868989888\\
-0.00466304996071698	-0.00414132631872111\\
0.00127094621924906	-0.00283564136478209\\
0.00807302249303077	0.0290831194771128\\
-0.00319247048042559	0.00227242241636771\\
0.000944458796862779	0.00336730830470544\\
0.00159210991669934	1.43781925998621e-05\\
0.000189966107745132	-0.00196198621978738\\
0.00309707529371339	0.00949105808003092\\
0.0012712512706079	0.00286387986120684\\
-0.000189102832403254	0.00113562956146818\\
0.00144365490939347	0.00395999779321543\\
-0.00376188778827347	-0.00529484773222089\\
-0.00089764371357326	-0.000497470302699199\\
0.00098363433830384	0.00112700480393357\\
-0.00230960941364045	-0.00356861874929504\\
-0.00552567572366813	-0.00560570246026459\\
0.00819605830144865	0.012907578945337\\
-0.00262596311976071	-0.00388544352509879\\
0.00127568291098101	-0.000934340624788921\\
0.00654113901643671	0.0110991203324631\\
-0.000750261050562084	0.00117427004503945\\
0.00205609524528505	0.00362816883733852\\
-0.00038425774086126	-0.000517970630100839\\
0.00984851264624087	0.00258974728935724\\
-0.000369225815214147	-0.0173108814831872\\
0.00819095052417307	0.0085845214524225\\
-0.00202453064386598	-0.00434996162841172\\
-0.00105215690991378	-0.0117026573013945\\
-0.0037393815432909	0.000225053555584331\\
-0.000114105345071946	0.00445960926639222\\
0.00293235762829713	0.00369137075342119\\
0.00550630649839068	0.00447890846557875\\
0.00320195738261364	0.0077736260090373\\
0.0015489219942515	-0.00160156567920089\\
-0.000472956803357905	0.00172628827852028\\
-0.00407085926772277	-0.00074433333667856\\
0.00897434357723248	0.0119578626513634\\
0.00536280703173797	0.00726072902019825\\
-0.00323026930898673	-0.0050051814065587\\
-0.000827893291369564	-0.00126374928804773\\
0.00198565568722597	3.24390342880676e-05\\
-0.000458166473157662	-0.0011618073889752\\
-0.00105841522388495	-0.00495110977533253\\
0.000790940869240808	0.000309295268849263\\
0.00183399877188051	0.000974339385326761\\
-0.00518315329180474	-0.00698603732483072\\
};

\end{axis}

\end{tikzpicture}%

%% file: tikz/OMXH_kurtosis.tex
\begin{tikzpicture}

\definecolor{darkgray176}{RGB}{176,176,176}
\definecolor{darkorange25512714}{RGB}{255,127,14}
\definecolor{forestgreen4416044}{RGB}{44,160,44}
\definecolor{steelblue31119180}{RGB}{31,119,180}

\begin{axis}[
width=0.95\fwidth,
height=0.44\fwidth,
at={(0\fwidth,0\fwidth)},
tick align=outside,
tick pos=left,
x grid style={darkgray176},
xmin=-0.35, xmax=7.35,
xtick style={color=black},
xtick={0,1,2,3,4,5,6,7},
xtick={0,1,2,3,4,5,6,7},
xtick={0,1,2,3,4,5,6,7},
xtick={0,1,2,3,4,5,6,7},
xticklabels={2015,2016,2017,2018,2019,2020,2021,2022},
xticklabels={2015,2016,2017,2018,2019,2020,2021,2022},
xticklabels={2015,2016,2017,2018,2019,2020,2021,2022},
xticklabels={2015,2016,2017,2018,2019,2020,2021,2022},
y grid style={darkgray176},
ymin=3.67450692162609, ymax=16.0800954087213,
ytick style={color=black},
xmajorgrids,
ymajorgrids,
legend style={anchor=north east, legend cell align=left, align=left,,font=\scriptsize, draw=white!15!black,at={(0.6,1.0)}}  
]
\addplot [semithick, steelblue31119180, mark=*, mark size=2.3, mark options={solid}]
table {%
0 9.1761725162033
1 9.08259271529115
2 11.0834519147217
3 8.31067343219424
4 10.983936400014
5 5.30284656035024
6 13.5870600689427
7 10.4874226508474
};
\addlegendentry{$\hat \nu$ of portfolio using TWE \cite{ollila2023affine}}
\addplot [semithick, darkorange25512714, mark=*, mark size=2.3, mark options={solid}]
table {%
0 9.91856822040722
1 9.95698102436118
2 12.2732761325827
3 9.0424865592671
4 12.1625477587604
5 5.61229267196173
6 15.5162050229442
7 11.6888302592554
};
\addlegendentry{$\hat \nu$ of portfolio using OPP \cite[Algorithm~1]{ollila2021shrinking}}

\end{axis}

\end{tikzpicture}